\documentclass[varenna]{cimento}

\usepackage{wrapfig} 

\usepackage{amsmath,amsfonts,amssymb,graphics,graphicx,epsfig,color,times,bbm}

\usepackage[sort&compress]{natbib}
\usepackage{bm} 
\usepackage{shadow,fancybox,amsmath} 
\usepackage{amscd} 
\usepackage{supertabular}

\bibpunct{[}{]}{,}{u}{}{}

\def\be{\begin{equation}}
\def\ee{\end{equation}}
\def\bea{\begin{eqnarray}}
\def\eea{\end{eqnarray}}
\newtheorem{proposition}{Proposition}
\newcommand{\proofend}{\hfill\fbox\\\medskip }

\begin{document}


\title{Entanglement in Graph States and its Applications}

%

\author{Marc Hein}

\institute{Institut f{\"u}r Theoretische Physik, Universit{\"a}t Innsbruck,
Technikerstra{\ss}e 25, A-6020 Innsbruck, Austria}

\author{Wolfgang D\"ur}

\institute{Institut f{\"u}r Theoretische Physik, Universit{\"a}t Innsbruck,
Technikerstra{\ss}e 25, A-6020 Innsbruck, Austria \\
Institut f\"ur Quantenoptik und Quanteninformation der \"Osterreichischen Akademie der Wissenschaften, Technikerstra{\ss}e 21a, A-6020 Innsbruck, Austria}

\author{Jens Eisert}

\institute{Institute for Mathematical Sciences,
Imperial College, 48 Prince's Gardens, SW7 2PE London, UK}

\author{Robert Raussendorf}

\institute{Institute for Quantum Information, California Institute of Technology, Pasadena, CA 91125, USA}

\author{Maarten Van den Nest}

\institute{Departement Elektrotechniek (ESAT-SCD), K.U. Leuven, Kasteelpark Arenberg 10, B-3001 Leuven-Heverlee, Belgium}

\author{Hans J\"urgen Briegel}

\institute{Institut f{\"u}r Theoretische Physik, Universit{\"a}t Innsbruck,
Technikerstra{\ss}e 25, A-6020 Innsbruck, Austria \\
Institut f\"ur Quantenoptik und Quanteninformation der \"Osterreichischen Akademie der Wissenschaften, Technikerstra{\ss}e 21a, A-6020 Innsbruck, Austria}

\shortauthor{M. Hein, W. D\"ur, J. Eisert, R. Raussendorf, M. Van den Nest \atque  H.-J. Briegel}


\maketitle

\clearpage

\tableofcontents


\section{Introduction}\label{Introduction}

The recognition of the key role entanglement plays in the
understanding of the radical departure of quantum from classical
physics came historically remarkably late. In the early years of
quantum mechanics starting from the mid twenties of the last
century, often referred to as the `golden years', this aspect was
not quite in the center of activities: Researchers were occupied
with successfully applying the new theory to a wide range of
physical phenomena, continuously adding to the impressive list of
theoretical studies matching experimental findings. It was not
until the year 1935, when Einstein, Podolsky and Rosen expressed
their dissatisfaction with the state of the theory, constructing a
Gedanken experiment involving a measurement setup in a `separated
laboratories paradigm' that should identify the description
provided by quantum mechanics as incomplete \cite{EPR35}. This
Gedanken experiment involved local measurements on constituents of
a composite quantum system prepared in an entangled state in a
distributed setup. In the same year, Schr{\"o}dinger, also one of
the major contributors to the theory, formulated some of the
mathematical implications of entanglement on the statistics of
measurement outcomes; and actually coined the term `entanglement'
both German and in English \cite{schroedinger35}. The program envisioned by Einstein and
colleagues -- to demonstrate the incompleteness of a quantum
mechanical description -- may be fairly said to have essentially
failed. They nevertheless could pinpoint the aspect of quantum
theory in which it would crucially depart from a local classical
statistical theory.

This was fully realized in the 1960ies, when Bell reconsidered the
situation discussed by Einstein, Podolsky and Rosen, restated in a
setting involving spin-$1/2$ degrees of freedom due to Bohm
\cite{Bohm51}. He demonstrated the validity of bounds to
correlation functions of measurement outcomes of dichotomic
measurements, provided that they would be resulting from a `local
realistic model', meaning from a {\it local classical statistical
theory} \cite{Bell64}. These bounds are indeed violated by the
predictions of quantum mechanics. After the advent of reliable
sources of entangled states, many experiments were performed, all
consistent with the predictions of quantum theory, and none with
the bounds put forth in form of {\it Bell's inequalities} (see,
e.g., ref.~\cite{As81,Zei99}). It can be said that it is in the
role of entanglement where the departure of
quantum from classical physics is most manifest,
indicating that the intrinsic
randomness in quantum theory can
{\it not} be thought of as resulting from mere classical ignorance
in an underlying classical statistical theory.

In the meantime, it has become clear that entanglement plays a
central role also from a different perspective: it can serve as an
essential ingredient in applications of {\it quantum information
processing} \cite{NielsenBook}. For example, entanglement is
required for an established key to be unconditionally secure in
quantum key distribution \cite{BB84,Ekert91,Shor00,Curty04}.
Entanglement is also believed to be responsible for the remarkable
speedup of a quantum computer compared to classical computers
\cite{Braunstein99,Lloyd00,JL02,Miy01,Vidal03}, the underlying logic of which being based on the laws of classical
physics \cite{Feynman,Deutsch,Shor97,Grover}.

Formally, entanglement is defined by what it is not: a quantum
state is called entangled, if it is not classically correlated.
Such a {\it classically correlated state} is one that -- in the
distant laboratories paradigm -- can be prepared using physical
devices locally, where all correlations are merely due to shared
classical randomness. The kind of correlations in such
preparations are hence of the same origin as ones that one can
realize in classical systems by means of communicating over
telephone lines. This is in sharp contrast to the situation in
entangled states, which cannot be produced using local physical
apparata alone. This facet of entanglement hence concentrates on
the preparation procedure. The concept of {\it distillable
entanglement} \cite{IBMPure} in turn grasps directly entanglement
as a resource, and asks whether maximally entangled states can be
extracted, distilled, within a distant laboratories paradigm.

Part of the theoretical challenge of understanding entanglement
lies in the fact that is also (at least partly) responsible for
the quantum computational speedup: {\it state space is big}. The
dimension of state space, the set of all quantum states
corresponding to legitimate preparation procedures, grows very
rapidly with the number of constituents in a composite quantum
system. In fact, it grows exponentially. Consequently, in the
whole development of quantum information theory, it has been a
very useful line of thought to investigate situations where the
involved quantum states could be described with a smaller number
of parameters, while still retaining the essential features of the
problem at hand. So certain `theoretical laboratories' facilitated
the detailed investigation of phenomena, properties, and protocols
that arise in quantum information theory, while keeping track of a
set of states that is at most polynomially growing in dimension.
So-called stabilizer states \cite{Gottesman,NielsenBook}, Werner
states \cite{We89}, matrix-product states \cite{MPS}, or
quasi-free states \cite{Gaussian} are instances of such
`laboratories'. In the center of this review article are the {\it
graph states}, the structure of which can be described in a
concise and fruitful way by mathematical graphs. They have been
key instrumental tools in the development of models for quantum
computing, of quantum error correction, and of grasping the
structure of bi- and multi-partite entanglement.

{\it Graph states} are quantum states of a system embodying
several constituents, associated with a graph\footnote{Note that
in the literature one finds several inequivalent concepts of
quantum states that are in one way or another associated with
graphs \cite{Graph,Graph2}. For example, entanglement sharing
questions have been considered in a multi-partite quantum system
based on quantum states defined through mathematical graphs, see
refs.~\cite{Rings,Buzek,Zanardi,Parker}.}. This graph may be
conceived as an interaction pattern: whenever two particles,
originally spin-$1/2$ systems, have interacted via a certain
(Ising) interaction, the graph connecting the two associated
vertices has an edge. Hence, the adjacency matrix of a simple
graph, a symmetric $N\times N$ matrix for a system consisting of
$N$ qubits with entries taken from $\{0,1\}$, fully characterizes
any graph state at hand \cite{Briegel01,OneWay3,
He04,Du03a,Nest04a}.
In this sense the graph can be understood as a summary of the
interaction history of the particles. At the same time, the
adjacency matrix encodes the stabilizer of the states, that is, a
complete set of eigenvalue equations that are satisfied by the
states \footnote{In some sense, this graphical representation
plays a similar pedagogical role as Feynman diagrams in quantum
electrodynamics: The latter provide an intuitive description of
interaction processes in spacetime, but, at the same time, they
have a concise mathematical meaning in terms of the corresponding
propagator in an expansion of the scattering operator.}. Thus
graph states are actually stabilizer states \cite{Gottesman}. This
class of graph states play a central role in quantum information
theory indeed.

To start with, graph states form a universal resource for quantum
computing based on measurements \cite{OneWay1,OneWay2,OneWay3,
OneWay5}. In such {\it one-way computing}, one starts off with a
cluster state, which is a specific instance of a graph state, and
performs von-Neumann measurements at single sites associated with
vertices. In fact, it was in the form of such {\it cluster states}
\cite{Briegel01}, when graph states have first been considered, in
a narrower sense, with respect to a graph reflecting a cubic
lattice. On the subspace that is retained any unitary can be
implemented, thereby realizing universal computation without the
need of making use of any controlled two-system quantum gates. The
cluster state hence forms a universal resource for quantum
computation. The performed measurements introduce a probabilistic
aspect to the scheme; yet, the overall set-up of the one-way
computer is deterministic, as the process can be de-randomized by
means of appropriate local unitaries taken from the Pauli group
applied before the readout step \cite{OneWay1,OneWay3}.

Such one-way computing is interesting at least from two
perspectives: on the one hand, it provides a {\it computational
model} \cite{OneWay2} different from the original gate-based
model, which resembles more closely the gate model of classical
computation. In fact, questions of complexity and simulatability
often become more transparent in the picture of the one-way
computer than in the gate model. For instance, operations from a
certain set of operations, the so-called {\it Clifford operations}
\cite{NielsenBook}, can be kept track of classically in an
efficient manner. When measuring Pauli operators in the course of
the computation, the states of the whole system will be a sequence
of graph states that can be described by a sequence of adjacency
matrices \cite{OneWay3}. However, if in the first step of the
gate-based model, a non-Clifford operation is applied, unlike in
the picture of the one-way computer, it is no longer obvious in
the gate-model that the dynamics of the rest of the network can be
described efficiently.

On the other hand, there are good reasons to believe that the
undoubtedly tremendous challenges associated with actually
realizing a quantum computer can be lessened when relying on an
{\it architecture based on graph states}. In many quantum systems
in a lattice, nearest-neighbor interactions are natural, rendering
the preparation of cluster states through one dynamical process a
relatively feasible step. Furthermore, and maybe more importantly,
one realizes a certain distinction between the process of creating
entanglement and the process of consuming it. Hence, even if one
exploits a lossy or even probabilistic process in order to prepare
the graph states, in many set-ups one can, in principle, end-up
with graph states free of errors, then to be used as a resource of
the actual computation. Even if the entangling operations are
themselves faulty, e.g. subject to photon loss in a linear optical
implementation, fault tolerant schemes can be devised, in
principle up to a loss rate of several tens of percents.

In quantum-gate-based quantum computation, graph states also play
a prominent role as {\it codewords in quantum error correction},
allowing for reliable storage and processing of quantum
information in the presence of errors. This is done by
appropriately encoding quantum information in quantum states of a
larger number of quantum systems. This is the second branch how
the concept of graph states originally came into play, namely in
form of {\it graph codes} \cite{Schlinge02a,
SchlingeHabilschrift}. These instances of quantum codes are determined by the underlying
graph.

Finally, the idea of the `theoretical laboratory' itself allows
for a wide range of applications. Aspects of bi-partite and, in
particular, {\it multi-partite entanglement} are typically
extraordinarily hard to grasp \cite{Du99}. This is again partially
due to the fact that state space is so rapidly increasing in
dimension with a larger number of constituents. The {\it decision
problem} whether a mixed state is entangled or classically
correlated is already provably a computationally hard problem in
the dimension of the constituents. Apart from the {\it
classification} of multi-particle entanglement, even for pure
states, a complete meaningful {\it quantification} of
multi-particle entanglement is yet to be discovered \cite{LiPo98}.
Already at the origin is the question in what units to phrase any
result on quantification matters: there is no multi-particle
analog of the maximally entangled pair of spin-$1/2$-particles, to
which any pure bi-partite entanglement is equivalent: any pure
state can be transformed into such maximally entangled pairs in an
asymptotically lossless manner, one of the key results of
entanglement theory \cite{pureState,Th02}. This means that the
achievable rate of distillation and formation is the same. The multi-partite analogue of such a {\it reversible
entanglement generating set} (as it is formed by the maximally entangled qubit pair in the bi-partite setting) has not yet been identified, at least none that has a finite
cardinality\footnote{For a brief review on multi-particle
entanglement, see, e.g., ref.~\cite{Eisert05}.}.

Within the setting of graph states, some of the intriguing
questions related to {\it aspects of multi-particle entanglement}
can be addressed, expressing properties of entanglement in terms
of the adjacency matrix. Central tasks of interconverting
different forms of multi-particle entanglement, in particular of
{\it `purifying'} it, can and have been studied: here, the aim is
to extract pure graph states from a supply of noisy ones, having
become mixed through a decoherence process \cite{Du03a}. Based on
such purification protocols, even protocols in quantum
cryptography have been devised, making use of the specific
structure of this class of quantum states \cite{Lo04,Du05c}. In
turn, graph states form an ideal platform to study the {\it
robustness of multi-particle entangled states} under such in
realistic settings anyway unavoidable decoherence processes. In a
nutshell, multi-particle entangled states may exhibit a surprising
robustness with respect to decoherence processes, independent from
the system size \cite{Du04b}.
Finally, questions of inconsistency of quantum mechanics with
local classical statistical theories become very transparent for
graph states \cite{Gue04}. In all of the above considerations, the
class of graph states is sufficiently restricted to render a
theoretical analysis feasible, yet often complex enough to
appropriately grasp central aspects of the phenomenon at hand.

Departing slightly from the original formulation, {\it weighted
graph states} have been considered \cite{CDHB05}, where the interaction is no
longer a fixed Ising interaction with a constant weight. Such
states can be thought of as resulting from a {\it semi-classical
Boltzmann gas}: classical particles carrying a quantum spin-$1/2$
degree of freedom would in an idealized description give rise to
such a graph state through collision of particles. Such weighted
graph states find numerous applications in the description of
random quantum systems. They can also be taken as a basis set of
states to approximate ground states of many-body systems in a
variational principle. Owing the structural similarity to the fact
that mathematically, discrete and continuous Weyl systems have so
much in common, graph states in {\it harmonic infinite-dimensional
quantum systems} have finally been studied that resemble very much
the situation of graph states for finite-dimensional quantum
systems \cite{HC,Pl04}.

This review article aims at providing a self-contained
introduction to the theory of graph states in most of these
aspects, both conceived as a tool and as a central resource in
quantum information theory in their own right. We will introduce
the basic notions of graph states, discuss possible descriptions
of their entanglement content, their interconversion and
purification, and various applications in the context of quantum
information science.

 \subsection{Outline}

We start with a detailed introduction of graph states, whose entanglement properties are analyzed in the following chapters. After setting basic notations in sec.~\ref{GS_Notations} that are frequently used throughout this article, we give essentially two alternative definitions for graph states in sec.~\ref{DefOfGS}, namely, in terms of the underlying interaction pattern and in terms of the stabilizer. We illustrate how elementary properties of a graph state and basic operations, such as (Pauli) measurements, on these states, can be phrased concisely in terms of the underlying graph. It is shown that the action of Clifford operations on graph states (sec.~\ref{Pauli measurements}) and the reduced states for graph states (sec.~\ref{Reduced_GS}) can be determined efficiently from the underlying graph. These relations will allow for a classification of graph states in sec.~\ref{Local_Equivalence} and for an efficient computation of entanglement properties in sec.~\ref{EntanglementGS}. We discuss some examples and applications to quantum error correction, multi--party quantum communication, and quantum computation in sec.~\ref{GS_Examples}. We briefly discuss some generalizations of
graph states in the language of discrete Weyl systems in
sec.~\ref{DefOfGS}.
Sec.~\ref{Implementations} contains also a short review about possible realizations of graph states in
physical systems.

In sec.~\ref{Local_Equivalence} we will then discuss the classification of graph states in terms of equivalence classes under different types of local operations and provide a complete classification for graph states with up to seven vertices. The results indicate that graph states form a {\em feasible and sufficiently rich class of multi-party entangled states}, which can serve as good starting point for studies of multi-party entanglement.

In  sec.~\ref{EntanglementGS} we discuss various aspects of entanglement in graph states. We briefly review some results about the `non-classicality' of graph states and how their entanglement can be detected with Bell inequalities. The genuine multi-particle entanglement of graph states  is characterized and quantified in terms of the Schmidt measure, to which we provide upper and lower bounds in graph theoretical terms.

Finally, we introduce two possible extensions of graph states. On the one hand, in  sec.~\ref{WeightedGS} the class of {\em weighted graph states} is introduced, which comprises particles that interact for different times and provides an interesting model for the study of the entanglement dynamics in many--particle systems. Here, we consider $N$ initially disentangled spins, embedded in a ring or $d$-dimensional lattice of arbitrary geometry, which interact via some long--range Ising--type interaction. We investigate relations between entanglement properties of the resulting states and the distance dependence of the interaction in the limit $N \to \infty$ and extend this concept to the case of spin gases.

On the other hand, in sec.~\ref{GS_decoherence}, {\em graph diagonal states} serve as standard forms for mixed states and occur naturally whenever pure graph states are exposed to decoherence. We show that the lifetime of (distillable) entanglement for GHZ-type superposition states decreases with the size of the system, while for a class of other graph states the lifetime is independent of the system size. These results are largely independent of the specific decoherence model. Finally, the concept of entanglement purification is applied to graph states and possible applications to quantum communication are described.


\subsection{Notations}\label{GS_Notations}

\begin{wrapfigure}[9]{r}{0.27\textwidth}
\vspace{-0.5cm}
\hspace{0.05cm}\includegraphics[width=0.25\textwidth]{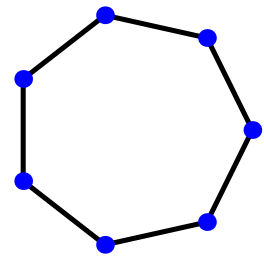}
\caption{This ring depicts a graph with $7$ vertices}\label{Ring}
\end{wrapfigure}
At the basis this review article lies the concept of a graph \cite{Graph,Graph2}. A graph is a collection of vertices and a description which of the vertices are connected by an edge. Each graph can be represented by a diagram in a plane, where a vertex is represented by a point and the edges by  arcs joining two not necessarily distinct vertices.
In this pictorial representation many concepts related to graphs can be visualized in a transparent manner. In the context of the present article, vertices play the role of physical systems, whereas edges represent an interaction.
Formally, an (undirected, finite) {\em graph}\index{graph $G$} is a pair
\begin{equation}
    G=(V,E)
\end{equation}
of a finite set $V=\{ 1,\ldots ,N \}$ and a set $E\subset [V]^2$, the elements of which are subsets of $V$ with two elements each \cite{Graph}. The elements of $V$ are called {\em vertices}\index{vertex set $V$}, the elements of $E$ {\em edges}\index{edge set $E$}. In the following, we will mainly consider {\em simple} graphs\index{simple graph}. A simple graph contains neither loops (edges connecting vertices with itself) nor multiple edges. We also regard a generalization of these simple graphs, where each edge $\{a,b\}$ is associated with a weight $\varphi_{ab}$ representing the strength of the respective interaction. Although the concept of a {\em weighted graph}\index{weighted graph} is more general than that of a simple graph, we will use the notion of a graph in the more narrow sense of a simple graph, unless we explicitly mention that a certain section is devoted to weighted graphs.

Since there are in general
\begin{equation}
 \tbinom{N}{2}= \tfrac{N(N-1)}{2}
 \end{equation}
 different possibilities for choosing set of edges $E$ in a graph of $|V|=N$ vertices, the number of distinct graphs is $2^{\tbinom{N}{2}}$. Graph theory is mostly interested in problems that are invariant under permutations of the vertices, when these permutations respect the neighborhood relation, i.e., map neighbored vertices onto neighbored vertices. Such permutations are called {\em graph isomorphisms}\index{graph isomorphism}. Two graphs $G_1=(V_1,E_1)$ and $G_2=(V_2,E_2)$ are called {\em isomorphic} if there exists a bijection $f:V_1\mapsto V_2$ such that \be \{a,b\} \in E_1 \hspace{1cm} \Longleftrightarrow \hspace{1cm} \{f(a),f(b)\} \in E_2\; .\ee
Note that the number of non-isomorphic graphs still grows exponentially with the number $N$ of vertices \cite{Harary73}.

Vertices $a,b\in V$ that are the endpoints of an edge are referred to as being {\em adjacent}\index{adjacent}. The adjacency relation gives rise to an {\em adjacency matrix}\index{adjacency matrix $\mathbf{\Gamma}$}
$\mathbf{\Gamma}_G=\mathbf{\Gamma}$ associated with a graph. $\mathbf{\Gamma}$ is a symmetric $N\times N$-matrix, with elements
\begin{equation}
    \mathbf{\Gamma}_{ab} =
    \left\{
    \begin{array}{ll}
    1,& \text{ if $\{a,b\}\in E$,}\\
    0 & \text{otherwise}.
    \end{array}
    \right.
\end{equation}
In the case of weighted graphs, the adjacency matrix also specifies the weights of the edges, i.e., $\mathbf{\Gamma}_{ab}=\varphi_{ab}$.
We will make repeated use of the {\em neighborhood}\index{neighborhood $N_a$} \be N_a:=\left\{ b\in V\,| \, \{a,b\} \in E\right\} \ee of a given vertex $a\in V$. The neighborhood is the set of vertices adjacent to a given vertex. The number $|N_a|$ of neighbors is called the {\em degree}\index{degree of a vertex} of the vertex $a$. A vertex $a\in V$ of degree $|N_a|=0$ will be called an {\em isolated vertex}\index{isolated vertex}.
An $\{a,b\}$-path is an ordered list of vertices $a=a_1,a_2,\ldots,a_{n-1},a_n=b$, such that $a_i$ and $a_{i+1}$ are adjacent for all $i$.  A {\em connected graph}\index{connected graph (state)} is a graph that has an $\{a,b\}$-path for any two $a,b\in V$. Otherwise it is referred to as {\em disconnected}\index{disconnected graph (state)} .

When a vertex $a$ is deleted in a graph $G$, together with all edges incident with $a$, one obtains a new graph, denoted by $G\setminus a$. For a subset of vertices $U\subset V$ of a graph $G=(V,E)$ let us denote with $G\setminus U$ the graph that is obtained from $G$ by deleting the set $U$ of vertices and all edges which are incident with an element of $U$. In a mild abuse of notation, we will also write $G\setminus F$ for the graph that results from a deletion of all edges $e\in F$, where $F\subset E\subset [V]^2$ is a set of edges.
For a set of edges $F\subset [V]^2$ we will write $G\cup F:= (V,E \cup F)$  and $G + F := (V, E + F)$, where
\begin{equation}\label{+}
    E+F= (E \cup F) \setminus (E \cap
    F)
\end{equation}
is the symmetric difference of $E$ and $F$. Note that the symmetric difference corresponds to the addition modulo $2$ or the component-wise XOR if the sets are considered as  binary vectors over the integer field $\mathbb{F}_2$ modulo two \index{binary field $\mathbb{F}_2$}.
Similarly, an induced {\em subgraph} $G[A]$\index{induced subgraph $G[A]$} of a graph $G=(V,E)$, where $A\subset V$, is obtained by deleting all vertices (and the incident edges) that are not contained in $A$.

Graphs may be colorable. A proper {\em two-coloring of a graph}\index{coloring}\index{two-colorable graph (state)}\index{bi-partite graph (state)} is a labeling $V\longrightarrow  \{1,2\}$, such that all adjacent vertices are associated with a different element from $\{1,2\}$, which can be identified with two colors.
In graph theory these graphs are also called `bi-partite graphs', since the set of vertices can be partitioned into two disjoint sets,  often called {\em sinks} or {\em sources}, such that no two vertices within the same set are adjacent. It is a well known fact in graph theory that a graph is two-colorable if and only if (iff) it does not contain any cycles of odd length.

In the remainder of this article each vertex stands for a two--level quantum system $\mathbf{H}^a\simeq \mathbb{C}^2$ or qubit. The state vector of the single--qubit system $\mathbf{H}^a$  can be written as $|\psi\rangle^a = \alpha |0\rangle^a + \beta |1\rangle^a$ with $|\alpha|^2+|\beta|^2=1$. The vectors $|0\rangle$ and $|1\rangle$ are the eigenvectors of the Pauli matrix $\sigma_z$ with eigenvalue $+1$ and $-1$.  The matrices $\sigma^a_0=\mathbf{1}_a$, $\sigma^a_1=\sigma_x^a$, $\sigma^a_2=\sigma_y^a$ and $\sigma^a_3=\sigma_z^a$ are the Pauli matrices of this two--level system, where the upper index specifies the Hilbert space on which the operator acts. Note that these operators form an orthogonal basis of Hermitian operators with respect to the scalar product $\langle A,B\rangle := \text{tr} (A^\dagger B)$. Up to the phase factors $\pm 1$ and $\pm i$ they also generate the {\em Pauli group}\index{Pauli group $\mathcal{P}$} $\mathcal{P}:=\langle \{\pm 1, \pm i\}\times \{\sigma_0,\sigma_x,\sigma_y,\sigma_z\} \rangle$. We will frequently use the projectors onto the eigenvectors of the Pauli operators. For example,
\begin{equation}
P^a_{z,\pm} = \frac{1\pm \sigma_z^a}{2}
\end{equation}
denotes the projector onto the eigenvector $|z,\pm \rangle$ of
$\sigma_z^a$ with eigenvalue $\pm 1$ (similarly for
$\sigma_x^a$ and $\sigma_y^a$).

To simplify notations, we use subsets $U\subseteq V$ as an upper index for states, operators and sets. They denote the respective tensor product of a given state or sets,  e.g.
\be \label{PlusState} |+ \rangle^{V} = \bigotimes_{a \in V} |+\rangle^{a} \hspace{0.7cm}\text{or}\hspace{0.7cm} \mathcal{P}^V =  \bigotimes_{a \in V} \mathcal{P}^{a} \; ,\ee
 where $|+\rangle=\frac{1}{\sqrt{2}}\left(|0\rangle + |1\rangle \right)$. The subsets are also used to label those vertices where the operator acts non-trivially, for example
\be
\sigma_z^U=\bigotimes_{b \in U} \sigma_z^{b}\; .
\ee
Moreover, we identify sets $U$ and their corresponding {\em binary vectors}\index{binary vector $U$} $U=(U_b)_{b\in V} \simeq (U_1,\ldots, U_{N})$ over the {\em binary field } $\mathbb{F}_2^V$ with the same symbol. Finally, $a$ refers to both the vertex and the corresponding one-element set $\{a\}$ ensuring that $\sigma_x^a \equiv \sigma_x^{\{a\}}$.
These notations allow us to use set and binary operations in the same formula. For example, for $A,B \in \mathcal{P}(V) \cong \mathcal{F}_2^V$ we will write $A\cup B$, $A\cap B$ and $A\setminus B$ ($\bar{A}:= V\setminus A$) for the union, intersection and difference (complement) as well as $A+B$ and $\langle A, B\rangle$ for the addition and the scalar product modulo $2$.

In the multi-partite case one can group the vertices into different partitions and, for example, study the entanglement with respect to these partitions. Here, any tuple $(A_1,...,A_M)$ of disjoint subsets $A_i \subset V$ with $\bigcup^M_{i=1} A_i =V$ will be called a {\em partition}\index{partition} of $V$. We will write
\begin{equation}
    (A_1,...A_M) \leq     (B_1,...,B_{M'}),
\end{equation}
if $(A_1,...A_M)$ is a {\em finer partition}\index{finer partition} than $(B_1,...,B_{M'})$. which means that every $A_i$ is contained in some $B_j$. The latter is then a {\em coarser partition}\index{coarser partition}.


\section{Definitions for graph states}\label{DefOfGS}

With the notations introduced in the previous section we can provide some definitions for graph states. Throughout this article, we mainly consider two alternative descriptions. Most naturally, graph states can be regarded as the result of an interaction of particles initially prepared in some product state. Certainly not all imaginable interaction patterns can be represented reasonably by a simple graph. In sec.~\ref{DefOfGS_Int} we introduce the description of graph states in terms of the interaction pattern and show that such a definition is also meaningful if all particles interact with the same Ising-type interaction but possibly for different interaction times. This description generalizes to so called weighted graph states, which are introduced in sec.~\ref{WeightedGS}. The alternative definition proposed in sec.~\ref{DefOfGS_Stab}, on the other hand, is restricted to the class of states that correspond to a simple graph. Such states can be described  efficiently in terms of their stabilizer, which is a subgroup of the Pauli group. We briefly address the question of local unitary equivalence, discuss the relation to stabilizer states and illustrate an alternative representation of the stabilizer formalism in terms of its binary representation\footnote{Although the remainder of the article will not be based on the binary representation.}. We sketch a possible extension of the stabilizer formalism to $d$-level systems and finally summarize further alternative approaches to graph states in sec.~\ref{DefOfGS_Alternative}.


\subsection{Interaction pattern}\label{DefOfGS_Int}

In this subsection we give a careful motivation for the concept of graph states in terms of interaction patterns, concluding with a precise definition given at the end of this subsection.
Let us consider a set of $2$-level systems (qubits) that are labeled by the vertices $V$ of a graph $G=(V,E)$. The qubits are prepared in some initial state vector $|\Psi\rangle$ and then are coupled according to some interaction pattern represented by the graph $G$. For each edge $\{a,b\}\in E$ the qubits of the two adjacent vertices $a$ and $b$ interact according to some (non-local) unitary $U_{ab}=e^{-i\varphi_{ab} H_{ab}}$. Here, $H_{ab}$ denotes the interaction Hamiltonian and $\varphi_{ab}$ represents the coupling strength or (with appropriate physical units) the interaction time. The most general of such setups, in which the qubits can interact according to different $2$-body interactions $H_{ab}$, has to be described by graphs, whose edges carry a labeling that specifies both the different unitaries $U_{ab}$ as well as the ordering in which interactions occur.

Under which conditions can the outcome of this interaction pattern be completely specified by a {\it simple} graph $G$? If the graphs shall give a sufficient characterization for a large class of interaction patterns, we can pose the following constraints:
\begin{itemize}
\item[{\bf (1)}] Since the graph $G$ does not provide any ordering of the edges, all two--particle unitaries $U_{ab}$ involved must commute:
\be\label{Constraint1} [ U_{ab},U_{bc}] = 0 \hspace{1cm} \forall a,b,c \in V \; .\ee
\item[{\bf (2)}] Because we deal with {\em undirected} graphs\footnote{In a {\em directed} graph the set of edges $E$ is given by ordered pairs $(a,b)$. The order implies that vertices $a$ and $b$ are connected by a directed edge from $a$ to $b$.}, the unitaries must be symmetric:
\be\label{Constraint2} U_{ab}=U_{ba}  \hspace{1cm} \forall a,b \in V \; .\ee
\item[{\bf (3)}] If the edges are not further specified by weights, all qubits should interact through the same two--particle unitary $\mathbf{U}$:
\be\label{Constraint3} U_{ab}=\mathbf{U}^{\{a,b\}} \hspace{1cm} \forall a,b \in V \; . \ee

\end{itemize}
In the case of qubits condition {\bf (1)} is already sufficient to restrict the analysis of particles to the case where the qubits interact according to the same {\em Ising interaction}\index{Ising interaction $H^I_{ab}$, $U^I_{ab}$}, e.g. $H^I_{ab}=\sigma_z^a\sigma_z^b$. This statement is reflected in the following proposition.

{\proposition[\bf Standard form for commuting interactions] With an appropriate choice of the local basis in each individual qubit system, any set of commuting two--particle unitaries, i.e., the unitaries fulfill {\bf (1)}, does only contain interactions of the form \be\label{Ising+LU}\varphi_{ab}\,  H_{ab} =  \varphi_{ab}\, \sigma_z^a\sigma_z^b + \alpha_a\, \sigma_z^a + \alpha_b\, \sigma_z^b \; .\ee
In other words, any interaction pattern in which the qubits interact according to some two--particle unitaries chosen from a set of commuting interactions, is up to local $z$-rotations\footnote{I.e., $V^a=e^{i\beta_a \sigma_z^a}$ to be performed before or after the interaction pattern.} an Ising interaction pattern
\be\label{IsingUnitary} U^I_{ab}(\varphi_{ab}):= e^{-i \,\varphi_{ab} \,\sigma_z^a \sigma_z^b} \; .\ee
}

{\em Proof:} 
It suffices to consider condition {\bf (1)} for two different unitaries $U=e^{-iH}$ and $\tilde{U}=e^{-i\tilde H}$ in the two settings of three vertices $a$, $b$ and $c$: \\
(i) $U_{ab}=U^{\{a,b\}}$ and $U_{bc}=\tilde U^{\{b,c\}}$: $[ H^{\{a,b\}},\tilde H^{\{b,c\}}] = 0 $,\\
(ii) $U_{ab}= \tilde U^{\{a,b\}}$ and $U_{bc}= U^{\{b,c\}}$: $[ \tilde H^{\{a,b\}},H^{\{b,c\}}] = 0 $.\\
Note that here $H$ and $\tilde H$ denote the complete Hermitian generator that includes the interaction time or coupling strength $\varphi$. We have also used the fact that $[f(A),f(B)]=0$ iff $[A,B]=0$. Every such Hermitian operator $H$ allows for a real decomposition with respect to the basis of Pauli operators $\{\sigma_0, \sigma_x,\sigma_y,\sigma_z\}$, i.e., $H^{\{a,b\}}= \sum_{ij}\, A_{ij}\, \sigma_i^a \sigma_j^b$. Moreover,  a local unitary transformation at a single qubit system translates to an orthogonal transformation of the corresponding operator basis $\sigma_i \mapsto \sigma_i'$, i.e., $A'=OAO^T$ for some orthogonal matrix $O$. By local unitaries we can thus diagonalize one of the Hamiltonians, say $H^{\{a,b\}}=\sum_i A_i \sigma_i^a\sigma_i^b$ and represent the other Hermitian matrix $\tilde H$ with respect to this basis, i.e., $\tilde H^{\{b,c\}}= \sum_{jk}\, B_{jk}\, \sigma_j^b \sigma_k^c$. With these decompositions (i) reads
\be \sum_{ijk} \,A_i B_{jk} \, \sigma^a_i \otimes [\sigma_i,\sigma_j]^b \otimes \sigma_k^c = 0 \; ,\ee from which
\be \forall i,j=1,2,3 \;\text{with}\; i\neq j \; :\hspace{0.7cm}  A_i =0 \vee B_{jk}= 0 \;\; \forall k=0,1,2,3 \; \ee
follows. If $H$ corresponds to a non-trivial two-body interaction up to a (local) change of basis we can assume that $A_3\neq 0$. Rewriting (ii) with these decompositions one essentially arrives at two different cases:  If another component, say $A_2\neq 0$, then all components in $B$ except $B_{00}$ have to vanish, which would imply that $\tilde H$ is a trivial interaction. If instead $A_1=A_2=0$, then at least all components in $B$ apart from $B_{00}$, $B_{03}$, $B_{30}$ and $B_{33}$ have to vanish. Since the component $B_{00}$ correspond to some negligible global phase factor, we thus have shown that any two commuting interaction Hamiltonians have to be of the form eq.~(\ref{Ising+LU}). Any terms due to $B_{03}$ or $B_{30}$ correspond to local $z$-rotations and all these rotations commute with the Ising interaction terms $H^I_{ab}=\sigma_z^a\sigma_z^b$. Thus the interaction pattern can alternatively be described by an interaction pattern with pure Ising interaction according to the same graph and some local $z$-rotations to be applied before or after the coupling operation.
\proofend

The remainder of this article is largely devoted to the entanglement properties of states that result from an interaction pattern described by a simple or weighted graph. We can omit the $z$-rotations, since they do not change these entanglement properties. In the following we thus consider an interaction pattern of qubits that are coupled only by pure Ising interactions. Note that the Ising interaction $H^I_{ab}=\sigma_z^a\sigma_z^b$ is already symmetric and hence {\bf (2)} does not yield an additional constraint.
Without condition {\bf (3)} the state, which results from the application of the interaction pattern, is determined by (a) the initial state vector $|\Psi\rangle$ and (b) by a weighted graph. This weighted graph identifies the pairs $\{a,b\}$ of qubits which interact together with the interaction strength $\varphi_{ab}$ (interaction time) of the respective interactions.
The resulting states are {\em weighted graph states} \index{weighted graph state \texttt{"|}$G\rangle$} as they are introduced in sec.~\ref{WeightedGS}. However, in the remainder of this section we will restrict to states that can be described by simple graphs. Now {\bf (3)} implies that we have to fix the interaction strength $\varphi$ in eq.~(\ref{IsingUnitary}). For graph states according to simple graphs we will from now on choose $\varphi=\frac{\pi}{4}$. Together with the choice of
\be |\Psi\rangle = |+\rangle^V \ee for the initial state this ensures that this interaction creates maximal entanglement between to qubits in the state with state vector $|+\rangle$, i.e., $U^I_{ab}|+\rangle|+\rangle$ is {\em maximally entangled}\footnote{A state vector $|\Psi\rangle^{ab}$ is maximally entangled\index{maximally entangled} iff the reduced state at one qubit is maximally mixed, i.e., $\text{tr}_a |\Psi\rangle^{ab}\langle \Psi| = \frac{1}{2}\mathbf{1}_b$.}. In sec.~\ref{DefOfGS_Stab} we will see that this choice also allows for an efficient description of the resulting states in terms of their stabilizer.

To further simplify notations we will not use the Ising interaction in eq.~(\ref{IsingUnitary}) but rather the {\em (controlled) phase gate}\index{phase gate $U_{ab}$}
\be\label{PhaseGate} U_{ab}(\varphi_{ab}):= e^{-i \varphi_{ab} H_{ab}} \hspace{0.7cm} \text{with} \hspace{0.7cm} H_{ab}:= |1\rangle^a\langle 1| \otimes |1\rangle^b\langle 1|\ee
as the elementary two-qubit interaction. Note that the corresponding interaction strength now is $\varphi_{ab}=\pi$, because from
\be H_{ab}=\frac{\mathbf{1}_a-\sigma_z^a}{2}\,\frac{\mathbf{1}_b-\sigma_z^b}{2}= \frac{1}{4}\left(\mathbf{1}_{ab}  - \sigma_z^a -\sigma_z^b + H^I_{ab} \right)\ee we find
\be U_{ab}(\varphi_{ab})= e^{-i\frac{\varphi_{ab}}{4}}\, e^{i\frac{\varphi_{ab}}{4} \sigma_z^a}\, e^{i\frac{\varphi_{ab}}{4} \sigma_z^a}\, U^I_{ab}(\frac{\varphi_{ab}}{4})\; .\ee
In other words, the phase gate corresponds to the Ising interaction up to some additional $\frac{\pi}{4}$--rotations around the $z$-axes at each qubit. For simple graphs, i.e., $\varphi_{ab}=\pi$, we find that
\begin{equation}\label{CPhase}
U_{ab}:=U_{ab}(\pi)=\;P^a_{z,+}\otimes {\mathbbm{1}}^b + P^a_{z,-}\otimes \sigma_z^b\; .
\end{equation}
This gate corresponds to a controlled $\sigma_z$ on qubits $a$ and $b$, i.e.
\begin{equation}\nonumber
U_{ab} \; \stackrel{\cdot}{=} \; \left(
\begin{array}{cccc}
1 & 0 & 0 & 0 \\
0 & 1 & 0 & 0 \\
0 & 0 & 1 & 0 \\
0 & 0 & 0 & -1
\end{array}
\right)\, .
\end{equation}
The choice $\varphi_{ab}=\pi$ ensures not only that the state vector
\be\label{Graph_Bell_State}  U_{ab} |+\rangle^a |+\rangle^b = \frac{1}{\sqrt{2}}\, \left( |0\rangle^a |+\rangle^b \, +\, |1\rangle^a |-\rangle^b \right) \ee
is maximally entangled ({\em Bell state}) but also that $U^2_{ab}=\mathbf{1}_{ab}$ or $U_{ab}={U_{ab}}^\dagger$. Consequently, the phase gate $U_{ab}$ {\em creates} as well as {\em deletes} the edge $\{a,b\}$ in a graph $G$ depending on whether the edge is already contained in $G$ or not.
We conclude the above findings into our first definition for graph states:

\begin{wrapfigure}[11]{r}{0.4\textwidth}
\vspace{-0.1cm}\includegraphics[width=0.4\textwidth]{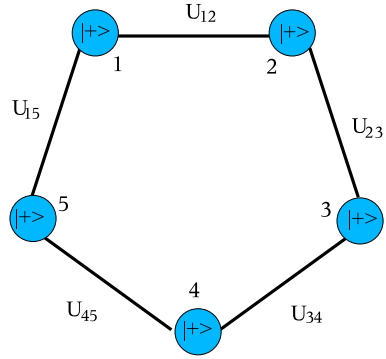}
\caption{The preparation procedure to obtain a graph state that corresponds to a ring graph with $5$ vertices.}\label{Ring1}
\end{wrapfigure}
\hspace{2.5cm}{\bf Graph states (I)} \\
{\em Let $G=(V,E)$ be a graph. The {\em graph state}\index{graph state \texttt{"|}$G\rangle$} $|G\rangle$ that corresponds to the graph $G$ is the pure state with state vector
\be \label{GS_Preparation} |G\rangle =  \prod_{\{a,b\}\in E} U_{ab}\, |+ \rangle^{ V} \; .\ee
We will also refer to the state
vector $|G\rangle$ of the pure state as a graph state.
The {\em preparation procedure}\index{preparation procedure for graph states}\index{graph state preparation} reads:
\begin{itemize}
\item[1.] Prepare the qubits at each vertex in the pure state with state vector $|+\rangle$ as eigenvector  of $\sigma_x$ with eigenvalue $+1$.
\item[2.] Apply the phase gate $U_{ab}$ to all vertices $a,b$ that are adjacent in $G$.
\end{itemize}
}

Since $U_{ab}$ is the unitary two-qubit operation on the vertices $a,b$, which adds or removes the edge $\{a,b\}$, the {\em initial state} with state vector $|+\rangle^V$ of the preparation procedure can also be regarded as the graph state that corresponds to the {\em empty graph}\index{empty graph (state)}.


\subsection{Stabilizer formalism}\label{DefOfGS_Stab}\index{stabilizer formalism|(}

\begin{wrapfigure}[11]{r}{0.45\textwidth}
\vspace{-0.7cm}\hspace{0.0cm}\includegraphics[width=0.45\textwidth]{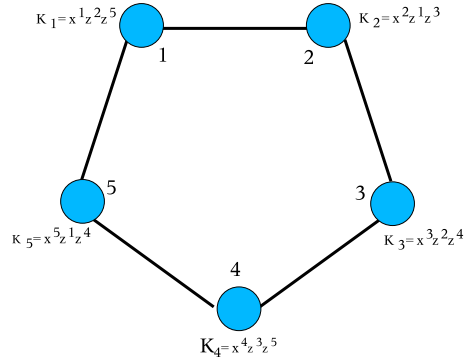}
\caption{The correlation operators for a graph state that
corresponds to a ring with $5$ vertices}\label{Ring3}
\end{wrapfigure}
It is often more convenient to work with the stabilizer of a
quantum state (or subspace) than with the state (or subspace)
itself. Quantum information theory uses the stabilizer formalism
in a wide range of applications. Among those, quantum error
correcting codes (stabilizer codes) are a very prominent example
\cite{Gottesman}. Here, the {\em stabilizer}\index{stabilizer
$\mathcal{S}$}\footnote{We refer the reader to ref.~\cite{NielsenBook}
for an introduction to the stabilizer formalism.} $\mathcal{S}$ is
a commutative subgroup of the Pauli group $\mathcal{P}^V$ that
does not contain $-\mathbf{1}_V$ (and thus not $\pm i
\mathbf{1}_V$). Apart from the interaction pattern, graph states
can also be defined uniquely in terms of their stabilizer:

\begin{proposition}{\bf Graph states (II)}\label{GS_Def2}
 Let $G=(V,E)$ be a graph. A {\em graph state vector }\index{graph state \texttt{"|}$G\rangle$}  $|G\rangle$ is the unique, common eigenvector in $({\mathbb{C}}^2)^{ V}$ to the set of independent commuting observables:
\be\label{GS_Stabilizer} \nonumber K_{a} = \sigma_x^a \,
\sigma_z^{N_a} := \sigma_x^{a}\prod_{b \in
N_a}\sigma_z^{b},\ee where the eigenvalues to the {\em correlation
operators}\index{correlation operator $K_a$} $K_{a}$ are $+1$ for
all $a\in V$. The Abelian subgroup $\mathcal{S}$ of the local
Pauli--group $\mathcal{P}^V$ generated by the set $\{K_a\, |\, a
\in V\}$ is called the stabilizer of the graph state.
\end{proposition}

{\em Proof:}  The fact that $|G\rangle$ is actually uniquely
defined by its correlation operators, follows from the subsequent
Proposition~\ref{Graph state basis}, since the set of eigenstates
to all possible eigenvalues for $K_{a}$ is a basis for
$({\mathbb{C}}^2)^{V}$. Nevertheless, Proposition~\ref{GS_Def2}
provides also an alternative definition for graph states. Hence,
we have to proof that this definition is equivalent to the
definition in the previous section. Note that the graph state for
the empty graph actually is stabilized by the set of independent
Pauli matrices $\{\sigma_x^a \,|\, a \in V\}$. Hence, by induction
over the set of edges $E$ it suffices to show the following: Given
a graph state vector $|G\rangle$ stabilized by $K_a$, the application of
the phase gate $U_{ab}$ in eq.~(\ref{CPhase}) leads to a graph
state vector  $|G'\rangle$ with a new stabilizer generated by $K'_a$,
which corresponds to the graph $G'$ that is obtained after the
edge $\{a,b\}$ is added (or removed). This certainly holds for all
vertices $c\in V\setminus \{a,b\}$, since $K_{c}$ commutes with
$U_{ab}$. For the remaining vertex $b$, we find \be
 U_{ab}\,K_a\,{U_{ab}}^\dagger = U_{ab}\, \left( P^a_{z,-}\, +\, P^a_{z,+}\sigma_z^b \right) \; K_a
= \sigma_z^b\,K_a \; , \ee because $\sigma_x  P_{z,\pm} =
P_{z,\mp} \sigma_x $. Due to $U_{ab}= U_{ba}$, we similarly obtain
for $a$
\begin{equation}
U_{ab}K_b{U_{ab}}^\dagger =\sigma_z^a\,K_b\; ,
\end{equation}
so that the transformed stabilizer corresponds indeed to a graph
$G'$, where the edge $\{a,b\}$ is added modulo $2$. \proofend

The generators $K_a$ of the stabilizer $\mathcal{S}$ have a
straightforward interpretation in terms of {\em correlation
operators}: Consider a graph state vector $|G\rangle$ that is measured
according to the measurement pattern given by
$K_a=\sigma_x^a\sigma_z^{N_a}$, i.e., the qubit at vertex $a$ is
measured in $x$-direction and the vertices $b$ in $N_b$ in
$z$-direction. Then $K_a$ provides constraints to the {\em
correlations} between the measurement outcomes $m^a_x=\pm1$ and
$m^b_z=\pm1$, namely \be\label{Correlations} m_x^a \prod_{b \in
N_a} m_z^b = 1 \; .\ee Since all elements $\sigma \in \mathcal{S}$
stabilize $|G\rangle$ they give rise to different constraints to
the correlations of the spin statistics for graph states.

That the set of correlation operators has a unique common
eigenstate, is most easily seen by considering the {\em graph
state basis}:

{\proposition[{\bf Graph state basis}]\label{Graph state
basis}\index{graph state basis \texttt{"|}$U\rangle_G$} \em Given
a graph state vector $|G\rangle$, the set of states \be|W\rangle =
\sigma_z^W |G\rangle \ee is basis for $({\mathbb{C}}^2)^{V}$. The
states $|W\rangle$ are the eigenstates for the correlation
operators $K_a$ according to different eigenvalues $W_a$ where
$W=(W_1,\ldots,W_N)$, i.e., \be\label{GS_Basis} K_a |W\rangle_G
=(-1)^{W_a} \, |W\rangle  \;. \ee The projector onto the graph
state\index{graph state \texttt{"|}$G\rangle$}  has a direct
representation in terms of the corresponding stabilizer
$\mathcal{S}$: \be\label{GS_Projector} |G\rangle\langle G | =
\frac{1}{2^N}\,\sum_{\sigma \in \mathcal{S}} \sigma \ee }

{\em Proof:} For the verification of eq.~(\ref{GS_Basis}) it
suffices to consider a single $\sigma_z^a$ operator at some vertex
$a$. $\sigma_z^b$ commutes with all correlation operators $K_a$
for $a\neq b$ and anti-commutes with $K_b$. For all $a\in V$, we
obtain \be K_a |W\rangle = K_a \prod_{b \in W} \sigma_z^b
|G\rangle =  (-1)^{\delta_{a\in W}} \prod_{b\in W} \sigma_z^b
|G\rangle = (-1)^{W_a} \, |W\rangle \; , \ee
where $\delta_{a\in U}=1$ if $a\in U$ and zero otherwise. Thus, any two distinct sets $W,W'\subseteq V$ correspond to  eigenvectors $|W\rangle, |W'\rangle$ for the set of generators $\{K_a \, |\, a\in V\}$ but with eigenvalues that differ in at least one position $a\in V$. Hence $\langle W |W' \rangle = \delta_{WW'}$, where $\delta_{WW'}=1$ if $W=W'$ and zero otherwise. Since there are  $2^N$ possible sets, the eigenvectors $\left\{|W\rangle\right\}_{W\subseteq V}$ form a basis of $({\mathbb{C}}^2)^{V}$.\\
A similar calculation verifies eq.~(\ref{GS_Projector}): \be
\langle W |\, \sum_{\sigma \in \mathcal{S}} \sigma\,|W' \rangle
\;=\; 2^N\, \delta_{W\emptyset} \, \delta_{W'\emptyset}\ee for any
basis vectors $|W\rangle$ and $|W'\rangle$. The normalization
constant $\frac{1}{2^N}$ is due to $\text{tr} (|G\rangle\langle
G|) = 1$ and because the number $|\mathcal{S}|$ of stabilizer
elements is $2^N$. \proofend

In the following we will briefly address local equivalence  for
the class of graph states and relate this class to the more
general concept of stabilizer states and codes. We also present an
alternative description of the stabilizer of a graph state in
terms of its binary representation and review a possible
generalization of the stabilizer formalism to $d$-level systems.

\subsubsection{\it \small Stabilizer states and codes}\label{Def_Stab_States}

There exists a natural generalization of the description of graph
states within the stabilizer formalism. Each stabilizer
$\mathcal{S}$, i.e., any commutative subgroup of the Pauli group
$\mathcal{P}^V$ that does not contain $-\mathbf{1}_V$, uniquely
corresponds to some {\em stabilized subspace}\index{stabilizer
$\mathcal{S}$} $\mathbf{H}_\mathcal{S}\subseteq
({\mathbb{C}}^2)^{V}$, which is the largest subspace satisfying
$\mathcal{S}\, \mathbf{H}_\mathcal{S} = \mathbf{H}_\mathcal{S}$.
The minimal number
 \be\label{GroupRank}
 r_\mathcal{S} := \text{rank}(\mathcal{S})
 = \min\left\{ n\,|\, \langle\{s_1,...,s_n\}\rangle = \mathcal{S}\, ,\; s_i\in \mathcal{S} \right\} \index{rank of a stabilizer}\leq N \ee
of generators for a stabilizer is a well-defined quantity and is
called the {\em rank} of the stabilizer. Thus, a necessary
requirement for some stabilizer $\mathcal{S}$ to represent a graph
state is that it has rank $N$, or, equivalently, that
$\mathcal{S}$ is generated by $N$ independent stabilizer elements.
More generally, any full rank stabilizer ${\cal S}$ stabilizes
exactly one (up to an overall phase factor) pure state, which is
called a \emph{stabilizer state} and which is in short denoted by
$|{\cal S}\rangle$. This stabilizer state is the pure state with the unique common
eigenvector with eigenvalue 1 of all elements of ${\cal S}$, which
is denoted by ${\cal S} |{\cal S}\rangle= |{\cal S}\rangle$. Thus,
every graph state is a stabilizer state; however, the class of
stabilizer states is strictly larger than the class of graph
states.

It is clear that an $N-$qubit stabilizer state vector $|{\cal S}\rangle$
is completely determined by a set of $N$ independent generators
$\{s_a\}_{a\in V}$ of ${\cal S}$. Note that, for computational
reasons, it is often much more efficient to deal with such a set
of independent generators rather than with the
complete stabilizer itself. 
However, this leads to an ambiguity in the description of a
stabilizer state, since there are many independent generating sets
for every stabilizer. Therefore,  the question frequently arises
whether two given sets of generators $\{s_a\}_{a\in V}$ and
$\{s'_a\}_{a\in V}$ generate the same stabilizer $\mathcal{S}$. In
section \ref{BinaryRepr} we will see an efficient approach to
answer this question.

If a stabilizer $\mathcal{S}$ does not have full rank $r<N$, it
only stabilizes an $N-r$ dimensional subspace $\mathcal{C}$ of
$({\mathbb{C}}^2)^{V}$ \cite{Gottesman,NielsenBook}. In principle,
such a subspace corresponds to an $[r,N-r]$--{\em stabilizer
code}\index{stabilizer code} encoding $N-r$ into $N$ qubits. For a
decent stabilizer code the $N-r$ degrees of freedom are used to
detect possible errors. The main idea is to arrange the
code\index{quantum error correcting code (QEC)} in such a way
that, under the influence of errors, the complete
$N-r$-dimensional space containing the encoded information is
mapped onto an orthogonal eigenspace of $\mathcal{S}$. The
coherent information encoded in this $r$-dimensional space can
then be maintained by some error correction procedure. More
precisely, suppose that some error operator $\sigma \in
\mathcal{P}^V$ occurs, i.e., the underlying noise process has a
Kraus representation with $\sigma$ as one of its Kraus operators
\footnote{Note that a restriction in the error considerations to
Pauli errors is legitimate, since error correction capabilities of
a code $\mathcal{C}$ can be determined w.r.t. any basis of
operation elements $E_i$ (see e.g. Theorem 10.1 and 10.2 in
ref.~\cite{NielsenBook}).}. Then if $\sigma \in \mathcal{S}$ the
stabilized subspace $\mathcal{C}$, and thus also any encoded
quantum information, is not affected at all. On the other hand, if
$\sigma \in \mathcal{P}^V\setminus \mathbf{N}(\mathcal{S})$, where
$\mathbf{N}(\mathcal{S})=\{ \sigma \in \mathcal{P}^V \, |\, \sigma
\mathcal{S} \sigma^\dagger \subseteq \mathcal{S}  \}$ denotes the
{\it normalizer} of the subgroup $\mathcal{S}$, then $\sigma$
anti-commutes with at least one element of the stabilizer
$\mathcal{S}$ and thus transforms the complete subspace
$\mathcal{C}$ into an orthogonal subspace. By measuring a
generating set of stabilizer elements $s_i$ the corresponding
error thus can be detected. Only if the error $\sigma \in
\mathcal{P}^V$ is an element of the normalizer
$\mathbf{N}(\mathcal{S})$ but not of the stabilizer itself, i.e.
$\sigma \in \mathbf{N}(\mathcal{S}))\setminus \mathcal{S}$, then
this transformation remains within the codespace $\mathcal{C}$ and
thus cannot be detected.  More generally, for a correction of a
set of possible errors $\{ E_i\}\subseteq \mathcal{P}^V$, i.e., a
noise process with Kraus operators $E_j$, the effect of different
errors has to be distinguishable by the error syndrome $\beta_i$
obtained through measuring the stabilizer generators $s_i$, i.e.
$s_i E_j\mathcal{C}=\beta_i E_j\mathcal{C}$.  One finds
\cite{Gottesman,NielsenBook} that the set of errors $\{ E_i\}$ is
correctable if $E_jE_k\notin \mathbf{N}(\mathcal{S}))\setminus
\mathcal{S}$ for all $j$ and $k$. If there is a unique error $E_j$
associated with a given error syndrome $\beta_i$, the error can be
corrected by applying $E_j^\dagger$. If, however, two errors $E_j$
and $E_k$ correspond to the same error syndrome $E_i$, which
implies $E_j\mathcal{C}=E_k\mathcal{C}$, both errors can be
corrected by applying either of the operators $E_j^\dagger$ and
$E_k^\dagger$, since if $E_j$ occurred but $E_k^\dagger$ is
applied for error correction one nevertheless finds $E_k^\dagger
E_j \mathcal{C} = \mathcal{C}$ by assumption.

Supplementing the $N-r$ generators of the stabilizer by $r$
additional elements $Z_1,\ldots,Z_r \in\mathcal{P}^V$ to form a
full rank stabilizer corresponds to the choice of a basis of
codeword vectors\footnote{Choose
$|W\rangle=|(W_1,\ldots,W_r)\rangle \in \mathcal{C}$ such that
$Z_i|W\rangle = (-1)^{W_i} |W\rangle$.} in the codespace
$\mathcal{C}$. This basis is frequently called the `logical
computational basis'. As we will discuss in
sec.~\ref{Application_QEC} graph states, or more generally
stabilizer states, also appear as codeword vectors in stabilizer
codes. Together with similarly defined logical $X$-operators
$X_i$, the logical $Z$-operations $Z_i$ allow for concise
manipulations\footnote{ For details we refer the reader to
ref.~\cite{Gottesman,NielsenBook}.}
of the underlying code space such
as error detection and correction, and the concatenation of codes
in order to improve error correction capabilities.

\index{stabilizer formalism|)}

\subsubsection{\it \small Local Clifford group and LC equivalence}\label{Def_LC}

Each graph state vector $|G\rangle$ corresponds {\em uniquely} to a graph
$G$. In other words, two different graphs $G=(V,E)$ and
$G'=(V,E')$ cannot describe the same graph state: the interaction
picture tells us that $|G\rangle=|G'\rangle$ would yield a
contradiction \bea |+\rangle^V  & = & \prod_{\{a,b\}\in
E'}U_{ab}\, |G'\rangle =  \prod_{\{a,b\}\in E'}U_{ab}\,  |G\rangle
\\ & =&  \prod_{\{a,b\}\in E'}U_{ab}\,\prod_{\{a,b\}\in
E}U_{ab}\,|+\rangle^V = \prod_{\{a,b\}\in E +
E'}U_{ab}\,|+\rangle^V \; .\nonumber \eea Here, $E+E'$ denotes the
symmetric difference eq.~(\ref{+}) of the edge sets, which is by
assumption not empty and thus yields a non-vanishing interaction.

However, graph states of two different graphs might be equal up to
some local unitary (LU) operation. We will call two graphs
$G=(V,E)$ and $G'=(V,E')$ {\em LU-equivalent}\index{local unitary
(LU)}\index{equivalence under!local unitaries (LU)}, if there
exists a local unitary $U\in \mathbf{U}(2)^{V}$ such that
\begin{equation}\label{LU_graphs}
    |G'\rangle = U\,|G\rangle.
\end{equation}
Locality here refers to the systems associated with vertices of
$G$ and $G'$. Denoting  \be \label{Sigma}\Sigma':=
U\mathcal{S}U^{\dagger} = \{UsU^{\dagger}\ |\ s\in
\mathcal{S}\},\ee where $\mathcal{S}$ is the stabilizer of the
state vector $|G\rangle$, one finds that $s'|G'\rangle = |G'\rangle$ for
every $s'\in\Sigma'$. In this sense the group $\Sigma'$ is a
'stabilizing subgroup' of the state vector $|G'\rangle$, being a group of
(local) unitary operators that have $|G'\rangle$ as a fixed point;
however, in general $\Sigma'$ is not equal to stabilizer of
$|G'\rangle$, since in general $\Sigma'$ is not a subgroup of the
Pauli group\footnote{This issue is closely related to the problem
of local unitary versus local Clifford equivalence of graph
states, which is discussed in sec.~\ref{Local_Equivalence}.}.
In view of this observation, it is interesting to consider the
subclass of those local unitary operators $U$ satisfying
$\mathcal{P}^V = U\, \mathcal{P}^V\, U^\dagger $, meaning that $U$
maps the whole Pauli group $\mathcal{P}^V$ onto itself under
conjugation. The set \be\label{Def_CliffU}
\mathcal{C}_1^V:=\{U\in\mathbf{U}(2)^{V}\, |\, U \mathcal{P}^{V}
U^\dagger =  \mathcal{P}^{V} \}\index{Local Clifford group
$\mathcal{C}_N$} \ee of such unitaries is a group, the so-called
{\em local Clifford group} (on $N$ qubits). If $|G\rangle$ and
$|G'\rangle$ are graph states  such that $U|G\rangle=|G'\rangle$
for some $U\in{\cal C}_1^V$, then the group $\Sigma'$ in
(\ref{Sigma}) is equal to the stabilizer of $|G'\rangle$.
Therefore, the action of local Clifford operations on graph states
can entirely be described within the stabilizer formalism -- and
this is one of the main reasons why the local Clifford group is of
central importance in the context of graph states. In the
following, we will call two graph states $|G\rangle$ and
$|G'\rangle$  {\em LC-equivalent}\index{equivalence under!local
Clifford unitaries (LC)} iff they are related by some local
Clifford unitary $U\in\mathcal{C}_1^V$, i.e., $|G'\rangle = U
|G\rangle$.

The local Clifford group is the $N-$fold tensor product of the
one-qubit Clifford group $\mathcal{C}_1$ with itself, where
$\mathcal{C}_1$ is defined by \be
\mathcal{C}_1:=\{U\in\mathbf{U}(2)\, |\, U \mathcal{P} U^\dagger =
\mathcal{P}\}.\ee One can show that, up to a global phase factor,
any one-qubit Clifford operation $U\in\mathcal{C}_1$ is a product
of operators chosen from the set $\{H, S\}$, where
\begin{gather}
H=\frac{1}{\sqrt{2}}\begin{pmatrix} 1 & 1\\ 1 & -1 \end{pmatrix}
\;\; \text{(Hadamard gate)}\index{Hadamard gate $H$} \hspace{1cm}
S=\begin{pmatrix} 1&0\\0&i \end{pmatrix} \; \;\text{(single-qubit
phase gate)}.
\end{gather}
The action of the Clifford group $\mathcal{C}_1$ under conjugation
permutes the Pauli matrices $\sigma_1$, $\sigma_2$ and $\sigma_3$
up to some sign $\pm 1$. This can be shown as follows: First, the
matrices $\pm \sigma_0$ and $\pm i \sigma_0$ are left unchanged
under conjugation. Secondly, the set $\{\pm\sigma_1, \pm\sigma_2,
\pm\sigma_3\}$ has to be mapped onto itself, since
$U\sigma_iU^\dagger$ is Hermitian iff  $\sigma_i$ is Hermitian.
Because the conjugation is invertible, the conjugation permutes
the matrices $\sigma_1$, $\sigma_2$  and $\sigma_3$ up to some
sign $\pm1$.
Also, note that it suffices to fix the action of $U$ for two
traceless Pauli matrices, say $\sigma_1$ and $\sigma_2$, since the
action for the other matrices follows from linearity of the
conjugation and the relation $\sigma_3=-i\sigma_1\sigma_2$.

 If one
disregards the overall phases of its elements, the one-qubit
Clifford group has finite cardinality. In Tab.~\ref{Tab_LC} we
have itemized all $24$ single-qubit Clifford unitaries,
disregarding such global phases. For each unitary we have also
included a possible decomposition in terms of Pauli operators and
the $\frac{\pi}{4}$-rotations \be\label{sqrt_sigma} \sqrt{\pm i
\sigma_j} = e^{\pm i \frac{\pi}{4} \sigma_j} \hspace{0.7cm}
j=1,2,3\; ,\ee that we frequently use throughout this article.
These rotations correspond to the elementary permutations
$\{1,2,3\}\mapsto \{1,3,2\}$, $\{1,2,3\}\mapsto \{3,2,1\}$ and
$\{1,2,3\}\mapsto \{2,1,3\}$ that only permute two indices.
Instead of $H$ and $S$ any two of these elementary permutations
can be used to generate the Clifford group $\mathcal{C}_1$.

\begin{table}
\begin{center}
\begin{minipage}{0.9\textwidth}
\begin{tabular}{ccc}
\begin{minipage}{0.45\textwidth}
\begin{tabular}{|ccc|c|}
\hline
$\sigma_1$ & $\sigma_2$ & $\sigma_3$ & Decomposition\\
\hline \hline
$\sigma_1$ & $\sigma_2$ & $\sigma_3$  & $\sigma_0$\\
$\sigma_1$ & $-\sigma_2$ & $-\sigma_3$  & $\sigma_1$\\
$-\sigma_1$ & $\sigma_2$ & $-\sigma_3$ & $\sigma_2$\\
$-\sigma_1$ & $-\sigma_2$ & $\sigma_3$  & $\sigma_3$\\
\hline
$\sigma_1$ & $-\sigma_3$ & $\sigma_2$  & $\sqrt{ i \sigma_1} $\\
$\sigma_1$ & $\sigma_3$ & $-\sigma_2$  & $\sqrt{- i \sigma_1}$\\
$-\sigma_1$ & $-\sigma_3$ & $-\sigma_2$ & $\sigma_3 \sqrt{ i \sigma_1}$\\
$-\sigma_1$ & $\sigma_3$ & $\sigma_2$  & $\sigma_3 \sqrt{- i \sigma_1}$\\
\hline
$\sigma_3$ & $\sigma_2$ & $-\sigma_1$  & $\sqrt{ i \sigma_2} $\\
$-\sigma_3$ & $\sigma_2$ & $\sigma_1$  & $\sqrt{- i \sigma_2}$\\
$\sigma_3$ & $-\sigma_2$ & $\sigma_1$ & $\sigma_3 \sqrt{ i \sigma_2}$\\
$-\sigma_3$ & $-\sigma_2$ & $-\sigma_1$  & $\sigma_3 \sqrt{- i \sigma_2}$\\
\hline
\end{tabular}
\end{minipage} & \hspace{-0.5cm} &
\begin{minipage}{0.45\textwidth}
\begin{tabular}{|ccc|c|}
\hline
$\sigma_1$ & $\sigma_2$ & $\sigma_3$ & Decomposition\\
\hline \hline
$-\sigma_2$ & $\sigma_1$ & $\sigma_3$  & $\sqrt{ i \sigma_3} $\\
$\sigma_2$ & $-\sigma_1$ & $\sigma_3$  & $\sqrt{- i \sigma_3}$\\
$\sigma_2$ & $\sigma_1$ & $-\sigma_3$ & $\sigma_1 \sqrt{ i \sigma_3}$\\
$-\sigma_2$ & $-\sigma_1$ & $-\sigma_3$  & $\sigma_1 \sqrt{- i \sigma_3}$\\
\hline
$-\sigma_2$ & $-\sigma_3$ & $\sigma_1$  & $\sqrt{ i \sigma_3}\sqrt{ i \sigma_1} $\\
$\sigma_2$ & $-\sigma_3$ & $-\sigma_1$  & $\sqrt{ i \sigma_3}\sqrt{- i \sigma_1}$\\
$-\sigma_2$ & $\sigma_3$ & $-\sigma_1$ & $\sqrt{- i \sigma_3} \sqrt{ i \sigma_1}$\\
$\sigma_2$ & $\sigma_3$ & $\sigma_1$  & $\sqrt{- i \sigma_3} \sqrt{- i \sigma_1}$\\
\hline
$\sigma_3$ & $\sigma_1$ & $\sigma_2$  & $\sqrt{ i \sigma_3}\sqrt{ i \sigma_2} $\\
$-\sigma_3$ & $\sigma_1$ & $-\sigma_2$  & $\sqrt{ i \sigma_3}\sqrt{- i \sigma_2}$\\
$\sigma_3$ & $-\sigma_1$ & $-\sigma_2$ & $\sqrt{- i \sigma_3} \sqrt{ i \sigma_2}$\\
$-\sigma_3$ & $\sigma_1$ & $-\sigma_2$  & $\sqrt{- i \sigma_3} \sqrt{- i \sigma_2}$\\
\hline
\end{tabular}
\end{minipage}
\end{tabular}

\end{minipage}
\end{center}
\caption{All $24$ single-qubit Clifford unitaries and their
decomposition into elementary permutations.}\label{Tab_LC}
\end{table}

An important result in the theory of graph states and stabilizer
states is that any stabilizer state is LC-equivalent to some graph
state. This statement was first proven in ref.~\cite{Schlinge02b} and
\cite{Grassl02} for the more general setup of stabilizer codes
over $d$-level systems. {\proposition[{\bf Stabilizer states}] Any
stabilizer state vector $|\mathcal{S}\rangle$  is LC-equivalent to some
graph state vector $|G\rangle$ , i.e., $|\mathcal{S}\rangle = U |G\rangle$
for some LC-unitary $U\in \mathcal{C}^{V}$. This unitary can be
calculated efficiently. }

{\em Proof:} A proof for the qubit case  in terms of the binary
framework (see sec.~\ref{BinaryRepr}) can be found in
ref.~\cite{Nest04a}. \proofend

A similar statement  holds more generally for all stabilizer
codes: Any stabilizer code is LC-equivalent to some {\em graph
code}.
Thus, graph states can be regarded  as standard forms\footnote{In ref.~\cite{Auden05} some normal forms for stabilizer states are suggested that do not rely on graph states, but which also allow for an efficient calculation of various (entanglement) properties.}  for
stabilizer states, since many properties, such as entanglement,
are invariant under LC operations. Note that this standard form is
however not unique. A stabilizer state vector $|\mathcal{S}\rangle$ can
be LC-equivalent to several  graph states  $|G_1\rangle= U_1
|\mathcal{S} \rangle$ and $|G_2\rangle = U_2 |\mathcal{S}
\rangle$, whenever these graph states are LC-equivalent
$|G_1\rangle = U_1 U_2^\dagger |G_2\rangle$. Thus, the study of
local equivalence of stabilizer states reduces to that of local
equivalence of graph states.

Note that in general there are $24^N$ different Clifford unitaries
(up to global phases) to relate two graphs states with $N$
vertices. Therefore, the difficulty to decide whether two graph
states are LC-equivalent seems to increase exponentially with the
number of parties. However in sec.~\ref{BinaryRepr} we will
briefly mention a method due to ref.~\cite{Nest04b} that scales only
polynomially with the number of vertices.

Interestingly, the action of local Clifford operations on graph
states can be described in terms a simple graph transformation
rule, called \emph{local complementation} \cite{Bouchet}: letting
$G=(V, E)$ be a graph and $a\in V$, the local complement of $G$ at
$a$, denoted by $\tau_a(G)$, is obtained by complementing the
subgraph of $G$ induced by the neighborhood $N_a$ of $a$ and
leaving the rest of the graph unchanged: \be \tau_a:\, G\mapsto
\tau_a(G):=G+N_a   \; .\ee \index{local complementation $\tau_a$}
With this notation the following result can be stated \cite{Glynn02,Nest04a}:
\begin{proposition}[{\bf LC-rule}]\label{loc}\index{LC-rule}
 By local complementation of a graph $G$ at some vertex $a\in V$ one obtains an LC-equivalent graph state $|\tau_a(G)\rangle$:
\be |\tau_a(G)\rangle = U^\tau_a(G)\,|G\rangle \; ,\ee where
\be\label{LU_Rule_U} U^\tau_a(G)= e^{-i\frac{\pi}{4}\sigma_x^a}
e^{i\frac{\pi}{4}\sigma_z^{N_a}}\propto \sqrt{K_a} \ee is a local
Clifford unitary. Furthermore, two graph states $|G\rangle$ and
$|G'\rangle$ are LC-equivalent\index{equivalence under!local
Clifford unitaries (LC)} iff the corresponding graphs are related
by a sequence of local complementations, i.e.
$G'=\tau_{a_1}\circ\ldots\circ\tau_{a_n}(G)$ for some
$a_1,\ldots,a_n\in V$.
\end{proposition}

\begin{wrapfigure}[12]{r}{0.45\textwidth}
\vspace{-1cm}\hspace{0.0cm}{\includegraphics[width=0.45\textwidth]{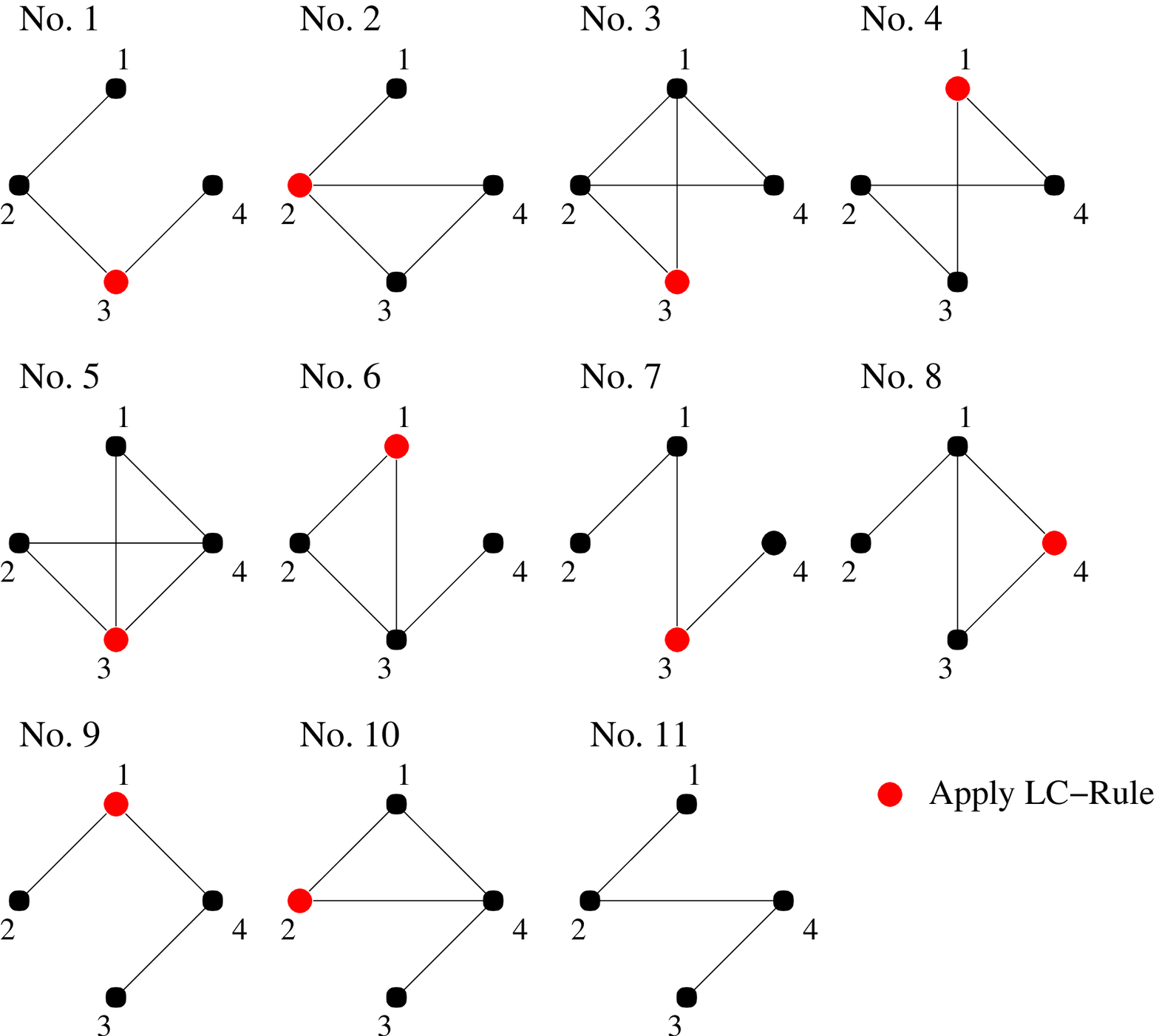}}
\caption{\label{fig:LUruleExample1} An example for a successive
application of the LC-rule, which exhibits the whole equivalence
class associated with graph No.\ 1. The rule is successively
applied to the vertex, which is colored red in the figure.}
\end{wrapfigure}
Fig.~\ref{fig:LUruleExample1} depicts an example for such a
successive application of the LC-rule. Starting with the first
graph the complete orbit can be obtained by applying the LC-rule
to the vertices in the preceding graph that appear above the arrow
of the following diagram:
\[ \begin{CD}
  \text{No.\ }1  @>3>> \text{No.\ }2  @>2>>  \text{No.\ }3  @>3>> \\
  \text{No.\ }4  @>1>> \text{No.\ }5 @>3>>   \text{No.\ }6  @>1>>  \\
  \text{No.\ }7  @>3>> \text{No.\ }8  @>4>>  \text{No.\ }9  @>1>> \\
  \text{No.\ }10  @>2>>   \text{No.\ }11
\end{CD} \]

{\em Proof of Proposition~\ref{loc}:}

Let $G$ be a graph with correlation operators $K_b$ and
$G'=\tau_a(G)$ the  corresponding graph under local
complementation at vertex $a$ with correlation operators $K'_b$.
For $c \in V\setminus N_a$ we find
\begin{equation}
    U^\tau_aK_c (U^\tau_a)^\dagger=K_c=K'_c \; .
\end{equation}
For $b \in N_a$, we compute \bea U^\tau_a\,K_b\,(U^\tau_a)^\dagger
& = &  \left(- i\sigma_x^a\right) \; \left( i \sigma_z^{b}\right)
\,\sigma_x^b \;  \,\sigma_z^a \sigma_z^{N_b \setminus a }
\nonumber\\ & = & \sigma_x^a\, \sigma_z^{N_a}\; \cdot \;
\sigma_x^b\,\sigma_z^{N_b+N_a} \nonumber\\ & = &  K'_a\; \cdot \;
K'_b  \, .\nonumber \eea Thus the stabilizer $ U^\tau_a\,
\mathcal{S}\,(U^\tau_a)^\dagger $ is generated by $\{K'_c\}_{c\in
V\setminus N_a} \cup \{K'_a K'_b\}_{b\in  N_a}$. By multiplication
of the generators $K'_a K'_b$ with $K'_a$ (since $a\in V\setminus
N_a$) it follows that $ U^\tau_a\, \mathcal{S}\,(U^\tau_a)^\dagger
$ is also generated by $\{K'_a\}_{a\in V}$ and therefore
stabilizes the graph state $|G'\rangle$.
This proves that a sequence LC-rule applications yields an
LC-equivalent graph state. That the action of any unitary within
the Clifford group on graph states can be decomposed into a
sequence of LC-rule applications is however more involved and we
refer to refs.~\cite{Glynn02,Nest04a} for a proof. \proofend

\subsubsection{\it \small Clifford group}\label{Def_Cliffordgroup}

Whereas the LC group ${\cal C}_1^V$ is defined to consist of all
\emph{local} unitary operators mapping the Pauli group to itself
under conjugation, one is also often interested in the group of
all unitary operators with this property, i.e., the group \be
\mathcal{C}_N:=\{U\in\mathbf{U}(2^N)\, |\, U \mathcal{P}^V
U^\dagger = \mathcal{P}^V\},\ee which is called the \emph{Clifford
group} (on $N$ qubits). By definition, Clifford operations map
stabilizer states to stabilizer states. Up to a global phase
factor any  Clifford operation $U$ can be decomposed into a
sequence of ${\cal O}(N^2)$ one- and two-qubit gates
\cite{NielsenBook,Gottesman} in the set $\{H, S, CNOT\}$, where
$H$ and $S$ are defined as before and
\begin{gather}
\text{CNOT} = \begin{pmatrix} 1&0&0&0\\0&1&0&0 \\0&0&0&1\\0&0&1&0
\end{pmatrix} \; \;\text{(controlled-NOT gate)}\; .
\end{gather}
For example, the controlled phase gate $U_{ab}$\index{controlled
NOT gate (CNOT)}\index{phase gate $U_{ab}$} can be decomposed into
$U_{ab}=H^b\text{CNOT}_{ab} H^b$.

\subsubsection{\it \small Binary representation}\label{BinaryRepr}\index{binary representation|(}

\begin{wrapfigure}{r}{0.45\textwidth}
 \vspace{-0.0cm}
 \fbox{
$\left(\mathbf{X}|\mathbf{Z}\right) =\left(
\begin{array}{ccccc|ccccc}
 1 & 0 & 0 & 0 & 0 & 0 & 1 & 0 & 0 & 1 \\
 0 & 1 & 0 & 0 & 0 & 1 & 0 & 1 & 0 & 0 \\
 0 & 0 & 1 & 0 & 0 & 0 & 1 & 0 & 1 & 0 \\
 0 & 0 & 0 & 1 & 0 & 0 & 0 & 1 & 0 & 1 \\
 0 & 0 & 0 & 0 & 1 & 1 & 0 & 0 & 1 & 0
\end{array} \right)
$ } \caption{The generator matrix for a graph state that
corresponds to a ring with $5$ vertices.}\label{Ring4}
\end{wrapfigure}
We now briefly review  an alternative representation of the
stabilizer formalism in terms of its binary representation
\cite{Gottesman,Nest04a}. This description is frequently used in
the literature, since it allows one to treat the properties of the
stabilizer in terms of a symplectic subspace of the vector space
$\mathbb{F}_2^{2N}$.

Any element $u$ of the Pauli group $\mathcal{P}^{V}$ can be
represented uniquely, {\em up to some phase factor}, by a vector
$\mathbf{U}=(\mathbf{U}_x,\mathbf{U}_z)\in \mathbb{F}_2^{2N}$,
where $\mathbf{U}_x,\mathbf{U}_z\in\mathbb{F}_2^N$: \be
u=\sigma_x^{\mathbf{U}_x}\sigma_z^{\mathbf{U}_z}\equiv \prod_{a
\in V} \sigma_x^{U_x^a} \, \prod_{a \in V} \sigma_z^{U_z^a}\; ,\ee
where $\mathbf{U}_x^a,\mathbf{U}_z^a\in \mathbb{F}_2$ for every
$a\in V$. At each qubit we have used the following encoding of the
Pauli matrices as pairs of bits:
 \be \sigma_0 \mapsto (0|0),\hspace{0.7cm} \sigma_x \mapsto (1|0),\hspace{0.7cm} \sigma_y \mapsto (1|1),\hspace{0.7cm} \sigma_z \mapsto (0|1) \; . \ee
For example, the correlation operator $K_1$ of the ring in
fig.~\ref{Ring3} has the binary representation \be K_1=\sigma_x^1
\sigma_z^2 \sigma_z^5 \longmapsto
(10000|01001)\in\mathbb{F}_2^{10}\; .\ee

It is important to note that the binary representation captures
the features of a Pauli operator only up to a phase factor.

The binary representation has the following two important
properties: letting $u,v,x \in \mathcal{P}^{V}$ with corresponding
binary vectors $\mathbf{U},\mathbf{W},\mathbf{X} \in
\mathbb{F}_2^{2N}$, one finds that \bea
\text{(i)} & u\, w \sim x  & \ \longleftrightarrow \hspace{0.5cm} \mathbf{U} + \mathbf{W}\, =\, \mathbf{X} \hspace{0.5cm}\text{(mod $2$)}\; , \nonumber \\
\text{(ii)} & [u,w]=0 & \ \longleftrightarrow \hspace{0.5cm}
\mathbf{U}^T\, \mathbf{P}\, \mathbf{W}\, = \,\mathbf{0}
\hspace{0.5cm} \text{(mod $2$)}\; ,\eea where $\sim$ denotes
equality up to a global phase factor and where the $2N\times 2N$
matrix \be
\mathbf{P} = \left(\begin{array}{c|c} \mathbf{0} & \mathbf{1} \\
\hline \mathbf{1}& \mathbf{0}
\end{array}\right) \ee defines a \emph{symplectic inner product} on
the binary space $\mathbb{F}_2^{2N}$. Property (i) shows that the
encoding $u\in \mathcal{P}^{V}\mapsto
\mathbf{U}\in\mathbb{F}_2^{2N}$ is a homomorphism of groups. Note
that the multiplicative structure of the group $\mathcal{P}^{V}$
is mapped to the additive structure of $\mathbb{F}_2^{2N}$, where
addition has to be performed modulo $2$. Property (ii) shows that
two Pauli operators commute if and only if the corresponding
binary vectors are orthogonal with respect to the symplectic inner
product.

It follows from (i) and (ii) that, within the binary
representation, the stabilizer $\mathcal{S}$ of any stabilizer
state $|\mathcal{S}\rangle$ on $N$ qubits is an
\emph{$N$--dimensional, self-dual linear
subspace}\index{self-orthogonal subspace} $\mathbf{S}$ of
$\mathbb{F}_2^{2N}$. By self-duality it is meant that
\begin{itemize}
\item $\mathbf{U}^T \mathbf{P} \mathbf{V} = 0$ for every
$\mathbf{U}, \mathbf{V}\in \mathbf{S}$, and \item if
$\mathbf{X}\in \mathbb{F}_2^{2N}$ and
$\mathbf{X}^T\mathbf{P}\mathbf{U}=0$ for every
$\mathbf{U}\in\mathbf{S}$, then $\mathbf{X}\in \mathbf{S}$.
\end{itemize}
The subspace $\mathbf{S}$ is usually presented in terms of a {\em
generator matrix} $\left(\mathbf{X}|\mathbf{Z} \right)$ (where
$\mathbf{X}$ and $\mathbf{Z}$ are $N\times N$ matrices), which is
a full rank $2N\times N$ matrix, the rows of which form a basis of
$\mathbf{S}$; a generator matrix is obtained by assembling the
binary representations $\{\mathbf{S}_a^T\}_{a\in V}$ of a set of
independent stabilizer generators $\{s_a\}_{a\in V}$ as the rows
of a $2N\times N$ matrix. Note that any generator matrix
$\left(\mathbf{X}|\mathbf{Z} \right)$ of a self-dual subspace
$\mathbf{S}$ satisfies \be \left(\mathbf{X}|\mathbf{Z} \right)
\mathbf{P}\left(\mathbf{X}|\mathbf{Z} \right)^T = \mathbf{0}\; \ee
from the self-duality of $\mathbf{S}$.
The generator matrix for a {\em graph state} $|G\rangle$ has the
standard form \be \nonumber \left(\mathbf{X}|\mathbf{Z}
\right)=\left(\mathbf{1}|\mathbf{\mathbf{\Gamma}} \right) \; ,\ee
where $\mathbf{\mathbf{\Gamma}}$ is the {\em adjacency matrix} of
the graph $G$. Fig.~\ref{Ring4} displays the generator matrix for
the ring on five qubits (see also fig.~\ref{Ring3}).

Choosing a different set of generators in $\mathcal{S}$
corresponds to a transformation of the generator matrix of the
form \be\label{GenMatrix_Trafo} \left(\mathbf{X}|\mathbf{Z}
\right)\mapsto \left(\mathbf{X'}|\mathbf{Z'} \right) =
\mathbf{R}\left(\mathbf{X}|\mathbf{Z} \right),\ee where
$\mathbf{R}$ is some $\mathbb{F}_2$--invertible $N\times
N$-matrix. From the definition of self-duality,  two generator
matrices $\left(\mathbf{X}|\mathbf{Z} \right)$ and
$\left(\mathbf{X'}|\mathbf{Z'} \right) $ correspond to the same
stabilizer iff \be\label{Equivalence_of _stab_bin}
\left(\mathbf{X'}|\mathbf{Z'} \right)
\mathbf{P}\left(\mathbf{X}|\mathbf{Z} \right)^T = \mathbf{0}\;
.\ee

In section \ref{Def_Stab_States} we encountered the problem of
recognizing whether two sets of commuting Pauli operators are
generating sets of the same stabilizer. This issue has a simple
solution within the binary representation as follows. First we
must test eq.~(\ref{Equivalence_of _stab_bin}) for the
corresponding generator matrices. In addition we have to compute
the transformation matrix $\mathbf{R}=(R_{ab})$ in
eq.~(\ref{GenMatrix_Trafo}), which can be achieved by Gaussian
elimination over $\mathbb{F}_2$. Finally, we ensure that, at the
level of the actual stabilizer rather than its binary
representation, $\{s_a\}_{a\in V}$ is indeed transformed into
$\{s'_a\}_{a\in V}$, i.e., whether \be s_a' \;=\; \prod_{b \in V}
\, (s_b)^{R_{ab}} \hspace{0.5cm} \forall a \in V \; .\ee

The action of (local) Clifford operations on stabilizer states
also has an elegant translation in terms of the binary stabilizer
framework. Let $U \in\mathcal{C}_N$ be an arbitrary (possibly
non-local) Clifford operation and let $f_U:\mathbb{F}_2^{2N}\to
\mathbb{F}_2^{2N}$ be the unique function such that
$f(\mathbf{X})$ is the binary representation of $UxU^{\dagger}$
when $x\in{\cal P}^V$ with binary representation
$\mathbf{X}\in\mathbb{F}_2^{2N}$. First, it follows from the
property  \be Uxx'U^{\dagger}= (UxU^{\dagger})(Ux'U^{\dagger})\ee
for every $x, x'\in{\cal P}^V$,
 that \be f_U(\mathbf{X}+\mathbf{X'})=
f_U(\mathbf{X})+f_U(\mathbf{X'})\ee for every $\mathbf{X},
\mathbf{X'}\in\mathbb{F}_2^{2N}$. In other words, $f_U$ is a
linear transformation of $\mathbb{F}_2^{2N}$ and we write
$f_U(\mathbf{X})= \mathbf{Q} \mathbf{X}$, for some (nonsingular)
$2N\times 2N$ matrix $Q$ over $\mathbb{F}_2$. Secondly, the
property  $[UxU^{\dagger},Ux'U^{\dagger}] = [x, x']$ implies that
\be \mathbf{X}^T\mathbf{Q}^T\mathbf{P}\mathbf{Q}\mathbf{X'}=
\mathbf{X}^T\mathbf{P}\mathbf{X'}\ee for every $\mathbf{X},
\mathbf{X'}\in\mathbb{F}_2^{2N}$, showing that $\mathbf{Q}$ is a
\emph{symplectic transformation}, i.e., $
\mathbf{Q}^T\mathbf{P}\mathbf{Q} = \mathbf{P}.$ One can also prove
the reverse statement, i.e., that every symplectic transformation
can be realized as a Clifford operation.

It follows that conjugation of the stabilizer $\mathcal{S}'= U
\mathcal{S}U^\dagger$ corresponds to the linear transformation \be
\left(\mathbf{X}|\mathbf{Z} \right) \mapsto
\left(\mathbf{X}|\mathbf{Z} \right) \mathbf{Q}^T \; .\ee
\index{equivalence under!local Clifford unitaries (LC)} Letting
${\cal S}$, ${\cal S}'$ be full rank stabilizers on $N$ qubits
with generator matrices $\left(\mathbf{X}|\mathbf{Z} \right)$,
$\left(\mathbf{X}'|\mathbf{Z}' \right)$, respectively, it is
straightforward to prove the following chain of equivalent
statements: \bea & & \mathcal{S}'= U \mathcal{S}U^\dagger
\hspace{0.3cm}\text{for some}
\hspace{0.2cm} U \in \mathcal{C}_N \mbox{ with corresponding symplectic matrix } \mathbf{Q}\hspace{0.3cm} \nonumber \\
 & \Leftrightarrow & \left(\mathbf{X'}|\mathbf{Z'} \right) =
 \mathbf{R}\left(\mathbf{X}|\mathbf{Z} \right) \mathbf{Q}^T
 \hspace{0.3cm} \text{for some invertible}\hspace{0.1cm}
 \mathbf{R}  \\
 & \Leftrightarrow & \left(\mathbf{X'}|\mathbf{Z'} \right)
 \mathbf{P}\mathbf{Q} \left(\mathbf{X}|\mathbf{Z} \right)^T =
 \mathbf{0} \hspace{0.3cm} \label{LC_Equiv_bin}\; .
\eea

In the special case where $U$ is a local Clifford operation, i.e.,
$U\in{\cal C}_1^V$, the corresponding symplectic matrix
$\mathbf{Q}$ has the following particular structure: \be\mathbf{Q}
= \left[ \begin{array}{cc} \mathbf{A}&\mathbf{B}\\
\mathbf{C}&\mathbf{D}\end{array}\right],\ee where $\mathbf{A},
\mathbf{B}, \mathbf{C}, \mathbf{D}$ are diagonal $N\times N$
matrices. The property $ \mathbf{Q}^T\mathbf{P}\mathbf{Q} =
\mathbf{P}$ is then equivalent to
$\mathbf{A}\mathbf{D}+\mathbf{B}\mathbf{C}=\mathbf{I}$. This is in
turn equivalent to stating that the $N$ $2\times 2$ matrices
\be\mathbf{Q}_a = \left[ \begin{array}{cc} \mathbf{A}_{aa}&\mathbf{B}_{aa}\\
\mathbf{C}_{aa}&\mathbf{D}_{aa}\end{array}\right]\ee are
nonsingular (over $\mathbb{F}_2$) for every $a\in V$. Note that
the matrices $\mathbf{Q}_a$ correspond to the one-qubit tensor
factors of $U$ and,  up to a simultaneous permutation of rows and
columns, the matrix $\mathbf{Q}$ is equal to $ \mathbf{Q}_1 \oplus
\ldots \oplus \mathbf{Q}_N$.

Now, suppose that $G$ and $G'$ are two graphs with adjacency
matrices $\mathbf{\Gamma}$ and $\mathbf{\Gamma}'$, respectively.
Then, from eq.~(\ref{LC_Equiv_bin}),  the graph states $|G\rangle$
and $|G'\rangle$ are LC-equivalent iff there exist $N\times N$
diagonal matrices $\mathbf{A}, \mathbf{B}, \mathbf{C}, \mathbf{D}$
satisfying $\mathbf{A}\mathbf{D}+\mathbf{B}\mathbf{C}=\mathbf{I}$,
such that
\be\label{LC_linear_system}\mathbf{\Gamma}'\mathbf{B}\mathbf{\Gamma}
+ \mathbf{D}\mathbf{\Gamma} + \mathbf{\Gamma'}\mathbf{A} +
\mathbf{C}= \mathbf{0}.\ee
Thus, in order to check whether $|G\rangle$ and $|G'\rangle$ are LC-equivalent, one has to decide
whether the linear system of equations (\ref{LC_linear_system}),
together with the additional quadratic constraints
$\mathbf{A}\mathbf{D}+\mathbf{B}\mathbf{C}=\mathbf{I}$, has a
solution. This approach leads to an efficient algorithm to
recognize LC-equivalence of graph states \cite{Nest04a, Bouchet}
as follows. First, note that the set ${\cal V}$ of solutions
$(\mathbf{A}, \mathbf{B}, \mathbf{C}, \mathbf{D})$ to the linear
equations (\ref{LC_linear_system}), with disregard of the
constraints, is a linear vector space. A basis $B=\{b_1,\dots,
b_d\}$ of ${\cal V}$ can be calculated efficiently in ${\cal O
}(N^4)$ time by standard Gauss elimination over $\mathbb{F}_2$. Then
we can search the space ${\cal V}$ for a vector which satisfies
the constraints. As (\ref{LC_linear_system}) is for large $N$ a
highly overdetermined system of equations, the space ${\cal V}$ is
typically low-dimensional. Therefore, in the majority of cases
this method gives a quick response. Nevertheless, in general one
cannot exclude that the dimension of ${\cal V}$ is of order ${\cal
O}(N)$ and therefore the overall complexity of this approach is
non-polynomial. However, it was shown in ref.~\cite{Bouchet} that it is
sufficient to enumerate the subset \be{\cal V'}:= \left\{ b +b' \
|\ b, b'\in B \right\}\subseteq {\cal V}\ee
  in order to find a solution which
satisfies the constraints, if such a solution exists, where one
observes that $|{\cal V}'| = {\cal O}(N^2)$. This leads to a
polynomial time algorithm to detect LC-equivalence of graph
states. The overall complexity of the algorithm is ${\cal
O}(N^4)$. We note, however,  that it is to date not known whether it is
possible to compute the LC orbit of an arbitrary graph state
efficiently.

\index{binary representation|)}

\subsubsection{\it \small Generalizations to $d$-level systems}\index{qudits|see{d-level system}}\index{d-level system}\label{GS_dlevel}

 The stabilizer formalism can be generalized to $d$-level systems, where
 $d$ is a prime power, see
 refs.~\cite{Schlinge02a,schlinge04,ZZXS03,Hostens04,David}.
 In such a generalizations to
 systems where the Hilbert spaces of constituents
 are ${\mathbbm{C}}^d$, a lot
 of the intuition developed for binary systems carries over.
 We will here
 sketch the situation only in which
 the individual constituents take a prime
 dimension. Ironically, in this more general
 framework, the case of
 binary stabilizer states even constitutes a special case, which has to be
 treated slightly differently than other prime dimensions. Actually,
 the language reminds in many respects of
 the setting of `continuous-variable systems' with
 canonical coordinates \cite{Gaussian}.
 The familiar real phase space in the latter setup is then
replaced by a discrete phase space. Also
the Weyl operators, so familiar
in the quantum optical context, find their equivalent
in the discrete setting.

At the foundation of this construction is the
unique {\it finite field}  ${\mathbbm{F}}_d$ of prime
order $d$. All arithmetic
operations are defined modulo $d$.
We may label a basis of ${\mathbbm{C}}^d$ as usual as
$|0\rangle,...,|d-1\rangle$.
The {\it shift operators} and the {\it clock} (or  {\it phase} or
{\it multiplier}) {\it operators}
are then defined as
\begin{eqnarray}
    U_{x} |m\rangle &:=& | m + x\rangle,\\
    V_{p} |m\rangle &:=& e^{i \frac{2\pi}{d} p  m }
    |m\rangle .
\end{eqnarray}
for $x,p\in {\mathbbm{F}}_d$. The number
$\omega:= e^{i \frac{2\pi}{d}}$ is a primitive $d$-th root
of unity.
Let us assume that $d$ is prime but
exclude the case $d=2$.
Using the above shift and clock
operators, one can associate with each point
$(x,p)\in{\mathbbm{F}}_d^2$ in
phase space a
{\it Weyl operator} according to
\begin{equation}
    w(x,p ) :=
    V_p U_x.
\end{equation}
These Weyl operators correspond to translations
in phase space. In analogy to the
previous considerations, we may define the
elementary operators
\begin{equation}
    X= U_1,\,\,\, Z=V_1,
\end{equation}
satisfying $X^d={\mathbf{1}}$ and $Z^d={\mathbf{1}}$.
The operators
\begin{equation}
    \omega^{v}\omega^{-2^{-1} px}
    w(x,p)
\end{equation}
for $(x,p,v)\in {\mathbbm{F}}_d^3$
form a representation of the {\it Heisenberg  group}
with its associated group composition law.
%
The Weyl operators in this sense
can be conceived as generalized
Pauli operators familiar from the binary setting.
They satisfy the {\it Weyl commutation
relations}
\begin{equation}
    w(x,p) w(x',p') =
    \omega^{ - p' x}w(x+x', p+p'),
\end{equation}
as can be readily verified using the above
definitions,
so the product of two Weyl operators
is up to a number again a Weyl operator.
This is the discrete analog of the familiar
canonical commutation relations for position and
momentum for Weyl operators in continuous
phase space, which takes essentially
the same form. It follows that
two Weyl operators $w(x,p)$ and $w(x',p')$
defined in this way commute if and only if
\begin{equation}
    [(x,p),(x',p')] =0,
\end{equation}
so if and only if the {\it standard symplectic
scalar product} vanishes, which is defined as
\begin{equation}
    [(x,p), (x',p')] := (x,p)\left(
    \begin{array}{cc}
    0 & -1\\
    1 & 0
    \end{array}
    \right)
    (x',p')^T = p x' - x p'
\end{equation}
for $(x,p)\in{\mathbbm{F}}_d^2$.
This is a antisymmetric bi-linear form in that
\begin{equation}
    [(x,p), (x',p')] = - [(x',p'), (x,p)].
\end{equation}
Hence, the discrete phase space is a
symplectic space over a finite field.
In turn,
the linear combinations of all Weyl
operators form an algebra, the full
observable algebra of the system.

The composition of $N$ constituents of a composite systems, each of dimension $d$,
can be incorporated in a natural fashion. We now encounter coordinates in phase space
$(x_1,p_1,...,x_N, p_N)$ with $\mathbf{x}\in {\mathbbm{F}}^N_d$ and $\mathbf{p} \in {\mathbbm{F}}^N_d$.
The above symplectic scalar product is then replaced by the one defined as
\begin{equation}
    [(\mathbf{x},\mathbf{p}),(\mathbf{x}',\mathbf{p}')] = (\mathbf{x},\mathbf{p}) \cdot {\pmb \sigma}
    \cdot (\mathbf{x}',\mathbf{p}')^T,
\end{equation}
with
\begin{equation}
    {\pmb\sigma} := \bigoplus_{j=1}^N
    \left(
    \begin{array}{cc}
    0 & -1\\
    1 & 0
    \end{array}
    \right).
\end{equation}
Similarly, the Weyl operators become
\begin{equation}
    \mathbf{w}(\mathbf{x},\mathbf{p}) = w(x_1,p_1)
     \otimes ...\otimes w(x_N,p_N),
\end{equation}
and let us set
\begin{equation}
    W(\mathbf{x},\mathbf{p}) := \omega^{-\frac{\mathbf{p}\cdot \mathbf{x}}{2} } \mathbf{w}(\mathbf{x},\mathbf{p}).
\end{equation}

We now turn to the actual definition of stabilizer codes and
stabilizer states \cite{Schlinge02a,schlinge04,David}.
At the foundation here is the
definition of an isotropic
subspace. An {\it isotropic space}
$\mathbf{S} \subset {\mathbbm{F}}_d^{2N}$
is a subspace on which the symplectic
scalar product vanishes
for all pairs of its vectors, so where
\begin{equation}
    [(\mathbf{x},\mathbf{p}),(\mathbf{x}',\mathbf{p}')]=0
\end{equation}
for all $(\mathbf{x},\mathbf{p}),(\mathbf{x}',\mathbf{p}')\in \mathbf{S}$.
Now, let $\chi$ a character as a map from $\mathbf{S}$ into the circle
group (for example, the map mapping all elements of $\mathbf{S}$ onto
$1$).
Let us denote the dimension of
$S$ with $k$. Then, the projector onto the stabilizer
code associated with the isotropic subspace $\mathbf{S}$ and
the character $\chi$ can be written as
\begin{equation}
    P = \frac{1}{d^k} \sum_{(x,p)\in \mathbf{S}} \chi^*(\mathbf{x},\mathbf{p})
    W(\mathbf{x},\mathbf{p}).
\end{equation}
In particular, the state vectors $|\psi\rangle$
from this stabilizer code are exactly those
that satisfy
\begin{equation}
      \chi^*(\mathbf{x},\mathbf{p})
      W(\mathbf{x},\mathbf{p}) |\psi\rangle  = |\psi\rangle ,
\end{equation}
for all $(\mathbf{x},\mathbf{p})\in \mathbf{S}$. In other words, this state vector
$|\psi\rangle$ is an eigenvector of
all of
the operators $\chi^*(\mathbf{x},\mathbf{p})  W(\mathbf{x},\mathbf{p})$ with the same eigenvalue $+1$, as we encountered it in the binary
setting. Again, it is said that $|\psi\rangle$ is {\it stabilized} by these operators. The above Weyl operators are, notably, no longer Hermitian. Hence, they do per se allow for an interpretation in terms of natural constraints
to the correlations present in the state.
This setting can be naturally been generalized to
prime power dimension $d=p^r$ with $p$ being prime
and $r$ being an integer. If, however, the underlying integer
ring is no longer a field, one loses the vector
space structure of ${\mathbbm{F}}_d$, which
demands some caution with respect to the concept
of a basis\footnote{If $d$ contains multiple prime factors the stabilizer, consisting of $d^N$ different elements, is in general no longer generated by a set of only $N$ generators. For the minimal generating set more elements $N\leq m\leq 2N$ of the stabilizer might be needed \cite{Hostens04}.}

Similarly, a {\it stabilizer code} can be conceived in this picture as
the image of an isotropic subspace $\mathbf{S}$ under the
Weyl representation. If a stabilizer code is
one-dimensional, it is a {\it stabilizer state}.

Again, any stabilizer state can be represented as
a graph state, up to local Clifford operations.
This has been shown in refs.~\cite{Grassl02,Schlinge02b}.
The notion of a {\it Clifford operation}
still makes sense,
as a unitary that maps Weyl operators onto
Weyl operators under conjugation,
\begin{equation}
    U W(\mathbf{x},\mathbf{p}) U^\dagger \propto W({\mathbf{Q}}(\mathbf{x},\mathbf{p})),
\end{equation}
where ${\mathbf{Q}}$ is an element of the symplectic group,
so preserves the above symplectic form.

The respective graph state corresponds to a {\it weighted graph} with
weights ${\bf \Gamma}_{ab}\in {\mathbbm{F}}_d$,
with $a,b$ again being associated with the vertices
of the underlying graph. A graph state is now
a state stabilized by the operators
\begin{eqnarray}
    K_a =
    U^a_1 \prod_{b\in N_a} (V^b_1)^{{\Gamma}_{ab}}
    =
    X^a \prod_{b\in N_a} (Z^b)^{{\Gamma}_{ab}}.
\end{eqnarray}
The symmetric adjacency
matrix ${\bf \Gamma}$ contains elements
${\Gamma}_{ab} \in {\mathbbm{F}}_d$ and, thus, it has no longer
binary entries as in the case of a simple graph as in qubit systems.
The interaction is instead specified by a {\it strength} $r={\bf \Gamma}_{ab} $
given by the weight of the edge $\{a,b\}$ in the weighted
graph. Note, however, that this concept of a graph state
based on a weighted graph is different from the one used in the
remaining part of this review article (see sec.~\ref{WeightedGS}).
When conceiving the preparation of the graph state
via the successive application of {\it phase gates}, the
associated unitary $U_{ab}^r$ acting on the
Hilbert spaces of the systems labeled $a$ and $b$
is given by
\be U^r_{ab}\,
|m\rangle^a|n\rangle^b\, = \, \omega^{- r m n}
|m\rangle^a|n\rangle^b \; .\ee

This picture of graph states in discrete Weyl systems,
embodying the case of $d$-dimensional systems, as
well as their processing in the context of the one-way
computer, has been considered in detail in refs.~\cite{schlinge04}. Also,
quantum error correcting codes have in this setting
been described in ref.~\cite{schlinge04}.
This language of discrete
Weyl systems provides a clear-cut picture to describe
finite-dimensional systems in phase space.

\subsubsection{\it \small Remarks on harmonic systems}
\index{qudits|see{harmonic systems}}\index{harmonic systems}
\label{GS_harmonic}

Finally, it is worth noting that
the close analogy between discrete and continuous
Weyl systems suggests the existence of
similar structures as graph states in the setting
of quantum systems with canonical coordinates,
so systems in a real phase space with
position and momentum coordinates.
Variants of such an idea have been considered
in a number of publications \cite{HC,Pl04,Frust,Zhang}; to describe
them in detail, yet, would be beyond the scope of
this review article. Here, we rather note the
structural similarities to the previous
finite-dimensional setting.

 The phase space of a system with
 $N$ canonical degrees of freedom -- $N$ harmonic oscillators --
 is ${\mathbbm{R}}^{2N}$, equipped with an antisymmetric bi-linear form
defined by
 \begin{equation}
    {\pmb\sigma} =
    \bigoplus_{j=1}^N
    \left(
    \begin{array}{cc}
    0 & 1 \\
    -1 & 0\\
    \end{array}
    \right).
\end{equation}
This form originates from the canonical
commutation relations between the {\it canonical
coordinates} of position and momentum,
which can be collected in a row vector as
$(r_1,..., r_{2N})=
( x_1, p_1,..., x_N, p_N)$. These canonical coordinates satisfy the
{\it canonical commutation relations} between position and momentum,
although  these variables can, needless to say, correspond to
quadratures of field modes. As before (in this now real
phase space) we
may introduce {\it Weyl operators} embodying translations in
real phase space, defined as
\begin{equation}
     W(\mathbf{x},\mathbf{p}) = e^{i (\mathbf{x},\mathbf{p}) {\pmb\sigma} (\mathbf{x},\mathbf{p})^T}
\end{equation}
for $(\mathbf{x},\mathbf{p})\in {\mathbbm{R}}^{2N}$. This Weyl operator is, in a number
of different conventions, a frequently used tool in quantum
optics under the name of {\it displacement operator}.
These Weyl operators inherit the canonical commutation relations:
it is easy to see that they satisfy the Weyl relations
\begin{equation}
    W(\mathbf{x},\mathbf{p})  W(\mathbf{x}',\mathbf{p}') = e^{-i (\mathbf{x},\mathbf{p}) \sigma (\mathbf{x}',\mathbf{p}') }
    W(\mathbf{x}+\mathbf{x}',\mathbf{p}+\mathbf{p}').
 \end{equation}
The structural similarities are obvious.
The {\it characteristic function} is here, just
as in the discrete case, defined as the expectation value of the Weyl operator,
\begin{equation}
    f(\mathbf{x},\mathbf{p}) = \text{tr}( W(\mathbf{x},\mathbf{p}) \rho)
\end{equation}
for $(\mathbf{x},\mathbf{p})\in {\mathbbm{R}}^{2N}$.
This is a generally complex-valued function in phase space,
uniquely defining the quantum state. Hence, the description in terms
of Weyl systems serves also as a language appropriate for the
description of both the
discrete and as well as the infinite-dimensional setting.

A certain class of states for which the assessment of entanglement
is relatively accessible is the important class of {\it Gaussian states}.
They
are those quantum states for which the characteristic function is a Gaussian.
Then, the first moments, $d_j =
\text{tr}(r_j\rho)$ and the second moments fully characterize the quantum
state. The {\it second moments}, in turn,
can be embodied in the real symmetric $2N\times 2N$-matrix
${\pmb \gamma}$, the entries of which are given by
\begin{equation}
    \gamma_{j,k}=
    2 \text{Re}\,
    \text{tr}
    \left((r_j - d_j)
    (r_k - d_k)\rho \right),
\end{equation}
$j,k=1,...,N$. This matrix is typically referred to as
the {\it covariance matrix} of the state.
Similarly, higher moments
can be defined.

Analogues or  `close relatives'
of graph states in the Gaussian setting
now arise in several context: (i) They can be thought of as originating
from an interaction pattern, similar to the interaction pattern for Ising
interactions \cite{Zhang}. These interactions may
arise from squeezing and Kerr-like interactions. (ii)
They can also be resulting as ground states from
Hamiltonians which are specified by a simple graph, in turn
reflecting the interaction terms in the Hamiltonian
\begin{equation}
    H= \mathbf{p} \mathbf{p}^T/2 + \mathbf{x} \mathbf{V} \mathbf{x}^T,
\end{equation}
where again $\mathbf{p}=(p_1,...,p_N)$, $\mathbf{x}=(x_1,...,x_N)$, and the
real symmetric matrix $N\times N$-matrix  $\mathbf{V}$ incorporates
the interaction pattern as the adjacency matrix of a weighted graph
\cite{HC,Pl04,Frust,Gap}. Then, the resulting covariance matrix is
nothing but $\gamma = \mathbf{V}^{-1/2}\oplus \mathbf{V}^{1/2}$, when ordering
the entries in the convention of $(x_1,...,x_N, p_1,...,p_N)$.
(iii) Also, the
direct analog of stabilizer state vectors
(conceived as state vectors
`stabilized by a stabilizer')
in the setting of continuous
Weyl systems still makes sense, yet one has
to allow for singular
states \cite{Infinite} which can no longer be associated
with elements of the Hilbert space of square
integrable functions, but can conveniently be
described in an algebraic language (or within a Gelfand
triple approach).


\subsection{Alternative approaches}\label{DefOfGS_Alternative}

Due to the description in terms of their stabilizer, graph states can be represented, under suitable interpretations, by various mathematical structures, which connect these objects also to other areas of applications in classical cryptography and discrete mathematics. For example, a graph code can be described by a {\it self-dual additive code over the field $\mathbb{F}_4=GF(4)$} or by a (quantum) {\it set of lines of the projective space over $\mathbb{F}_2=GF(2)$} \cite{Calderbank98,Glynn02}. Graph states are also equivalent to {\it quadratic boolean functions} \cite{database}, which are used in classical cryptography.

\index{density matrix renormalization group (DMRG)|(}
\index{Valence Bond Solids (VBS)|(}
\begin{wrapfigure}[13]{r}{0.4\textwidth}
\vspace{-0.5cm}
\includegraphics[width=0.4\textwidth]{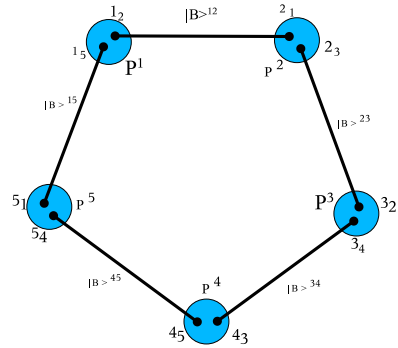}
\caption{\label{Fig:Ring2}A graph state for a ring with $5$ vertices as valence bond solid.}\label{Ring2}
\end{wrapfigure}
In the remainder of this section we focus on another description of graph states in terms of {\it Valence Bond Solids} (VBS), which does not rely on the stabilizer formalism and, hence, can be extended to weighted graph states (see sec.~\ref{WeightedGS}). This representation was recently introduced by Verstraete and Cirac \cite{Ve04} and has already found interesting applications in density-matrix renormalization group (DMRG)\footnote{For a review of these methods we refer the reader to ref.~\cite{Sc05}.} methods. The VBS picture has its roots in the Affleck-Kennedy-Lieb-Tasaki (AKLT) model \cite{AKLT}, which allows to find exact expressions for the ground states or exited states of some particular Hamiltonians. In DMRG variational methods are applied to a generalization of these AKLT-states, the so-called {\it matrix-product states} \cite{MPS}, in order to perform numerical studies of various physical systems, especially within the field of condensed matter physics. Lately, VBS states have attracted some attention, since they allow for a clear reformulation of DMRG algorithms leading to improvements of the DMRG methods for the simulation of many-body systems in two or higher dimensions or with periodic boundary conditions \cite{Ve042d,AdvancedDMRG}.

In the context of this article graph states can also be regarded as particular VBS states: Here, the {\em graph state} $|G\rangle$ arises from a set of Bell pairs (bonds) \be |B\rangle^{a^ib^j} = U_{a^ib^j} |+\rangle^{a^i} |+\rangle^{b^j} 
\ee  between some {\em virtual} qubits after some suitable projections
$P$ onto the {\em real} qubits (see fig.~\ref{Fig:Ring2}):
\be  |G\rangle =  \prod_{a \in V} P_a \prod_{\genfrac{}{}{0pt}{}{a^i,b^j}{\{a,b\}\in E}} |B\rangle^{a^ib^j} \ee

More precisely, the graph state can be obtained following the procedure:
\begin{itemize}
\item[1.] Replace the real qubits at each vertex $a$ by $d_a$ virtual qubits $a^1,\ldots, a^{d_a}$, where $d_a=|N_a|$ denotes the {\em degree} of the vertex $a$.
\item[2.] For each edge $\{a,b\}$ in $G$ create a Bell pair $|B\rangle^{a^ib^j}$ between some virtual qubit $a^i$ at vertex $a$ and some virtual qubit $b^j$ at vertex $b$ by using the Ising interaction $U$.
\end{itemize}

\begin{wrapfigure}[11]{r}{0.55\textwidth}
\vspace{-1cm}{\setlength{\unitlength}{1cm}
\begin{picture}(8,4)
\put(0.5,0){\includegraphics{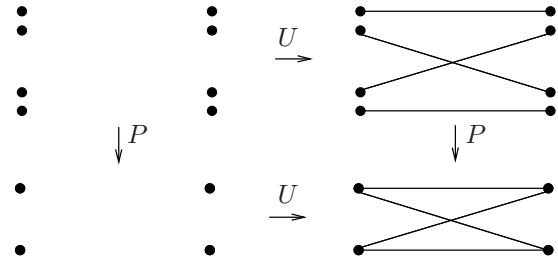}}
\put(6.5,1.5){$P$}
\put(4,0.7){$U$}
\put(2,1.5){$P$}
\put(4,2.8){$U$}
\end{picture}}
\caption{\label{CommutingDiagram} The phase gate on the level of the virtual qubits `commutes' with the projection onto the real physical qubits.}
\end{wrapfigure}
\begin{itemize}
\item[3.] Project all virtual qubits at each vertex $a$ into the real qubit (sub-)system by
\be P_a := ^a\hspace{-0.05cm}|\tilde 0\rangle\langle 0 |^{a^1}\ldots\langle 0 |^{a^{d_a}} + ^a\hspace{-0.05cm}|\tilde 1\rangle\langle 1 |^{a^1}\ldots\langle 1 |^{a^{d_a}} \, .  \ee
\end{itemize}
That this procedure provides an equivalent description for graph states, can be shown inductively using the fact that the phase gate $U_{a_ib_j}$ on the level of the virtual qubits `commutes' with the projection onto the real physical qubits, i.e.
\be [U_{a^ib^j},P_{c}]=0 \hspace{1cm} \forall a^i,b^j, c=(c^1,\ldots,c^{d_c}) \; .\ee For the twisted four-qubit ring this is depicted by the commutative diagram in fig.~\ref{CommutingDiagram}.
\index{density matrix renormalization group (DMRG)|)}
\index{Valence Bond Solids (VBS)|)}


\section{Clifford operations and classical simulation}\label{Pauli measurements}\index{Clifford operations|(}

The stabilizer formalism is not only suited to describe states (or codes), but also to calculate the action of Clifford operations on these states. {\em Clifford operations} are (possibly non-local) Clifford unitaries $U\in \mathcal{C}_N$ (see eq.~(\ref{Def_CliffU})) and projective measurements of a Pauli operator $s\in \mathcal{P}^V$, which we will call {\em Pauli measurements}. The restriction to projective measurements in the Pauli basis ensures that such measurements performed on stabilizer states [codes] yield again stabilizer states [codes] as measurement results \cite{Gottesman,NielsenBook}.

It is not necessary to consider measurements of arbitrary Pauli operators $s\in \mathcal{P}^V$. Since it is possible to efficiently decompose an arbitrary Clifford unitary $U\in \mathcal{C}_N$ in terms of the one- and two-qubit gates $H$, $S$ and CNOT (see sec.~\ref{Def_LC}),  any Clifford operation can be simulated by  a sequence of at most ${\cal O}(N^2)$ of these gates together with one Pauli measurement (say $s=\sigma_z$) at a single vertex. Thus a sequence of Clifford operations acting on some stabilizer state can be replaced by a sequence of one- and two-qubit gates $H$, $S$ and CNOT and single-qubit Pauli measurements with only a polynomial overhead ${\cal O}(N^2)$ in the number of gates. In the circuit model for quantum computation an equivalent scheme is often considered. The class of quantum computations that involve only
\begin{itemize}
\item state preparations in the computational basis,
\item the one- and two qubit gates $H$, $S$ and CNOT and\index{Hadamard gate $H$}\index{controlled NOT gate (CNOT)}
\item measurements of observables in the Pauli group $\mathcal{P}$, including the classical control of gates conditioned on the outcome of such measurements,
\end{itemize}
is called the class of {\em stabilizer circuits}\index{stabilizer circuit}. All the states of the `quantum register' in each step of such a stabilizer circuit are stabilizer states. These states can be characterized by their set of stabilizer generators. A formal representation of this set of generators in the memory of a classical computer\footnote{In the binary representation a computer has to store the generator matrix and additional phases at each qubit. The matrix requires ${\cal O}(N^2)$ memory size, whereas for the phases a register of size ${\cal O}(N)$ is sufficient. With this information the complete stabilizer can be recovered (see sec.~\ref{Def_Stab_States}).} allows one to efficiently keep track of all changes by pure classical computation. The effect of the one- and two-qubit gates as well as the one-qubit Pauli measurements to the generating set can be calculated using ${\cal O}(N^3)$\footnote{Note that the update of the stabilizer can be determined in only ${\cal O}(N^2)$, but the determination of the exact measurement result in the case of measuring a Pauli-matrix $\pm\sigma_i \in \mathcal{S}$ seems to require some Gaussian elimination, which needs ${\cal O}(N^3)$ time in practice \cite{Aaronson04}.} steps on a classical computer. In this way, any Clifford operation can be efficiently simulated on a classical computer, which is the content of the {\em Gottesman--Knill theorem}\index{Gottesman--Knill theorem} \cite{Gottesman99,NielsenBook}.
{\proposition[{\bf Gottesman--Knill theorem}]\label{GottKnillTh}
Any stabilizer circuit on a quantum register of $N$ qubits, which consists of $M$ steps, can be simulated on a classical computer using at most ${\cal O}(N^3 M)$ elementary classical operations.
}
\proofend

\begin{wrapfigure}[16]{r}{0.5\textwidth}
\vspace{-0.0cm}
\includegraphics[width=0.47\textwidth,clip]{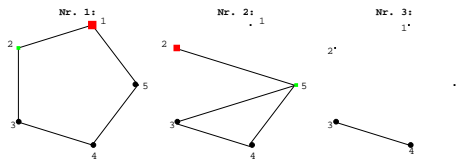}
\caption{\label{fig:xMeasurementExample1}\small Example for a $\sigma_x$-measurement at vertex $1$ in graph No.\ 1, which is followed by a $\sigma_z$-measurement at vertex $2$: In graph  No.\ 1 a $\sigma_x$-measurement is performed at the vertex $1$. For the application of the $x$-measurement rule, vertex $2$ was chosen as the special neighbor $b_0$, yielding the graph No.\ 2 up to a local unitary $U_{x,\pm}^{(1)}=  (\pm i \sigma_y^{(2)} )^{1/2}$. As stated in Table \ref{tabl}, the subsequent $\sigma_z$-measurement on the new graph state is therefore essentially another $\sigma_x$-measurement, now at vertex $2$ with a single neighbor $b_0=5$. The final graph is then graph No.\ 3.}
\end{wrapfigure}
Although this result was already known for a few years \cite{Gottesman99}, only very recently, such classical simulator was implemented \cite{Aaronson04} that actually requires only ${\cal O}(N^2)$ elementary operations on classical computer. In the remainder of this section we will see how graph states can provide an alternative algorithm. This algorithm is based on elementary graph manipulations and was proposed and implemented in ref.~\cite{Anders}. Its complexity for the elementary gate operation requires ${\cal O}(d^2)$ basic steps on a classical computer, where $d$ denotes the maximal degree of the graph representing the quantum register. This proposals is hence advantageous if this maximal degree remains comparably small to $N$ for the different register states throughout the computation.

According to sec.~\ref{Def_Stab_States} any stabilizer state $|\mathcal{S}\rangle$ can be represented as a graph state $|G\rangle$ up to some LC-unitaries $U\in\mathcal{C}_1^V$. Thus, in order to keep track\footnote{Usually a stabilizer circuit is required to start in some kind of standard input state $|0\rangle^V$ of the computational basis, which has a trivial representation in terms of graph states, i.e., $|0\rangle^V = H^V |+\rangle^V$. But the following argumentation will also hold for an arbitrary stabilizer state as the input, if one allows for a polynomial overhead at the beginning of classical simulation in order to determine the corresponding graphical representation.} of the different steps $i=1,\ldots,M$ in the stabilizer circuit computation, one has to store a local Clifford unitary $W_i$ as well as the graph $G_i$ of the graph state $|G_i\rangle$, which is LC-equivalent to the actual stabilizer state $|\mathcal{S}_i\rangle$ in the quantum register of step $i$, i.e., $|\mathcal{S}_i\rangle = W_i |G_i\rangle$. Note that the storage of a graph $G_i$ on $N$ vertices requires only $\frac{N(N-1)}{2}$ bits for the entries of the corresponding adjacency matrix. Moreover as discussed in sec.~\ref{Def_Stab_States} at each vertex $a$ the list of single-qubit LC-unitaries $W_i$ can be characterized by one of $24$ `permutations' $W_i^a$ of the Pauli matrices depicted in Table~\ref{Tab_LC}.

In order to be an efficient representation for the classical simulation of the stabilizer circuit, the graphical description remains to be provided with a set of graphical rules that account for the changes of the stabilizer when a one- or two-qubit gate or some single-qubit Pauli measurement is applied to it. The one-qubit Clifford unitary occurring at a vertex can easily be dealt with by updating the corresponding unitary according to some fixed `multiplication table'.  For the two-qubit unitaries we will, instead of the CNOT gate, consider the phase gate $U_{ab}$ in eq.~\ref{CPhase}, since on `pure' graph states it simply acts by adding or deleting the corresponding edge $\{a,b\}$. However, the case where $U_{ab}$ does not act directly on $|G\rangle$ but on $W |G\rangle$ for some LC-unitary $W$ that is non-trivial at the vertices $a$ and $b$, i.e., $W^a\neq \mathbf{1}_a$ or $W^b\neq \mathbf{1}_b$, requires a more careful treatment. Remember that each of the possible single-qubit unitaries has a decomposition in terms of elementary $\frac{\pi}{4}$--rotations given in Table~\ref{Tab_LC}. Since all unitaries $W_a = \mathbf{1}_a, z^a,  x^a, y^a , \sqrt{\pm i \sigma_z}$ can be `commuted through' the phase gate yielding at most some additional $\sigma_z$ on the vertex $a$ or $b$, e.g. $U_{ab}\sigma_x^a = \sigma_x^a \sigma_z^b U_{ab}$, we are left with the analysis of the $14$ additional cases, for which at least one unitary $W_a$ or $W_b$ is of the type $\sqrt{\pm i \sigma_x}$ or $\sqrt{\pm i \sigma_y}$.
The graphical rules for these cases can be obtained using the LC-rule derived in sec.~\ref{Def_LC} in order to remove these unitaries. For example, one finds that for an arbitrary graph $G$
\be U_{ab} \,\sqrt{\pm i \sigma_x}^a |G\rangle =  \sqrt{\mp i \sigma_z}^{N_a}\, U_{ab} |\tau(G) \rangle   \; ,\ee
since $\sqrt{+ i \sigma_x}^a =  \sqrt{- i \sigma_z}^{N_a} (U^\tau_a)^\dagger$ (similarly for  $\sqrt{- i \sigma_x}^a$), where $U^\tau_a$ denotes the LC-unitary in eq.~(\ref{LU_Rule_U}) for the LC-rule.
In fact in ref.~\cite{Anders} it is shown that, in this way, any of the remaining LC-unitaries $W_a$ at some vertex $a$ can be removed by means of at most five local complementations $\tau$ applied at this vertex or at one of its neighbors.

\begin{table}
\begin{center}
\begin{minipage}{0.8\textwidth}
\begin{tabular}{ccc}
\begin{minipage}{0.45\textwidth}
\begin{tabular}{|c|}
\hline
$P_{x,\pm} \sigma_z = \sigma_z P_{x,\mp}$,\\
$P_{y,\pm} \sigma_z = \sigma_z P_{y,\mp},$\\
$P_{z,\pm} \sigma_z = \sigma_z P_{z,\pm},$ \\
\hline 
$P_{x,\pm} (-i \sigma_z)^{1/2} = (-i\sigma_z)^{1/2} P_{y,\mp},$ \\
$       P_{x,\pm} (i \sigma_y)^{1/2} = (i \sigma_y)^{1/2} P_{z,\pm}$,\\
$       P_{x,\pm} (- i \sigma_y)^{1/2} = (- i \sigma_y)^{1/2} P_{z,\pm}$,\\
$    P_{x,\pm} (i \sigma_z)^{1/2} = (i\sigma_z)^{1/2} P_{y,\pm},$ \\
\hline
\end{tabular}
\end{minipage}

& \hspace{-1cm} &

\begin{minipage}{0.45\textwidth}
\begin{tabular}{|c|}
\hline
$P_{y,\pm} (-i\sigma_z)^{1/2}  = (-i\sigma_z)^{1/2} P_{x,\pm}, $ \\
$P_{y,\pm} (i\sigma_y)^{1/2}  =  (i\sigma_y)^{1/2} P_{y,\pm}, $ \\
$ P_{y,\pm} (-i\sigma_y)^{1/2}  =  (-i\sigma_y)^{1/2} P_{y,\pm},  $ \\
$ P_{y,\pm} (i \sigma_z)^{1/2}  =  (i\sigma_z)^{1/2} P_{x,\mp}, $ \\
\hline
$P_{z,\pm} (-i\sigma_z)^{1/2} = (-i\sigma_z)^{1/2} P_{z,\pm},  $ \\
$P_{z,\pm} (i \sigma_y)^{1/2} = (i \sigma_y)^{1/2} P_{x,\pm},$\\
$P_{z,\pm} (- i \sigma_y)^{1/2} = (- i \sigma_y)^{1/2} P_{x,\pm},$\\
$P_{z,\pm} (i \sigma_z)^{1/2} = (i\sigma_z)^{1/2} P_{z,\pm},  $ \\
\hline
\end{tabular}
\end{minipage}
\end{tabular}
\end{minipage}

\caption{ The relevant commutation relations for Pauli projections and Clifford
operators if a sequence of Pauli measurements is applied to a graph state.} \label{tabl}
\end{center}
\end{table}
Let us finally examine the effect of {\em single-qubit Pauli measurements} in more detail, since the graphical rules will be used in the subsequent secs. We will at first consider the case of a projective measurement of some Pauli operator $\sigma_x$, $\sigma_y$ or $\sigma_z$ at a singe vertex $a$ in a graph state without additional LC-unitaries at this vertex and will later mention how to cope with the general case.
For a Pauli measurement of the graph state $|G\rangle$ at a vertex $a$ we find that the graph $|G'\rangle$ on the remaining unmeasured vertices can be obtained from the initial graph $G$ by means of vertex deletion and local complementation:
\begin{description}\label{MeasRuleList}\index{Pauli measurements}\index{measurement rule}
\item[$\sigma_z:$] deleting the vertex $a$ from $G$;
\item[$\sigma_y:$] inverting $G[N_a]$ and deleting $a$;
\item[$\sigma_x:$] choosing any $b_0 \in N_a$, inverting $G[N_{b_0}]$, applying the rule for $\sigma_y$ and finally inverting $\tilde{G}[N_{b_0}]$ again.
\end{description}
This is the content of the following proposition  \cite{He04,schlinge04}.
{\proposition[{\bf Local Pauli measurements}] \label{Pauli_Measurement}
A projective measurement of $\sigma_{x}$, $\sigma_{y}$, or $\sigma_{z}$ on the qubit associated with a vertex $a$ in a graph $G$ yields up to local unitaries $U_{i,\pm}^a$ a new graph state $|G'\rangle$ on the remaining vertices.
The resulting graph $G'$ is
\begin{eqnarray}
 P^a_{z,\pm} |G\rangle  & = &  \frac{1}{\sqrt{2}}\, |z,\pm\rangle^a \otimes U_{z,\pm}^a |G-a\rangle ,\\
 P^a_{y,\pm} |G\rangle  & = &  \frac{1}{\sqrt{2}}\, |y,\pm\rangle^a \otimes U_{y,\pm}^a |\tau_a(G)-a\rangle, \\
 P^a_{x,\pm} |G\rangle  & = &  \frac{1}{\sqrt{2}}\, |x,\pm\rangle^a \otimes U_{x,\pm}^a |\tau_{b_0}\left(\tau_a\circ\tau_{b_0} (G)-a\right)\rangle\, ,
\end{eqnarray}
for any choice of some $b_0 \in N_a$, whenever the $\sigma_x$-measurement is not performed at an isolated vertex. If $a$ is an isolated vertex, then the outcome of the $\sigma_x^a$-measurement is $+1$, and the state is left unchanged.
The local unitaries $U_{i,\pm}^a$ are
\begin{eqnarray}
    U_{z,+}^a = &\mathbf{1},  & U_{z,-}^a = \sigma_z^{N_a}, \label{uz}\\
    U_{y,+}^a = &\sqrt{- i\sigma_z}^{N_a}, & U_{y,-}^a  = \sqrt{ + i \sigma_z}^{N_a} \label{uy}\\
    U_{x,+}^a = &\sqrt{+ i \sigma_y}^{b_0}  \sigma_z^{N_a\setminus(N_{b_0}\cup b_0)}, &
    U_{x,-}^a = \sqrt{- i \sigma_y}^{b_0}  \sigma_z^{N_{b_0}\setminus(N_{a}\cup a)}.\label{uxm}
\end{eqnarray}

}

For a measurement of $\sigma_{x}$ the local unitary $U_{x,\pm}$ depends on the choice of $b_0$. But the resulting graph states arising from different choices of $b_0$ and $b'_0$  will be equivalent via the LC-unitary $U_{b'_0}U^\dagger_{b_0}\in\mathcal{C}_1^V$.

For a sequence of local Pauli measurements, the local unitaries have to be taken into account, if the measured
qubit is affected by the unitary. We have summarized the necessary commutation relations in Table \ref{tabl}, which denote the transformation of the measurement basis, if a subsequent measurement is applied to a unitarily transformed graph state. Fig.~\ref{fig:xMeasurementExample1} shows two subsequent applications of
the rather complicated $\sigma_x$-measurement. An exhaustive table can also be provided for the general case that the Pauli measurement occurs at some vertex $a$ that has an arbitrary non-trivial LC-unitary $W_a$ attached to it.


{\em Proof} (of Proposition~\ref{Pauli_Measurement}):
The $\sigma_z$-measurement rule follows directly from the definition of graph states in terms of the underlying interaction pattern (see eq.~(\ref{GS_Preparation})):
\begin{eqnarray}
P_{z,\pm}^a |G\rangle & = & P_{z,\pm}^a \prod_{\{a,b\} \in E} U_{ab} \prod_{\genfrac{}{}{0pt}{}{\{c,d\} \in E}{c,d \neq a}} U_{cd} |+\rangle^V  \\
& = & P_{z,\pm}^a \,\left(P_{z,+}^a + P_{z,-}^a \sigma_z^{N_a}\right) \,|+\rangle^a \otimes|G\setminus a\rangle^{V\setminus a}  \nonumber \\
& = & \frac{1}{\sqrt{2}} \,\left\{ \begin{array}{ll} |z,+\rangle^a \otimes |G\setminus a\rangle^{V\setminus a} & \text{\small if measurement result is}\; m_z^a =+1 \\ |z,-\rangle^a \otimes \sigma_z^{N_a}|G\setminus a\rangle^{V\setminus a } & \text{\small if measurement result is}\; m_z^a =-1   \end{array} \right.  \nonumber
\end{eqnarray}
In other words, with probability $\tfrac{1}{2}$ a $\sigma_z$-measurement at a vertex of some graph state gives either $|G\setminus a \rangle$ as the graph state on the remaining vertices if the measurement outcome is $m_z=+1$ or $\sigma_z^{N_a}|G\setminus a \rangle$ as the graph state on the remaining vertices if the measurement outcome is $m_z=-1$.

With the LC-rule at hand (see Proposition~\ref{loc}) one can now derive the measurement rules for a $\sigma_x$- or $\sigma_y$-measurement from this $\sigma_z$-measurement rule.
For this, one can use commutation relations, which are similar to those in Table~\ref{tabl}, in order to show that
\begin{eqnarray} \label{ProofXMeas} P^{a}_{x,\pm}  & = &  U^\tau_{b_0}(G) P^{a}_{y,\pm} (U^\tau_{b_0}(G))^\dagger  \\
\label{ProofYMeas} P^{a}_{y,\pm} & = & U^\tau_{a}(G) P^{a}_{z,\mp} (U^\tau_{a}(G))^\dagger \, \end{eqnarray} where $b_0$ is a neighbor\footnote{Note that if $a$ is an isolated vertex the graph state is $|G\rangle = |+\rangle^a \otimes |G'\rangle^{V\setminus a}$ for some graph on the remaining vertices. In this case a $\sigma_x$-measurement yields $+1$ with probability $1$.} of $a$ in $G$.
With eq.~(\ref{ProofYMeas}) and using $e^{\pm i\frac{\pi}{4}\sigma_k}=\sqrt{\pm i\sigma_k}=\frac{\mathbf{1}\pm i\sigma_k}{\sqrt{2}}$ we can now compute
\begin{eqnarray}
P^{a}_{y,\pm} |G\rangle & = & U^\tau_{a}(G) P^{a}_{z,\mp} |\tau_a(G)\rangle  \nonumber \\
& = & U^\tau_{a}(G) \frac{1}{\sqrt{2}} \,\left\{ \begin{array}{ll} |z,-\rangle^a \otimes \sigma_z^{N_a}|\tau_a(G)\setminus a\rangle^{V\setminus a} & \text{\small if }\; m_y^a =+1 \\ |z,+\rangle^a \otimes |\tau_a(G)\setminus a\rangle^{V\setminus a } & \text{\small if}\; m_y^a =-1   \end{array} \right.  \nonumber  \\
& = &  \frac{1}{\sqrt{2}} \,\left\{ \begin{array}{ll} \sqrt{-i\sigma_x^a}|z,-\rangle^a \otimes \sqrt{i\sigma_z^{N_a}} \sigma_z^{N_a}|\tau_a(G)\setminus a\rangle^{V\setminus a} & \text{\small if }\; m_y^a =+1 \\ \sqrt{-i\sigma_x^a} |z,+\rangle^a \otimes \sqrt{i\sigma_z^{N_a}} |\tau_a(G)\setminus a\rangle^{V\setminus a } & \text{\small if }\; m_y^a =-1   \end{array} \right.  \nonumber  \\
& \propto &  \frac{1}{\sqrt{2}} \,\left\{ \begin{array}{ll}
|y,+\rangle^a \otimes \sqrt{-i\sigma_z^{N_a}} |\tau_a(G)\setminus a\rangle^{V\setminus a} & \text{\small if }\; m_y^a =+1 \\
|y,-\rangle^a \otimes \sqrt{i\sigma_z^{N_a}} |\tau_a(G)\setminus a\rangle^{V\setminus a} & \text{\small if }\; m_y^a =-1
\end{array} \right.
\end{eqnarray}
This is the $\sigma_y$-measurement rule, from which the $\sigma_x$-measurement rule now can be derived along the same lines using eq.~(\ref{ProofXMeas}).
\proofend

\index{Clifford operations|)}


\section{Examples and applications}\label{GS_Examples}

In this section we give some prominent, not necessarily distinct classes of examples for graph states. We also sketch important applications of these states in multi-party quantum communication, quantum computation and quantum error correction. These examples illustrate that graph states do not only provide an interesting model to study multi-party entanglement, as it will be carried out in the following sections, but can also be an important resource for quantum information processing.

\subsection{GHZ--states}\label{GHZ_GS}

\begin{wrapfigure}[9]{r}{0.45\textwidth}
\vspace{-0.5cm}
\includegraphics[width=0.2\textwidth]{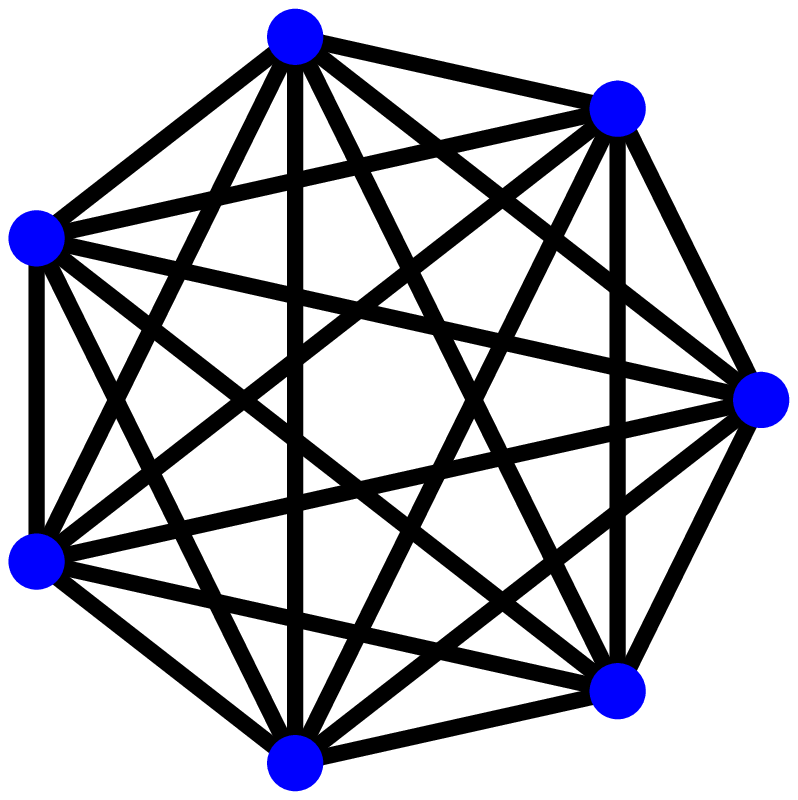}\hspace{0.03\textwidth}\includegraphics[width=0.2\textwidth]{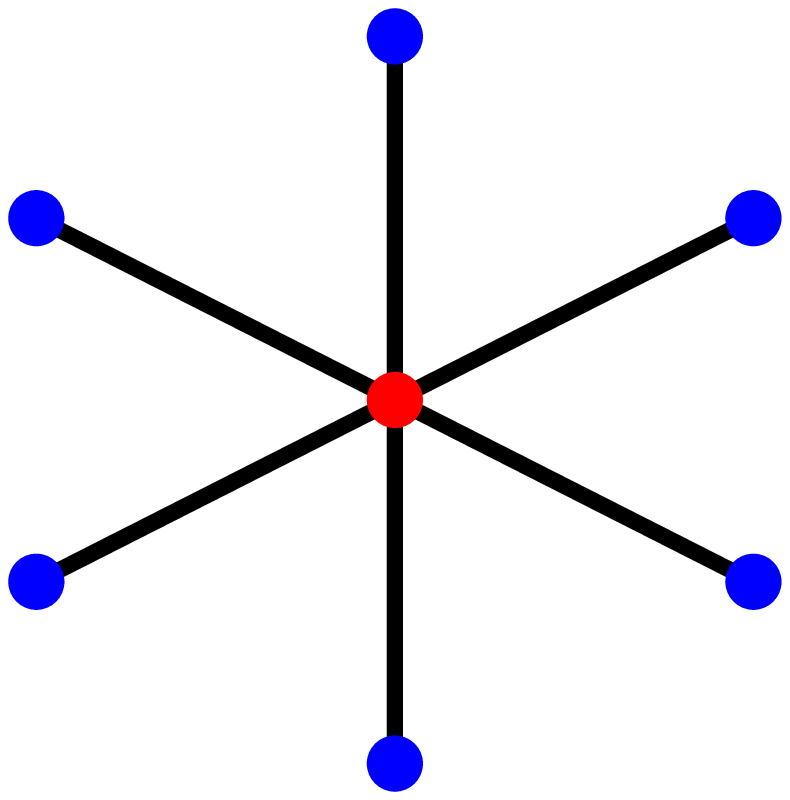}
\caption{\label{fig:GHZ} The GHZ-state is LU-equivalent to the graph state corresponding to a star graph or the complete graph.}
\end{wrapfigure}
We start by considering the $N$--qubit Greenberger-Horne-Zeilinger states\index{GHZ state}
\be
|GHZ\rangle=\frac{1}{\sqrt{2}}(|0\rangle^{\otimes N} + |1\rangle^{\otimes N}),\label{GHZstate}
\ee
which were introduced in ref.~\cite{GHZ89} for the case of three qubits and since serve as a `text book' example for  multi-party entangled `Schr\"odinger cat' states. These states are special examples of states that maximally violate multi-partite Bell inequalities \cite{BI} and can, for instance, be used to improve frequency standards \cite{Frequency}. GHZ states have also become an interesting resource for multi-party quantum communication, e.g. in the context of secret sharing and secure function evaluation \cite{Secure}. The multi--party {\em GHZ-state} corresponds to the {\em star graph}\index{star graph (state)|see{GHZ state}} and the {\em complete graph}\index{complete graph (state)}. This is easily seen by applying Hadamard unitaries $H^{V\setminus a}$ to all but one qubit $a$ in the GHZ-state, which yields the star graph state with $a$ as the central qubit. A further application of the LC-unitary $U_a^\tau$ for the LC-rule $\tau_a$ in Proposition~\ref{loc} then transforms the star graph state into the complete graph state. Thus the star graphs for different central vertices as well as the complete graph are LC-equivalent representations of the GHZ-state.

\subsection{Cluster states and the one-way quantum computer}\label{one-way-QC}\index{one-way quantum computer ($\text{QC}_G$)|(}\index{quantum computation}

The initial resource for quantum computation in the {\em one-way quantum computer} ($\text{QC}_G$), as it was introduced in refs.~\cite{Briegel01,OneWay1,OneWay2,OneWay3,OneWay5}, is the cluster state in two dimensions, which corresponds to a rectangular lattice. More generally, a {\em $d$-dimensional cluster state}\index{cluster (graph) state} is represented by a graph of the form of a $d$-dimensional lattice. But only cluster states of dimension two and higher can serve as a {\em universal resource} for quantum computation \cite{Nielsen05a}.
\begin{figure}
\vspace{-0.3cm}
\begin{center}
\includegraphics[width=0.25\textwidth]{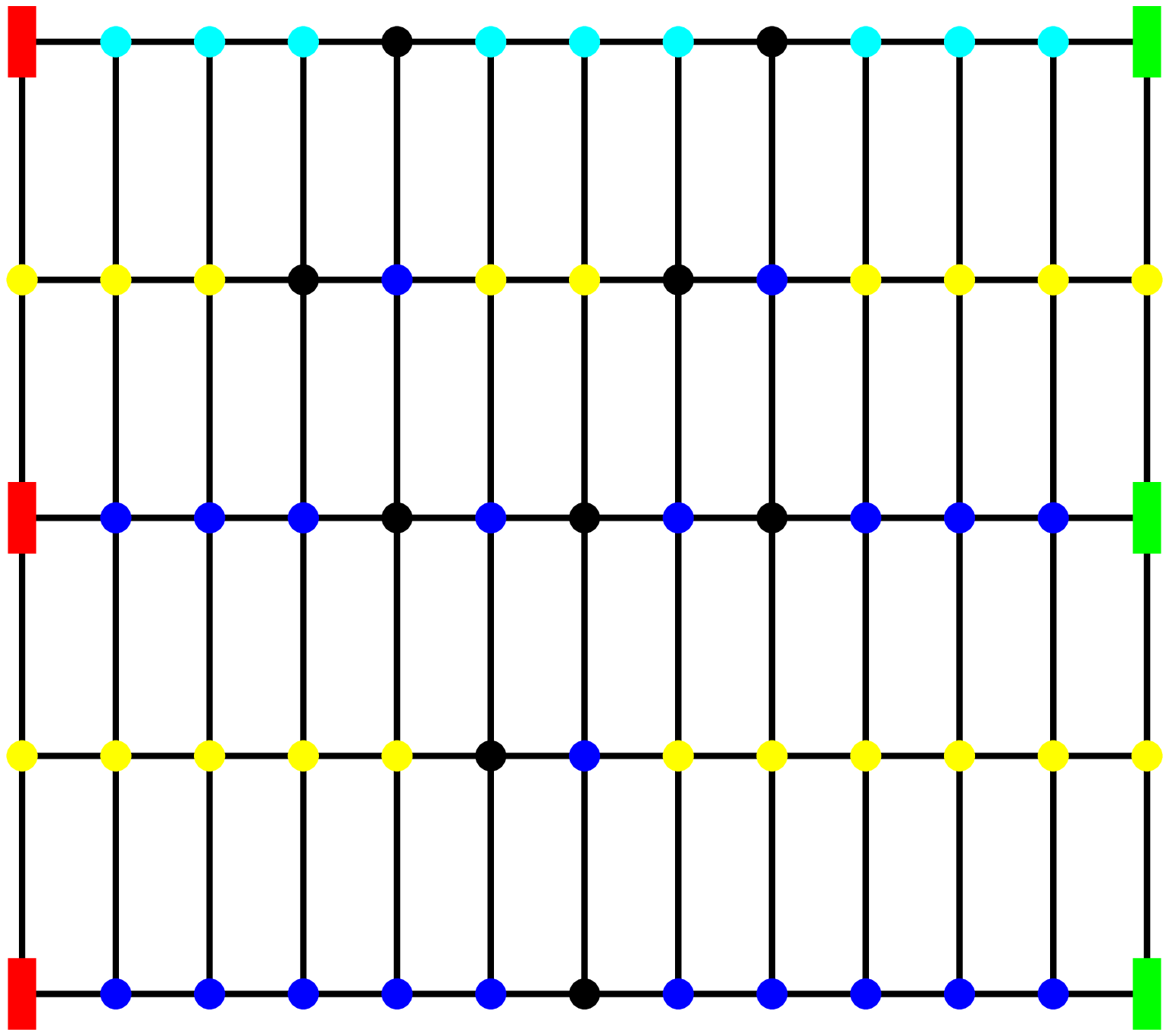}\hspace{0.05\textwidth} \includegraphics[width=0.35\textwidth]{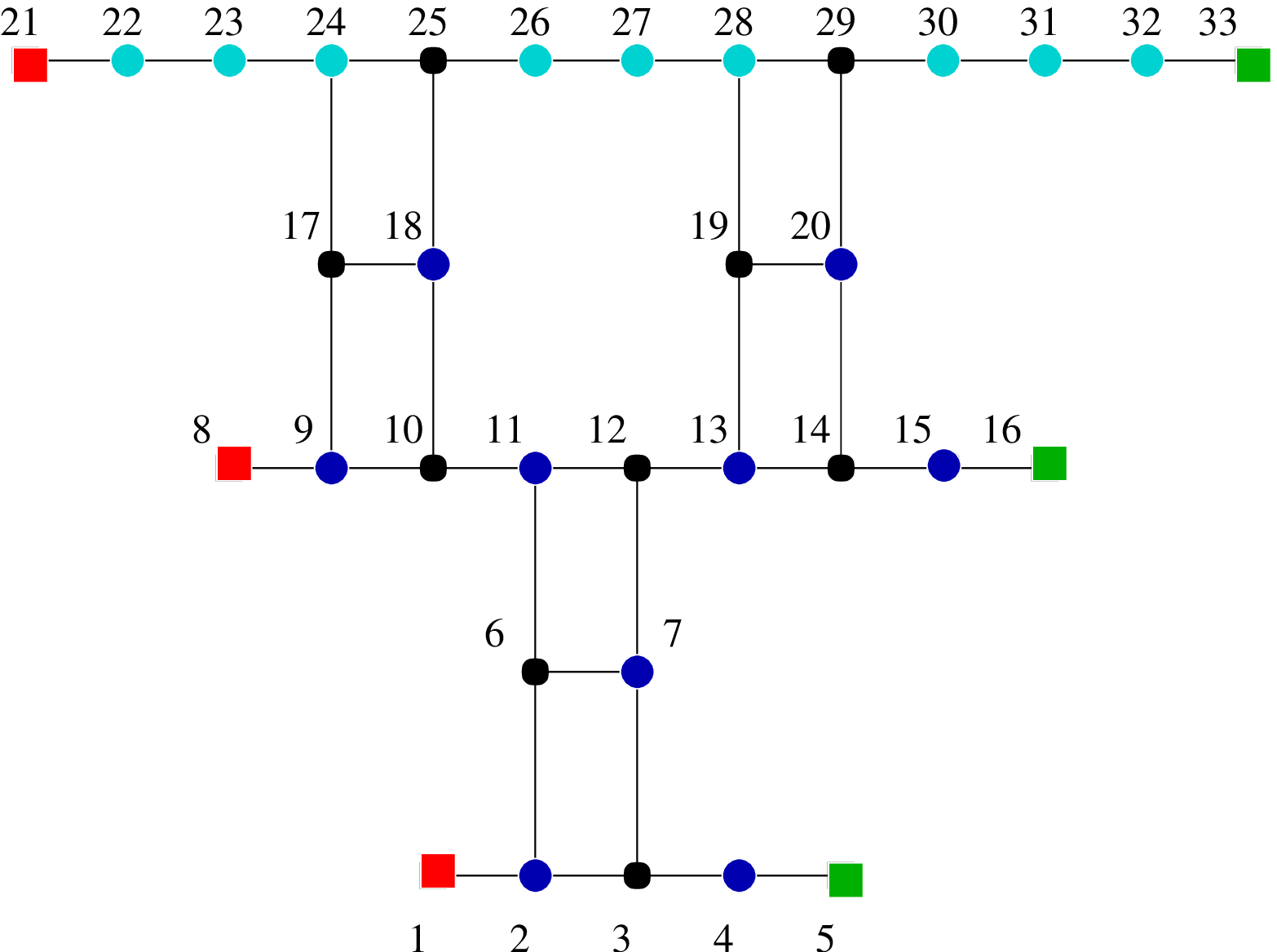} \\  \includegraphics[width=0.5\textwidth]{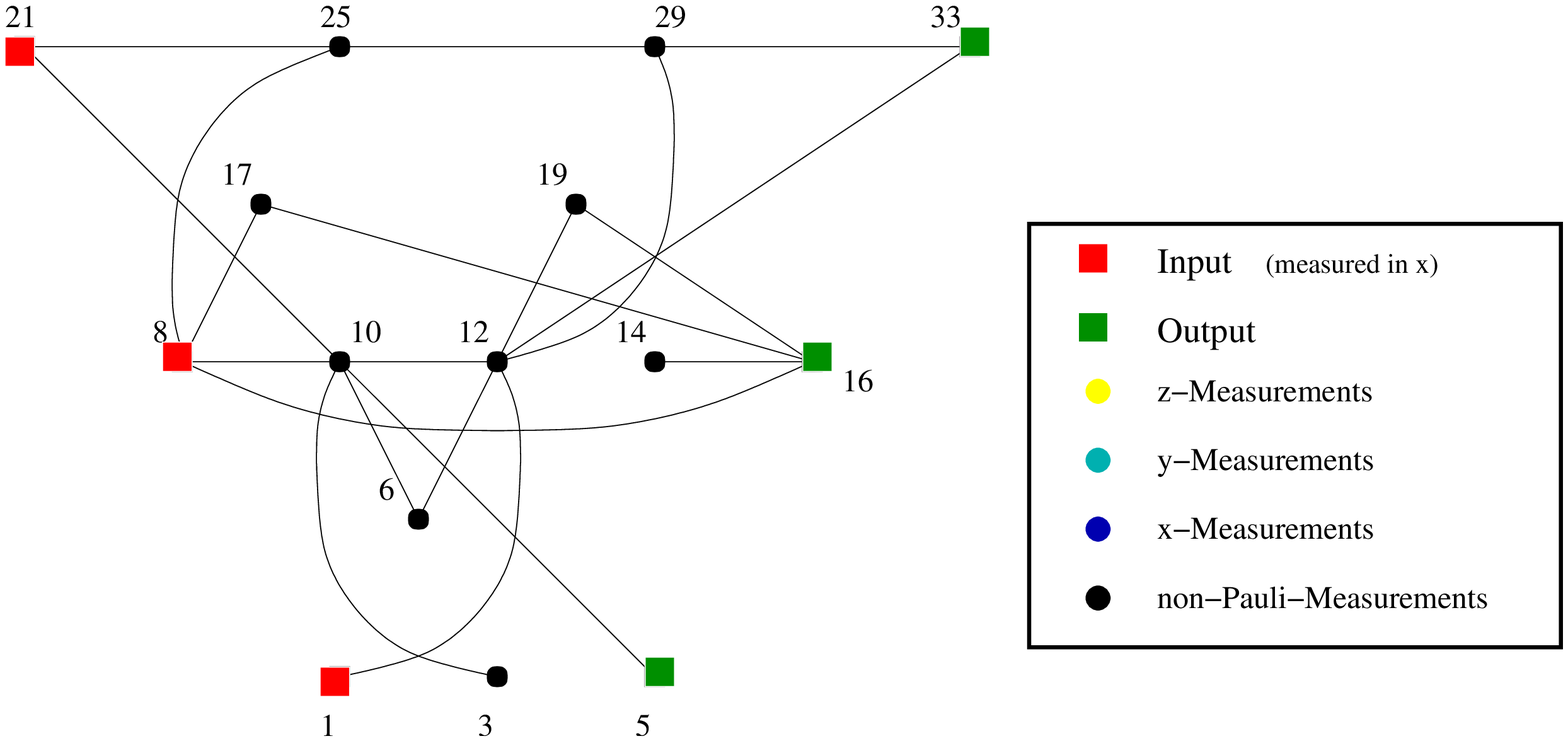}
\end{center}
\vspace{-0.5cm}
\caption{\label{Fig:QFT3_Appl} --{\em Quantum Fourier Transformation on $3$ qubits}\index{quantum Fourier transformation (QFT)}-- The upper left graph is a $13\times 5$-cluster state and represents the initial resource for implementing a QFT on three qubits in the framework of the one-way quantum computer. After removing some superfluous vertices by means of $\sigma_z$-measurements the right upper graph is obtained, whereas the lower graph is achieved after performing {\em all} Pauli measurements within the local measurement protocol associated with the QFT, except the measurements on the input and output vertices and corresponds to the `non-classical' resource for the QFT. The form and color of the vertices indicates the different measurement directions involved or whether the respective vertex corresponds to an input or output qubit.   }
\end{figure}

In the following we consider the example of the quantum Fourier transform (QFT) on three qubits as depicted in fig.~\ref{Fig:QFT3_Appl} to sketch the basic concept of the $\text{QC}_G$ model. In the standard framework of quantum computation\footnote{For details we refer e.g. to ref.~\cite{NielsenBook}.} this QFT is implemented by performing a sequence of elementary ($2$-qubit) gates operating on the $n=3$ input qubits. In the $\text{QC}_G$ model we start instead with the preparation of a highly entangled resource state via the same entangling procedure as for the cluster state, except that a subset of $n$ of these qubits representing the input qubits initially are prepared in the desired input state. Then the actual computation consists in a sequence of local measurements which turns the state on the remaining, non-measured qubits into the desired (pure) output state.

In general, the measurement outcomes of the local measurements have to be stored in a classical register ('information flow vector') of $2n$ bits, since the direction of subsequent measurements have to be adapted according to these results \cite{OneWay1,OneWay2,OneWay3,OneWay5}. Moreover, each measurement round requires an update of the information flow vector that contains the `algorithmic information', from which, at the end, the outcome of the quantum computation\footnote{E.g. as it would have been obtained by a corresponding network of quantum gates.} can be read off directly. The different dependencies of the measurement directions upon previous measurement results induce a temporal ordering on the set of vertices that can be measured simultaneously, whose total length represent the temporal complexity of the quantum computation in the $\text{QC}_G$ model. In a first measurement round all Pauli-measurements can be performed at the same time, since these do not depend on previous measurement results. This fraction of the quantum computation corresponds to the part of the network model that solely consists of Clifford operations. According to the Gottesmann-Knill theorem\footnote{See Proposition~\ref{GottKnillTh} in sec.~\ref{Pauli measurements}.}\index{Gottesmann--Knill theorem} this Clifford part can be efficiently simulated on a classical computer. Note that stabilizer circuits are not even universal for classical computation \cite{Aaronson04}.

In the framework of the $\text{QC}_G$ the stabilizer circuits can be dealt with by using only a single measurement round. If the input state is some stabilizer state, e.g. some state of the computational basis, then such a Pauli-measurement round still ends up in a stabilizer state, whereas the states obtained after some subsequent measurements can in general no longer be described within the stabilizer formalism.  Note that the first measurement round might as well contain the input and output\footnote{If the objective of the quantum computation is not only to compute a (coherent) quantum state on the output qubits but also to perform a read out in order to obtain some classical result.} vertices, which in general are measured in $\sigma_x$- and $\sigma_z$-direction if the quantum computation is carried out with respect to the computational basis. In other words, input and output qubits can also be measured long before the whole quantum computation is completed, since then the actual result of the quantum computation is given by the information flow vector after the last measurement round.

In fig.~\ref{Fig:QFT3_Appl} we have depicted the graphs that are obtained after performing all $\sigma_z$ measurements and after all Pauli measurements except those at the input and output vertices. The corresponding graph states can still be used to implement the QFT. For this the input state is teleported\footnote{Note that this teleportation can be described within the $\text{QC}_G$ picture. For this the qubit holding the input state at qubit $a_1$, some auxiliary qubit $a_2$ in the $|+\rangle$ state and the qubit at the input vertex $b$ are entangled to form a chain attached to the graph at $b$ . Then $a_1$ and $a_2$ are measured in $\sigma_x$-direction. The resulting state on the graph vertices now coincides with the state after some teleportation of the input state into the vertex $b$ (up to LC).} into the input vertices and the usual measurement protocol is applied. Alternatively one can regard the graph as a preparation procedure for an initial resource that already incorporates the input state. More precisely one prepares the input qubits in the input state and the remaining qubits in the $|+\rangle$ state, entangles the qubits according to the interaction pattern given by the graph and finally carries out the measurement protocol as described above.

In \cite{OneWay1,OneWay2,OneWay3,OneWay5} it is shown that any quantum algorithm within the network picture requiring only a polynomial\footnote{In $n$ input qubits.} amount of temporal, spatial and operational resources can be efficiently simulated by a $\text{QC}_G$ that requires (i) a polynomially overhead of elementary operations in the classical preprocessing to derive the equivalent setup and measurement protocol for the $\text{QC}_G$,  (ii) a polynomially bounded amount of classical and quantum resources, and finally (iii) a polynomial increasing time cost for the classical and quantum processing during the computation\footnote{The complexity of the quantum processing is given by the number of local measurements. For the classical processing the number of elementary classical steps for the update of the information flow vector and for the determination of subsequent measurement directions is logarithmic in $n$ and proportional to the number of measurement rounds.}. In this way the $\text{QC}_G$ can serve as a {\em universal} model for quantum computation that has quite promising {\em scaling behavior} for those practical implementations, 
 in which the resource cluster state or more generally the initial graph state can be prepared by a homogeneous Ising interaction, i.e., independently of the system size. Furthermore, the $\text{QC}_G$ is equivalent \cite{Ve04,1QCequiv} to other {\em measurement-based schemes for quantum computations}.

First results towards fault-tolerant quantum computation with the $\text{QC}_G$ were also obtained in refs.~\cite{OneWay1,OneWay2,OneWay3,OneWay5}. For a reasonable noise model, including noisy cluster state preparation as well as erroneous measurements, a quantum computation subjected to decoherence below a certain threshold can be implemented on a $\text{QC}_G$  in a fault-tolerant way. This means that for sufficiently small noise of a specific type the quantum computation can be implemented on a $\text{QC}_G$ with `any' desired accuracy if one allows for a reasonable overhead in the computational resources. But note that the derived error threshold seems rather unrealistic for practical purposes.
Current research focuses on an improvement of this threshold and a generalization of the underlying noise model combining standard concepts of quantum error correction with some intrinsic error detection possibilities of the $\text{QC}_G$ and purification methods for the underlying graph state as the computational resource (see sec.~\ref{CSS_GS}). Further results on fault tolerant quantum computation can also be found in ref.~\cite{FaultTolOneWay}.
\index{one-way quantum computer ($\text{QC}_G$)|)}

\subsection{Quantum error correcting codes}\label{Application_QEC}\index{quantum error correcting code (QEC)|(}\index{stabilizer code}

\begin{wrapfigure}[12]{r}[0.1\textwidth]{0.55\textwidth}
\vspace{-0.5cm}{
\raisebox{0.3cm}{\includegraphics[width=0.2\textwidth]{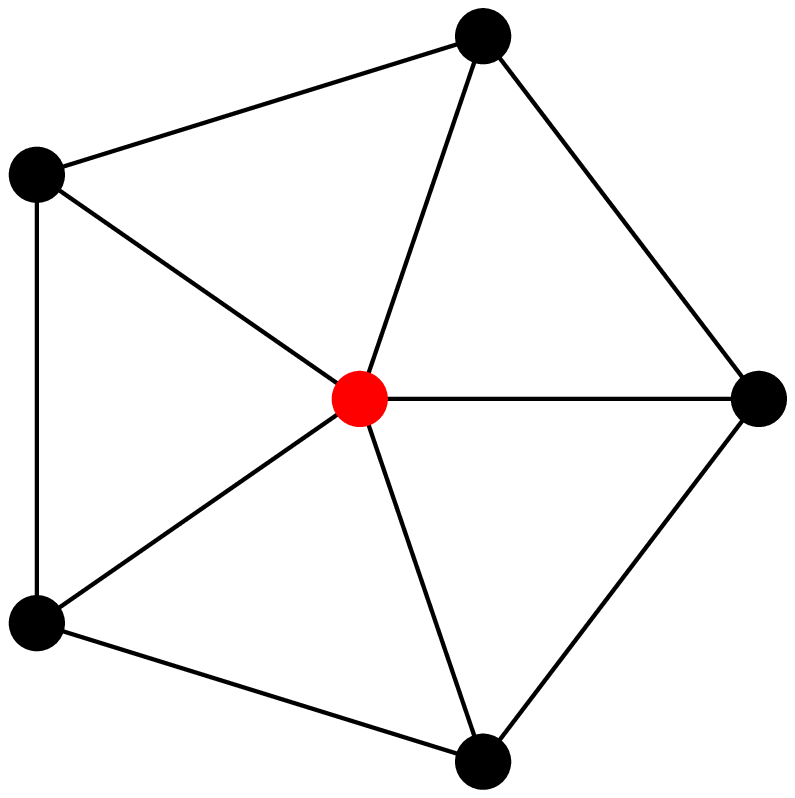}}\hspace{0.01\textwidth}
\includegraphics[width=0.4\textwidth]{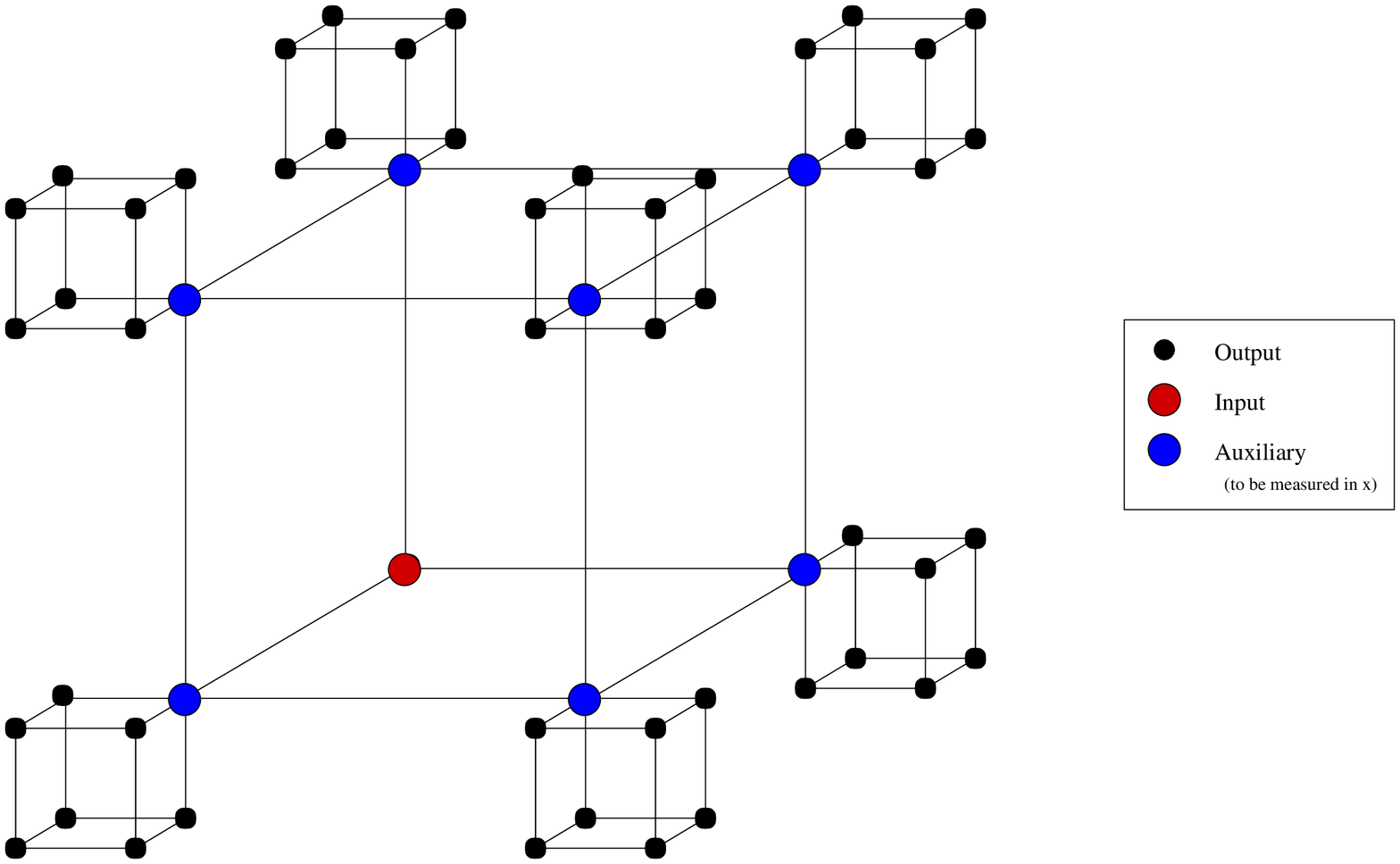}
\caption{\label{Fig:CodeAppl}--{\em Five-Qubit-Code and concatenated CSS-Code}\index{CSS code}-- The graphs representing the encoding procedure for the five-qubit and the concatenated $[7,1,3]$-CSS-code with input (red), auxiliary (blue) and output (black) vertices.}}
\end{wrapfigure}

For {\em quantum error correcting codes} based on stabilizer codes~\cite{Gottesman} the {\em codewords} as well as the {\em encoding procedures} can be represented as graphs \cite{Schlinge02a,Schlinge02b,Grassl02}. The latter can be understood along the lines of the previous subsection, because the graph state is the computational resource for implementing the encoding process in terms of the $\text{QC}_G$ model. The graphs depicted in fig.~\ref{Fig:CodeAppl} for example correspond to the encoding procedures for the five-qubit {\em Steane code}\index{Steane code} and the {\em concatenated $[7,1,3]$-CSS-code}\index{CSS code} that encode a state on one qubit into some state on five and $49$ qubits respectively.

The encoding consists in preparing the qubit at the (red) input vertex in the input state and the remaining qubits in the $|+\rangle$ state, entangling the qubits according to the graph and finally measuring the (blue) auxiliary vertices\footnote{Whereas the five-qubit code does not require any auxiliary qubits to be measured, the auxiliary vertices of concatenated CSS-code are located at the corners of the inner cube, where the open side indicates the input qubit.} and the input vertex in the $\sigma_x$-direction. In this way the input state is encoded into the state on the remaining (black) qubits. Alternatively one might as well prepare the graph states depicted in the figures as such, teleport the input state into the input vertex and then perform the same measurement protocol.

\begin{wrapfigure}[10]{r}{0.45\textwidth}
\vspace{-0.2cm}
\includegraphics[width=0.22\textwidth]{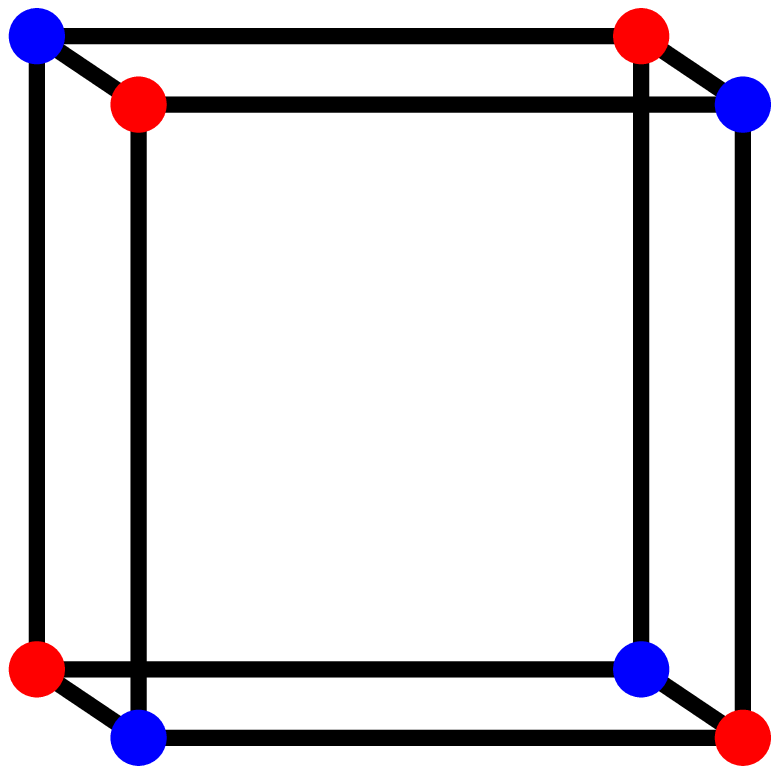}\hspace{0.01\textwidth}\includegraphics[width=0.22\textwidth]{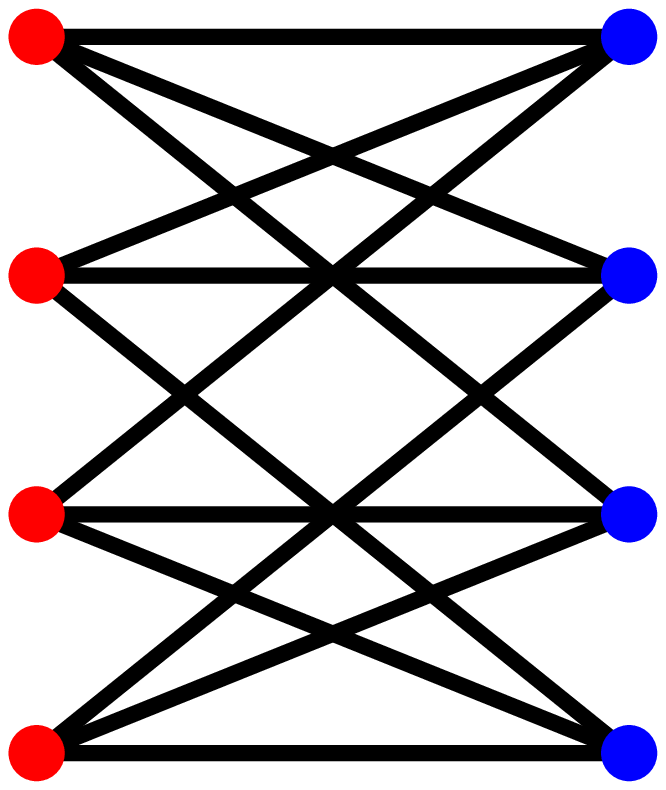}
\caption{\label{Fig:SteaneCodeAppl} The encoding graph for the Steane code is two-colorable.}
\end{wrapfigure}

When applying this procedure to an eigenstate of one of the Pauli-matrices $\sigma_x$, $\sigma_y$ or $\sigma_z$,  the encoded state is a graph state and can be regarded as a code word vector for the respective quantum code (see sec.~\ref{Def_Stab_States}). Note that the $[7,1,3]$-CSS-code is a concatenation of the Steane code depicted in fig.~\ref{Fig:SteaneCodeAppl}. This means that each of the $7$ output qubits of the first encoding level is encoded again by the same encoding procedure, such that the overall encoding procedure maps some input state into an encoded state on the $7\times 7=49$ output qubits of the second level. This is graphically reflected by the fact that the right graph in fig.~\ref{Fig:CodeAppl} is obtained from the left graph in fig.~\ref{Fig:SteaneCodeAppl} by attaching another cube at each `output'-corner in the initial cube and merging the former output and input vertex into one auxiliary vertex.
\index{quantum error correcting code (QEC)|)}

\subsection{CSS--states and secret sharing}\label{CSS_GS}
\index{bi-partite graph (state)}\index{two-colorable graph (state)}

The class of {\em CSS states}\index{CSS state} corresponds to the class of {\em two-colorable graphs}\index{two-colorable graph (state)|(}, i.e., graphs that allow for a coloring of the vertices with two colors such that no two adjacent vertices have the same color (see sec.~\ref{GS_Notations}). In general a CSS-code is a stabilizer code, whose stabilizer can be generated by stabilizer elements $\sigma$ that either consists of $\sigma_x$- or $\sigma_z$-matrices, i.e., $\sigma=\sigma_x^U$ or $\sigma=\sigma_z^U$. By performing Hadamard operation to the vertices of one coloring partition any two-colorable graph state is easily seen to be of CSS-type\footnote{I.e., a CSS code with stabilizer of full rank $\text{rank}(\mathcal{S})=N$.}. Conversely it is shown in ref.~\cite{Lo04} that any CSS-state is also LC-equivalent to some two-colorable graph state.

In \cite{Lo04} CSS states are used to design a `prepare and measure' protocol for quantum cryptography that can be used for {\em conference key agreement}\footnote{This protocol allows several parties to share a secure conference key in the presence of an eavesdropper.}\index{conference key agreement} and {\em quantum sharing of classical secrets}\footnote{This protocol allows a party to share a classical secret with $n$ parties, such that at least k other parties are needed to reveal this key (for some fixed $k>\frac{n}{2}$).}\index{quantum secret sharing}.

We remark that not all Pauli measurements preserve the two-colorability of the underlying graph. Whereas local $\sigma_x$- or $\sigma_z$-measurements in two-colorable graphs yield graph states according to two-colorable graphs, $\sigma_y$-measurements of two-colorable graphs can lead to graph states which are not even locally equivalent to two-colorable graphs \cite{He04}. This might be important in the context of 1-way QC models, in particular in the context of fault tolerance. There, graph states corresponding to the non--clifford part of a computation may be purified via multi-particle entanglement purification, where current protocols require two--colorability of the underlying graph.
\index{two-colorable graph (state)|)}

\subsection{Entanglement purification and secure state distribution}
\label{EPP_Blind}

Despite of their central use for quantum information processing purposes, in reality, multi-partite entangled pure states, such as graph states, will not be available with unit fidelity. The reasons for this are manifold. For instance, the operations to create these states are always noisy, the distribution of these states among distant parties occurs through noisy quantum channels, and the storage of states is subjected to decoherence. For the class of two-colorable graph states {\em entanglement purification}\index{entanglement purification} procedures are known to maintain or purify these states. Some of these procedures \cite{Du03a} provably apply to the regime of noisy control operations. The basic idea in entanglement purification is to use several copies of low-fidelity entangled states in order to distill a few highly entangled states. Thereby, one copy of the state with low fidelity is measured in order to reveal information about either one copy (recurrence and pumping schemes) or several other copies (hashing protocols) of the imperfect graph state. For the purification step each of the imperfect copies (mixed states) is transformed\footnote{This purification is done by means of some local twirling operations (see Proposition~\ref{GraphTwirling} in sec.~\ref{GS_decoherence}), so with respect to the average with respect
to an appropriate local symmetry group.} into a mixed state that is diagonal in a graph state basis associated with the two-colorable graph state in question. Given that the initial fidelity is sufficiently high, an ensemble of these graph diagonal states can then be purified. In \cite{Du03a} genuine multi-party entanglement purification protocols based on a recurrence or pumping schemes and based on hashing procedures (see also \cite{Lo04}) were considered and compared with procedures using bi-partite purification\footnote{ Here one e.g. establishes some highly entangled pairs shared between different parties, in order to teleport the desired $N$-particle state prepared at one party to the remaining parties.}. The different purification schemes have their own advantages. But for local noise, multi-party purification protocols are, in most cases, not only more efficient but provide also better achievable fidelities for the distilled states in the case of noisy control operations. For practical purposes it is useful that the purification regime\footnote{I.e., the purification regime is given by the necessary fidelity of the initial low-entangled copies, such that some higher-entangled state can be distilled.} for the recurrence protocols as well as the error threshold for the noisy control operations involved seem not to depend on the number of particles in the graph state but rather on the maximal degree of the underlying graph. Thus two-colorable graph states provide a reservoir of entangled states between a large number of particles, which can be created and maintained even in the presence of decoherence. Entanglement purification schemes will be discussed in more detail in sec.~\ref{EPP}.

These purification schemes were also modified to {\em blind purification protocols}\index{blind purification}. That is, the purification takes place in such a way that even the involved parties (except of a central party) do neither know the identity of the state they are attempting to purify nor have a possibility to learn it. Such protocols allow for the secure distribution of two-colorable graph states, which remain unknown to the different parties and any potential eavesdropper \cite{Du05c}. Such {\em secure state distribution} is a quantum primitive in a real--world scenario (i.e., taking imperfections in local control operations and channel noise into account) which may be used as a basic building block for both quantum and classical security applications.


\section{Physical implementations}\label{Implementations}\index{graph state preparation}

 Let us briefly discuss a few proposals of how to
prepare graph states in real physical systems, aiming at realizing
some of the applications, such as the one-way quantum computer ($\text{QC}_G$).
In general, graph states do not appear as ground states of
physical systems, since, in most cases, they rely dominantly on two-body interactions \cite{Nielsen05a}.
Following the preparation
procedure in sec.~\ref{DefOfGS_Int},
graph states can nevertheless be obtained
in any physical system that allows to implement an Ising interaction
$H_{ab}^I = \sigma_z^a \sigma_z^b$. In particular, graph states
can be generated, of course, in all physical devices for {\it universal}
quantum computation. Clearly,
the complexity of the graph state preparation
depends on the respective computational primitives of the
physical realization \cite{Perdrix}.

One may distinguish two classes of physical systems where cluster
or graph state preparation very naturally reflects the underlying
architecture. The first one consists in lattice systems, such as
cold atoms in {\it optical lattices}. Here, a neighborhood
relation is inherited by the physical neighborhood of constituents
in the lattice itself, and by means of appropriate switching of
interactions, one can generate cluster states of some dimension
\cite{Briegel-HaenschBand}.
The second one comprises of physical
systems where the neighborhood relation can be freely chosen, such
as in purely optical systems or hybrid setups making use of
`flying' optical qubits for the entangling process and `static'
matter qubits for storage of the actual graph state. We will
briefly sketch both approaches, starting with the former.

The preparation of a two- or three-dimensional cluster state (of
`arbitrary' size) can be regarded as a fixed operational step, if
the Ising interaction can be implemented {\it homogeneously} (for
all directions) throughout the whole lattice
\cite{Briegel01,Briegel-HaenschBand}.
This globally tunable Ising interaction can be realized in optical
lattices with ultra-cold atoms via state-selective displacement of
the atoms and controlled cold collisions \cite{Jaksch99} or via
tunneling \cite{Duan03}. For instance, employing a superfluid-Mott
insulation quantum phase transition a Bose-Einstein condensate can
be loaded into an optical lattice achieving almost unit occupation
per lattice site \cite{Jaksch98,Greiner02}.
Interference experiments \cite{Greiner03}
indicate that graph states can be obtained within such setups.
Graph states prepared in optical lattices therefore are a
promising resource for scalable quantum computation in the
framework of the one-way quantum computer, since cluster states
can be created with an operational effort that is independent of
the system size. Existing implementations are currently still
facing at least two major challenges: On one hand, the number of
atoms at each lattice site has to be exactly controllable. In
particular, the presence of defects, i.e., empty lattice sites,
might spoil any quantum computation. \index{one-way quantum
computer ($1$-QC)} On the other hand, the one-way computer
requires the atoms at different sites to be {\it individually
addressable} by local measurements.
A number of proposals have already been made to overcome these
obstacles (see, e.g., refs.~\cite{Kay05,Vollbrecht04,Rabl03}), and
one can say that optical lattices remain one of the prime
candidates to study multi-particle entanglement and scalable
quantum computation. Alternative implementations have also been
suggested for physical systems, in which the underlying
Hamiltonian is given by an (isotropic) {\it Heisenberg
interaction} \cite{Clark04,Loss05}. These proposals can be
realized in XY-spin chains formed within an optical lattice of
neutral atoms as well as in solid-state systems, such as quantum
dots.

The second type of physical systems does not directly exploit
immediate adjacency of the respective constituents: In this class
of physical systems the key point is the fact that one introduces
a separation between the act of creating entanglement and the act
of performing the actual computation. This is of central
importance in particular in setups where the elementary gates
function on a probabilistic basis, or where erasure errors such as
photon losses constitute the predominant obstacle that has to be
overcome. This very much applies to architectures of quantum
computing based on {\it linear optics}. Linear optical setups are
attractive, as photons are relatively robust with respect to
decoherence, and accurate state control is possible using linear
optical elements \cite{KLM}. However, in order to achieve
universal quantum computing based on {\it dual rail encoding} --
where logical qubits are encoded in state vectors $|0\rangle |
1\rangle$ and $|1\rangle | 0\rangle$ of two modes -- measurements
are necessary, rendering any scheme probabilistic. Notably, the
scheme of Knill, Laflamme, and Milburn in ref.~\cite{KLM} employs
the non-linear sign shift gate, acting as
\begin{equation}
    x_0 |0\rangle + x_1 |1\rangle + x_2 |2\rangle\mapsto
    x_0 |0\rangle + x_1 |1\rangle - x_2 |2\rangle
\end{equation}
 as a primitive probabilistic gate, realized using additional
 modes and measurement. Here, $|0\rangle$, $|1\rangle$,
 and $|2\rangle$
 denote the state vectors of states containing $0,1,2$ photons.
 It has been shown that, although these
 gates operate on a probabilistic basis (only certain measurement
 outcomes are accepted as being successful), the overall computational
 scheme can be uplifted to almost unit success probability. This is
 done using teleportation, based on
 appropriate entangled resources which are
 in turn built up by invoking these probabilistic gates.
 This is a powerful theoretical idea, heavily exploiting
 the previously described stabilizer formalism. So in this way,
 near-deterministic quantum computation is possible in principle
 using linear optical elements, single-photon sources, photon-resolving
 detectors, and feedforward, meaning later action may depend on
 earlier measurement outcomes \cite{KLM}.
 Practically, the key
 obstacle is, however, that the required resources are tremendous
 (approximately $10^5$ beam splitters required for a single CNOT gate
 operating with a failure rate of $10^{-4}$), essentially originating from
 the fact that the elementary gates work with such a small probability
 of success \cite{NSBound}.

Graph state methods can indeed significantly reduce the
surmounting number of required resources (although still giving
rise to challenging prescriptions). In the remainder of this
section, we will thus briefly review recent work discussing graph
states and one-way quantum computation in the context of linear
optical setups.
Ref.~\cite{Reznik} is the first work making use of states related
to graph states to reduce the overhead in resources. In
ref.~\cite{Nielsen04}, the presented scheme has explicitly been
phrased in terms of cluster and graph states, making use of the
setting of the scheme by Knill-Laflamme-Milburn.

The number or necessary resources
was further reduced in refs.~\cite{Browne04,Ralph05},
employing two types of
fusing gates that glue pieces of linear one-dimensional clusters or
of two-dimensional graph states together. Essentially, the basic
resource here are Bell states using non-deterministic {\it parity-check
measurements} \cite{PC},
involving the combination of photons on polarizing
beam splitters, followed by measurements on the output modes.
The computational basis is again the one of dual
rail encoding, here specifically the basis
\be
|{H}\rangle := |0\rangle|1\rangle,\,\,\,\,
|{V}\rangle := |1\rangle|0\rangle
\ee
 corresponding
to a horizontally and vertically polarized photon.
A fusion of the first type amounts to mixing the input,
assumed to contain only one photon per spatial mode,
at a polarizing beam splitter, rotating the output by $45^o$
and measuring it with a polarization-discriminating photon
detector. This is the parity-check operation considered in ref.~\cite{PC}. In the case when one and only one photon is detected --
occurring with a probability of $1/2$ and considered the
successful event -- the state is
transformed according to a non-trace-preserving
completely positive maps with either of the two Kraus
operators,
\begin{eqnarray}
     K_1 &=&
     \left(|{H}\rangle\langle {H} | \langle {H} |  - |{V}\rangle\langle {V} | \langle {V} | \right)/\sqrt{2} ,\\
    K_2 &=&
    \left( |{H}\rangle\langle {H} | \langle {H} |  + |{V}\rangle\langle {V} | \langle {V} |
    \right)/\sqrt{2} .
\end{eqnarray}
Given such a successful event, one glues two pieces
of a graph state together \cite{Browne04}. From
maximally entangled two-dimensional resources,
one can build up linear cluster states.
For a first experimental
demonstration, see ref.~\cite{Zhang05}. To build up
higher-dimensional structures, a variant of a
destructive CNOT gate can be used \cite{PC}, in ref.~\cite{Browne04}
referred to as fusion of second type.
This fusion gate of the second type consists of a
polarizing beam splitters, two polarization rotators of
$45^o$ at the inputs and outputs each, followed again by
polarization resolving photon detection.
These gates are still
probabilistic, yet, failures can be tolerated by merely
rebuilding the affected graph section, assuming that failures
are heralded, meaning that one has classical knowledge of
inappropriate measurement outcomes. First steps towards
fault tolerance with respect to photon
losses have been taken \cite{Trees}.
In this way, graph states
can be prepared from essentially probabilistic ingredients, giving
rise to deterministic quantum computation with smaller overhead in
resources.

Quite recently, arbitrary graph states on four qubits
were experimentally generated in an entirely optical system, with
the four qubits being represented by the polarization state of
four photons \cite{Walther05}.
Although the feasibility of $1$-QC was demonstrated through a
set of one- and two-qubit operations, at the present stage such
experiments rather constitute a proof-of-principle than a clear-cut
route to scalable quantum computation. This is mainly because the
scalability of the preparation by means of parametric down conversion is currently
bound to only a few photons. Using such technology as source of
entangled photons, hence, the overall
success probability is exponentially decreasing.

An alternative route in the second framework is provided by hybrid
solutions, where the advantages of photons providing coupling
mechanisms and of {\it matter systems} as long-lived
storage devices are simultaneously exploited. Based on
schemes that allow for the preparation of entangled
states using matter systems in leaking
cavities using flying optical qubits
\cite{flyingQubits,Cabrillo},
one can immediately construct schemes that allow for
graph state preparation \cite{Sean,Almut,Benjamin}, even in
setups where in intermediate steps no further local
rotations of the state of the matter qubits
is required.
The matter qubits may be spatially separated, and the adjacency
pattern of the graph states to be prepared is in then, in principle,
completely arbitrary. Such schemes can even be made
essentially deterministic, in that if an entangling operation fails,
it can often be repeated until a successful operation
occurs, without
damage to the nascent graph state
\cite{Almut,Benjamin}.
Using multi-port linear
optical devices, the need of doing intermediate
local rotations can also be largely eliminated in the
preparation of the graph states.
Also, whole parts of graph states can be fused together,
in an essentially deterministic manner.
In architectures making use of optically active
electron-nuclear
systems, such as in N-V-centers in diamond,
 one has effectively a ${\mathbbm{C}}^2 \otimes {\mathbbm{C}}^2 $-system
 locally available, allowing for a way to entangle matter qubits in a way
 that is more robust with respect to unavoidable photon loss \cite{NV}.
 All these schemes have in common that the neighborhood
 relation between constituents is fairly arbitrary, and that arbitrary
 graph states can -- in principle -- directly be prepared.

One general lesson to learn from graph state approaches
from the perspective of implementations
is that there are probably
no generally valid uncompromising
requirements for physical systems
to allow for scalable quantum computation.
Instead, it very much depends
on the underlying physical architecture what computational model
is advantageous; hence rendering otherwise inappropriate architectures
useful and promising.
Schemes for quantum computing based on graph states
promise to significantly lessen the challenges
posed by the obvious requirement in any scheme for quantum
computation, that one needs extraordinarily good isolation against
decoherence effects and precise state manipulation at the same time.


\section{Reduced states of graph states}\label{Reduced_GS}\index{reduced state $\rho_G^A$|(}

As we will discuss in sec.~\ref{EntanglementGS}, for pure multi-partite states the reduced state obtained after tracing out (forgetting the information in) some part of the system captures many interesting entanglement properties of the state. More precisely, consider a pure state $|\psi\rangle^{AB}$ of a joint system of parties $A$ and $B$. On the one hand the reduced state
\be \rho^A:=\text{tr}_B (\rho)\ee
represents the information available for one subsystem $A$, if it is not provided with any information that corresponds to some measurement statistics on party $B$. The study of entanglement accessible to subsystem $A$ in this way is an interesting issue in itself. On the other hand, for pure states $\rho=|\psi\rangle\langle\psi|$ the state $|\psi\rangle$ is {\em entangled} with respect to the partitioning $(A,B)$ iff the reduced state $ \rho^A=\text{tr}_A (\rho)$ is mixed and thus cannot be written as a
product state $|\psi\rangle^{AB}= |\psi\rangle^{A} |\psi\rangle^{B}$. Moreover, the mixedness of this reduced state $|\psi\rangle^{AB}$, for example in terms of some entropic measure, allows one to quantify the amount of entanglement contained in this state between the parties $A$ and $B$.

The following proposition shows that for graph states the reduced density matrices can be represented efficiently in terms of their stabilizer elements or their adjacency matrix \cite{He04,Nest04c}.
{\proposition[{\bf Reduced state}]\label{reduced_GS} Let $A\subseteq V$ be subset of
vertices for a graph $G=(V,E)$ and $B=V\setminus A$ the
corresponding complement in $V$.  The reduced state
$\rho^A_G:=\text{tr}_B\left(|G\rangle\langle G|\right)$ is given
by \be\label{reduced_GS_1} \rho^A_G= \frac{1}{2^{|A|}}\,
\sum_{\sigma \in \mathcal{S}_A} \, \sigma \; ,\ee where  \be
S_A:=\{ \sigma \in \mathcal{S}\, | \, \text{supp} (\sigma)
\subseteq A \}\ee denotes the {\em subgroup} of stabilizer
elements $\sigma\in\mathcal{S}$ for $|G\rangle$ with {\em
support}\footnote{The support of a Pauli operator
 $\sigma=\sigma_{i_1}^1\otimes \ldots \otimes \sigma_{i_N}^N $ is
the set of all indices $a\in V$ for which $\sigma$ acts
non-trivially, i.e., $i_a\neq 0$.}\index{support $\text{supp}(\sigma)$} on the set of vertices within
$A$. $\rho_G^A$ is up to some factor a projection, i.e., \be\label{RhoGProj} \left(\rho_G^A\right)^2 = \frac{|\mathcal{S}_A|}{2^{|A|}}\, \rho_G^A\; .\ee It projects onto the subspace in
$\mathbf{H}^A$ spanned by the vectors \be |\mathbf{\Gamma}' B'\rangle_{G[A]} = \sigma_z^{\mathbf{\Gamma}' B'}\, |G[A]\rangle\hspace{1cm}
(B'\subseteq B) \; , \ee where $G[A]=G\setminus B$ is the subgraph
of $G$ induced by $A$ and
 $\mathbf{\Gamma}':= \mathbf{\Gamma}^{AB}$ denotes the $|A|\times
|B|$--off--diagonal sub-matrix of the adjacency matrix $\mathbf{\Gamma}$ for
$G$ that represents the edges between $A$ and $B$:
\begin{equation}\label{Gamma for bi-partition}\index{adjacency matrix $\mathbf{\Gamma}$}
\left(\begin{array}{cc}
  \mathbf{\Gamma}_{A} & \mathbf{\Gamma}_{AB} \\
   \mathbf{\Gamma}_{AB}^T & \mathbf{\Gamma}_{B} \\
\end{array}\right)
= \mathbf{\Gamma} \;.
\end{equation} In this basis, $\rho^A_G$ can be written as \be \label{reduced_GS_2}
\rho^A_G = \frac{1}{2^{|B|}}\, \sum_{B'\subseteq B}\, |\mathbf{\Gamma}' B'
\rangle_{G[A]}\langle \mathbf{\Gamma}' B' |\; . \ee }

{\em Proof:} Eq.~(\ref{reduced_GS_1}) immediately follows from
eq.~(\ref{GS_Projector}) and the fact that the partial trace of
$\sigma=\sigma_{i_1}^1\otimes \ldots \otimes \sigma_{i_N}^N
 \in\mathcal{S}$
can be taken successively over the different vertices in $b\in B$
and gives  $\text{tr}_{b} (\sigma_{i_b}^b) = 2 \delta_{i_b 0} $.
$\rho_G^A$ is proportional to a projection, i.e., \be \left(\rho_G^A\right)^2
= \frac{1}{4^{|A|}}\, \sum_{\sigma,\sigma' \in \mathcal{S}_A} \,
\sigma \sigma'  = \frac{|\mathcal{S}_A|}{4^{|A|}}\, \sum_{\sigma \in \mathcal{S}_A} \,
\sigma = \frac{|\mathcal{S}_A|}{2^{|A|}}\, \rho_G^A \ee follows, because $\mathcal{S}_A$ is a subgroup of $\mathcal{S}$. To show
eq.~(\ref{reduced_GS_2}), the partial trace over $B$ can be taken
in the basis of $B$ given by \be |B'\rangle^B_z \equiv
\bigotimes_{b\in B} |B'_b\rangle^b_z := \sigma_x^{B'}|0\rangle^B
\; . \ee This basis decomposition corresponds to successive local
$\sigma_z$-measurements of all vertices in $B$. The set
$B'\subseteq B$ or the corresponding binary vector determines the
measurement outcomes, i.e., if $b \notin B'$ or likewise the
corresponding component of the binary vector $B'_b=0$ then the
measurement outcome at this vertex is $+1$ and $-1$ otherwise.
According to Proposition~\ref{Pauli_Measurement}, after
measurement of $\sigma_z^{b}$ the state of the remaining vertices
is the graph state vector $| G \setminus b\rangle$ in the case of
the outcome $+1$ , and $\sigma_z^{N_b} | G \setminus b\rangle$ if the outcome is $-1$.
This can be summarized to
\begin{equation}
    \left(
    \sigma_z^{N_b}\right)^{B'_b} | G\setminus b
    \rangle,
\end{equation}
since $B'_b \in \{0,1\}$ represents the measurement result $\{+1,-1\}$. Because the subsequent measurements commute with the previous local unitaries, the final state vector according to the result
$B'=(B'_b)_{b\in B}\in {\mathbbm F}_2^B$ is
\be
 \prod_{b \in B} \left( \sigma_z^{N_b} \right)^{B'_b}\, | G\setminus B\rangle^A \otimes |B'\rangle_z^B
 \; =\;  \sigma_z^{\left(\sum_{b\in B} N_b B'_b\right)} \, | G[A]\rangle^A \otimes |B'\rangle_z^B
\; = \; \sigma_z^{\mathbf{\Gamma} B'}\, | G[A]\rangle^A \otimes
|B'\rangle_z^B \; ,\ee because the sum is performed modulo 2, and
for all $b\in B$ the binary vector corresponding to $N_b$ is the
$b$-th column of the adjacency matrix $\mathbf{\Gamma}$. Distinguishing the
two parts of the adjacency matrix $\mathbf{\Gamma}_B$ and $\mathbf{\Gamma}_{AB}$
that correspond to edges within $B$, and edges between $A$ and $B$,
one arrives at the state vector associated with the measurement
result given by $B'$
 \be\label{SchmidtdecompSummands} (-1)^{\langle B' | \mathbf{\Gamma}_{B} B'
\rangle}\,  \sigma_z^{\mathbf{\Gamma}_{AB} B'} \,| G[A]\rangle^A  \otimes
|B'\rangle^B_z   \; =\; (-1)^{\langle B' | \mathbf{\Gamma}_{B} B'
\rangle}\, | \mathbf{\Gamma}_{AB} B'\rangle^A_{G[A]} \otimes |B'\rangle^B_z
\; . \ee Because the possible measurement results are attained
with probability $1/2$, this proves eq.~(\ref{reduced_GS_2}).  \proofend
\index{reduced state $\rho_G^A$|)}


\section{Equivalence classes under local operations}\label{Local_Equivalence}
\index{equivalence under!local Clifford unitaries (LC)|(}
\index{equivalence under!local unitaries (LU)|(}
\index{equivalence under!stochastic local operations and classical
communication (SLOCC)|(}


For a characterization of the entanglement in graph states we  now
examine the equivalence classes under local operations. Concerning
locality we restrict to the finest partitioning, i.e., each vertex
$a\in V$ represents a single party  and the quantum operations
$\mathcal{E}$ in question are a certain subclass of the class of
all completely positive maps (CPM) that are {\em separable with
respect to this finest partitioning}. Since we are interested in
the equivalence classes of graph states under these local
operations, we can consider the situation in which a pure state
$|\psi_1 \rangle$ is mapped onto another pure state
$|\psi_2\rangle$ by the CPM $\mathcal{E}$ with non-zero
probability. It is generally quite a subtle problem to
characterize the class of all transformations $\mathcal{E}$ that
can be implemented by means of {\em local operations and classical
communication} (LOCC), and in the following we restrict to a
subclass of LOCC-protocols where $\mathcal{E}$ factors out as the
tensor product of a local operator $E_i$ for each party: \be
\mathcal{E}(\rho) = E_1^1\otimes \ldots \otimes E_N^N \, \rho \,
(E_1^1)^\dagger\otimes \ldots \otimes (E_N^N)^\dagger \; .\ee This
means that the pure state $|\psi_1 \rangle$ is converted (in
general: stochastically) into the state $|\psi_2 \rangle=
E_1^1\otimes \ldots \otimes E_N^N \, |\psi_1 \rangle $.
In the following we will consider three different classes of local
operations, namely
\begin{itemize}
\item {\bf SLOCC:} invertible {\em stochastic local operations and
classical communication}, i.e., the operation $E_a\in
\text{SL}(2,\mathbb{C})$ at each qubit is an arbitrary invertible
matrix; SLOCC-equivalence occurs typically with nonunit
probability; \item {\bf LU:} arbitrary local {\em unitaries}, i.e.
the operation $E_a\in \text{SU(2)}$ at each qubit is some unitary
operation; LU-equivalence occurs with unit probability; \item {\bf
LC:} {\em local Clifford unitaries}, i.e., the operation $E_a\in
\mathcal{C}_1$ at each qubit is one of the Clifford unitaries
introduced in sec.~\ref{Def_LC}; LC-equivalence occurs with unit
probability;
\end{itemize}
For $\text{X} \in \{\text{LC}, \text{LU}, \text{SLOCC}\}$, two
states $|\psi_1 \rangle$ and $|\psi_2 \rangle$ are called
$\text{X}-$equivalent if there exists an $\mbox{X}-$operator
$\mathbf{X}$ such that $\mathbf{X}|\psi_2 \rangle\sim|\psi_1
\rangle$. We will frequently write \be |\psi_1 \rangle
\longleftrightarrow_\text{X} |\psi_2 \rangle \ee if both states
are {\em X-equivalent}.

Let us first give a brief overview of entanglement classes for
arbitrary states, i.e., not necessarily graph or stabilizer
states. We start with the case of two $d$-level systems, i.e., we
consider bi-partite entanglement. Here the {\em Schmidt
decomposition} serves as a well-known standard form, from which
the conditions for different types of interconvertibility can be
read off. A bi-partite state $|\psi_1\rangle$ can be transformed
into the state $|\psi_2\rangle$ \be |\psi_1 \rangle
\longrightarrow_\text{X} |\psi_2 \rangle
\hspace{0.7cm}\text{for}\hspace{0.5cm} \text{X} \in \{\text{LU},
\text{LOCC}, \text{SLOCC}\} \ee
 by means of (i) SLOCC, (ii) LOCC or (iii) LU operations iff for  the
 corresponding Schmidt decomposition of
\be |\psi_1\rangle^{AB}=\sum_{i=1}^{R_1} \lambda^1_i |i\rangle^A
|i\rangle^B \hspace{0.7cm} \text{and}\hspace{0.7cm}
|\psi_2\rangle^{AB}=\sum_{j=1}^{R_2} \lambda^2_j |j\rangle^A
|j\rangle^B \ee (i) $R_1\geq R_2$, (ii) the coefficient list
$(\lambda^1_i)_{i=1}^{R_1}$ is a majorization of
$(\lambda^2_j)_{j=1}^{R_2}$ or (iii) the coefficient list
$(\lambda^1_i)_{i=1}^{R_1}$ coincides with
$(\lambda^2_j)_{j=1}^{R_2}$ up to permutations, respectively. It follows that in
the bi-partite case the SLOCC-equivalence leads to $d$ distinct
equivalence classes corresponding to the different values for the
{\em Schmidt rank}\index{Schmidt rank $\text{SR}_A$} $R$. On the
other hand, there is an infinite number of bi-partite
LU-equivalence classes, which are parameterized by the $d$ {\em
Schmidt coefficients} $\lambda_i$\index{Schmidt decomposition}.

Moving to multi-partite entanglement, in the case of three qubits
it was shown in ref.~\cite{Du00b} that there are $6$ orbits under
SLOCC-operations. But true three-qubit entanglement is only
contained in the two distinct classes represented by the GHZ-state
$|GHZ\rangle$ (see sec.~\ref{GHZ_GS}) and the W-state
\index{W-state} \be |W\rangle = \frac{1}{\sqrt{3}}\,\left(
|0,0,1\rangle + |0,1,0\rangle + |1,0,0\rangle \right) \; .\ee Note that
all two-qubit states are SLOCC-equivalent to a two-qubit graph
state corresponding to either the empty graph (product state) or
connected graph (Bell state). But the W-state is an example for a
pure state on three qubits that is not SLOCC-equivalent to some
graph state. This is because the GHZ-state $|GHZ\rangle$ is the
only `connected' graph state with three vertices, as we will see
below.

For $N$-qubit systems with $N\geq 4$ the number of orbits under
SLOCC-operations is infinite and is in fact specified by an
exponentially increasing number of parameters. The latter scaling
behavior for the number of required parameters is due to the fact
that a generic orbit is specified (up to some irrelevant complex
constant) by $3N$ complex parameters describing the group
$\text{SL}(2,\mathbb{C})^{\otimes N}$ in question whereas a pure
state on $(\mathbb{C}^2)^{\otimes N}$ is specified by $2^N-1$
complex parameters (neglecting an overall complex phase). For $4$
qubits it was shown in ref.~\cite{Verstraete02a} that there exists a
standard form for generic states under SLOCC-operations that is
determined by $3$ complex parameters and $8$ further standard
forms corresponding to classes of `degenerate' pure states.
Needless to say that the number of LU- or LC-equivalence classes
necessarily has to be specified by even more parameters since the
corresponding matrix groups are contained in
$\text{SL}(2,\mathbb{C})$.

In \cite{Verstraete03a} it is shown that any multi-partite state
can be  transformed by SLOCC-operations into some {\em standard form} that is unique up to LU, while
maximizing all entanglement monotones\footnote{More precisely the
{\em entanglement monotones} in question are linearly homogeneous
positive functions of the state that remain invariant under
determinant-1-SLOCC operations.}. In this way the problem of
deciding, whether two states are SLOCC-equivalent, can be reduced
to the problem of deciding, whether the standard forms of these
states are LU-equivalent. In this formalism all pure states
$|\psi\rangle$ with maximally mixed  single-qubit reduced density
matrices \be \rho_\psi^a = \text{tr}_{V\setminus a}(|\psi
\rangle\langle \psi |)= \frac{1}{2}\, \mathbf{1}_a \hspace{1cm}
\forall a\in V \ee
 are standard forms of this local filtering sequence and
they are {\em maximally entangled}\index{maximally entangled} in
that these states maximize all entanglement monotones.

Let us return to the question of equivalence under local
operations for graph states. We start with SLOCC-equivalence.
According to Proposition~\ref{reduced_GS}, any single-qubit
reduced state $\rho_G^a$ is maximally mixed for all graph states
corresponding to connected graphs. Therefore, `connected' graph
states are already in standard form under SLOCC in the above
sense. Thus, when restricting to connected graph states all
SLOCC-equivalence classes coincide with LU-equivalence classes;
this property can easily be extended  to the case of general (not
necessarily `connected') graph states by considering each
connected component separately. We arrive at the following result:

{\proposition[\bf SLOCC- equals LU-equivalence] Two graph states
$|G_1\rangle$ and $|G_1\rangle$ are SLOCC-equivalent iff they  are
LU-equivalent \cite{Nest04c}: \be |G_1 \rangle
\longleftrightarrow_\text{SLOCC} |G_2 \rangle \hspace{0.2cm}
\Longleftrightarrow \hspace{0.2cm} |G_1 \rangle
\longleftrightarrow_\text{LU} |G_2 \rangle  \; .\ee }

This result was first obtained in ref.~\cite{Nest04c}. We remark that
the question whether for graph states also \be |G_1 \rangle
\longleftrightarrow_\text{LU} |G_2 \rangle \hspace{0.7cm}
\Longrightarrow \hspace{0.7cm} |G_1 \rangle
\longleftrightarrow_\text{LC} |G_2 \rangle  \ee holds is still an
open question. Note that the backward implication is trivial since
the group of LC-unitaries is a proper subgroup of all LU. For a
large subset\footnote{A stabilizer element $\sigma \in
\mathcal{S}$ with {\em minimal} support\index{support
$\text{supp}(\sigma)$!minimal} $\text{supp}(\sigma)$, i.e., no
other stabilizer element $\sigma'\in \mathcal{S}$ has a support
$\text{supp}(\sigma')$ that is a proper subset of
$\text{supp}(\sigma)$, is called {\em minimal element} of
$\mathcal{S}$. Let $\mathcal{M}$ denote the subgroup generated by
all minimal elements in $\mathcal{S}$. Now the notion of LU- and
LC-equivalence coincide for all stabilizer states, for which
$\sigma_x$, $\sigma_y$ and $\sigma_z$ occurs at each qubit in
$\mathcal{M}$.} of graph states, however, it was shown in
\cite{Nest04d} that both notions of equivalence coincide. The
hypothesis of a general coincidence for all graph states is
further supported by results \cite{CliffordInvariants} about the
corresponding invariants under these operations, which will be
briefly reviewed later in this section, and the classification of
graph states with up to seven vertices that we will discuss below.
Note that a general coincidence of SLOCC-, LU- and LC-equivalence
for graph states would be particularly advantageous for the
following two reasons: since in this case all 3 local equivalences
would correspond to LC-equivalence, firstly, checking whether two
given graph states are locally equivalent could then be done
efficiently (see sec.~\ref{BinaryRepr}); secondly, all 3 locally
equivalences would entirely be described by the local
complementation rule, yielding a description of local equivalence
of graph states in simple, purely graph theoretic terms.

To distinguish the different  equivalence classes under LU
(or, equivalently,  SLOCC) we will first derive a simple (though
not complete) set of invariants that can be efficiently computed
for graph states from the underlying adjacency matrix of the
graph. As mentioned above, for any pure state $|\psi\rangle^{AB}$
in the joint system $\mathbf{H}_A\otimes\mathbf{H}_B$ of two
parties $(A,B)$ with arbitrary dimensionality
$d_A=\text{dim}_{\mathbb{C}} \mathbf{H}_A$ and
$d_B=\text{dim}_{\mathbb{C}} \mathbf{H}_B$ the Schmidt
rank\footnote{Note that from the Schmidt decomposition it follows
that $\text{SR'}^{A}(\psi) =\text{SR'}^{B}(\psi) $.}
$\text{SR'}^{A}(\psi) := \text{rank}\left(\rho_\psi^A \right)$ is
an entanglement monotone with respect to $(A,B)$-local operations.
In the case where $(A, B)$ is a bi-partition of a many-qubit system
we will for simplicity consider not the rank
$\text{SR'}^{A}(\psi)$ of the reduced states but rather the
logarithm of it with respect to the basis $2$, i.e., \be
\text{SR}_{A}(\psi) :=
\text{log}_2\left[\text{rank}\left(\rho_\psi^A \right)\right]\;
.\ee With this notational simplification the Schmidt rank
$\text{SR}_{A}(\psi)$ of pure state $|\psi\rangle$ has the
transparent interpretation\footnote{This follows straightforwardly
from statement (i) in this section about the interconvertibility
of pure bi-partite states under SLOCC-operations.} in terms of the
maximal number of Bell pairs that
\begin{itemize}
\item are required to prepare $|\psi\rangle$ and \item can be
extracted from $|\psi\rangle$
\end{itemize}
with finite probability of success (i.e., under SLOCC-operations).

\begin{wrapfigure}[20]{r}{0.5\textwidth}
\vspace{-0.5cm}
\hspace{0.25cm}\includegraphics[width=0.5\textwidth]{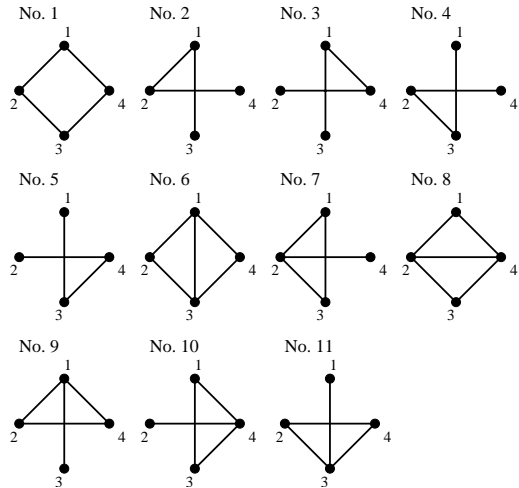}
\caption{\label{fig:LUclassExample2} An example of an equivalence
class which is a proper subset of the class No.\ 4 in List~{\bf A}
but which is  not LC-equivalent to any of the graphs depicted in
fig.~\ref{fig:LUruleExample1}. With the LC-rule in
sec.~\ref{Def_LC} it is straightforward to check that all graphs
within this class are LC-equivalent.}
\end{wrapfigure}

We will now study the entanglement in graph states $|G\rangle$ by
considering the Schmidt rank $\text{SR}_{A}(G)$ with respect to
different bi-partitions $(A,B)$. For a bi-partition
$(A,B)$\footnote{I.e., $A\cup B=V$ and $A\cap B=\emptyset$.} of a
graph $G=(V,E)$ we can again use the decomposition for the
adjacency matrix $\mathbf{\Gamma}$ of eq.~(\ref{Gamma for
bi-partition}) into $\mathbf{\Gamma}_A$, $\mathbf{\Gamma}_B$ and
$\mathbf{\Gamma}_{AB}$ according to edges within $A$, edges within
$B$ and those edges between $A$ and $B$. In this notation the
corresponding Schmidt rank $\text{SR}_{A}(G)$ is simply given by
the binary rank (i.e., the rank over GF(2)) of the
$|A|\times|B|$-off-diagonal sub-matrix
$\mathbf{\Gamma}'=\mathbf{\Gamma}_{AB}$.

{\proposition[\bf Schmidt rank]\label{Schmidt_rank} Let $(A,B)$ be
a bi-partition for some graph state $|G\rangle$. Then the Schmidt
rank of the graph state with respect to this bi-partition is given
by \be\label{SR1} \text{SR}_{A}(G) = \text{rank}_{\mathbb{F}_2}
\left(\mathbf{\Gamma}'\right) \; .\ee Alternatively, the Schmidt
rank is determined in terms of the rank\footnote{See
eq.~(\ref{GroupRank}) in sec.~\ref{Def_Stab_States}.} of the
subgroup $\mathcal{S}_A$ of stabilizer elements with support in
$A$ by the formula: \be\label{SR2}  \text{SR}_{A}(G) =  |A| -
\text{rank}\,(\mathcal{S}_A)\; .\ee }

{\em Proof:} Because of Proposition~\ref{Graph state basis}, the
linear independence of the vectors
$|\mathbf{\Gamma}'B'\rangle_{G[A]}$\footnote{In the graph state
basis for the graph $G[A]$ of Proposition~\ref{reduced_GS}.} is in
one-to-one correspondence to the linear independence of the
corresponding vectors $\mathbf{\Gamma}'B'$ over $\mathbb{F}_2^A$.
Hence we find for the rank of the reduced state $\rho_G^A$
\begin{eqnarray}
\text{rank}\left(\rho_G^A \right) & = & \text{dim}_\mathbb{C} \,
\text{span}\left\{|\mathbf{\Gamma}'B'\rangle_{G[A]}  \,|\,
B'\subseteq B \right\} \nonumber \\ & = &
\text{dim}_{\mathbb{F}_2} \, \text{span}\left\{\mathbf{\Gamma}'B'
\,|\, B'\subseteq B \right\} \nonumber \\ & = &
\text{rank}_{\mathbb{F}_2}\left(\mathbf{\Gamma}'\right) \; .
\end{eqnarray} Taking the $\text{log}_2$ we obtain the
eq.~(\ref{SR1}). Eq.~(\ref{SR2}) follows from the fact that
according to eq.~(\ref{RhoGProj})
$\frac{2^{|A|}}{|\mathcal{S}_A|} \rho_G^A$ is a projection and
thus we can compute the rank alternatively as \be
\text{rank}\left(\rho_G^A \right) \; = \;
\text{tr}\left(\frac{2^{|A|}}{|S_A|} \rho_G^A\right) \; = \;
\frac{2^{|A|}}{|\mathcal{S}_A|} \,\text{tr} (\rho_G^A) \; =\;
\frac{2^{|A|}}{|\mathcal{S}_A|} \; .\ee Taking again the
$\text{log}_2$ we obtain eq.~(\ref{SR2}), since the minimal number
of generators for $\mathcal{S}_A$ is given by
$\text{log}_2(|\mathcal{S}_A|)$. \proofend
\begin{table}
\begin{center}
\begin{tabular}{||c|cccccccc||}
\hline 
 $A$  &  $\emptyset$ & $\{1\}$  & $\{2\}$ & $\{3\}$ & $\{4\}$ & $\{1,2\}$ & $\{1,3\}$ & $\{1,4\}$\\
 $A$  &  $\{1,2,3,4\}$  & $\{2,3,4\}$  & $\{1,3,4\}$ & $\{1,2,4\}$ & $\{1,2,3\}$ & $\{3,4\}$ & $\{2,4\}$ & $\{2,3\}$\\
 \hline \hline
$\text{SR}_A(G_1)$ & 0 & 1 & 1 & 1 & 1 & 1 & 2 & 2 \\
$\text{SR}_A(G_1)$ & 0 & 1 & 1 & 1 & 1 & 2 & 1 & 2 \\
 \hline 
\end{tabular}
\end{center}
\caption{\label{SR_List_Comparison} The list of Schmidt ranks
$\text{SR}_A(G_1)$ and  $\text{SR}_A(G_2)$ for the graphs $G_1$ in
fig.~\ref{fig:LUruleExample1} and  for the graphs $G_2$ in
fig.~\ref{fig:LUclassExample2}.  }
\end{table}

Thus, for any partition $(A,B)$ the Schmidt rank $\text{SR}_A(G)$
is an invariant under arbitrary local unitaries that can be
formulated in purely graph theoretic terms. We now consider the
list of Schmidt ranks with respect to all possible
bi-partitions\footnote{Since $\text{SR}_A(G)=\text{SR}^B(G)$ the
different bi-partitions are fixed by choosing the smaller
partition, say $A$, of the bi-partition $(A,B)$. This gives
$2^{N-1}$ possible bi-partitions.}. This yields a set of
invariants which has already been considered in graph theory under
the name {\em connectivity function }\index{connectivity function}
\cite{Bouchet}. For example, a comparison of the invariants for
the graphs depicted in fig.~\ref{fig:LUruleExample1} and
fig.~\ref{fig:LUclassExample2} shows that the corresponding lists
of Schmidt ranks within each of these figures coincide but differ
between the two figures (see Tab.~\ref{SR_List_Comparison}). This
implies that the corresponding sets of graph states are not
equivalent neither under LC-operations nor under general local
unitaries.

\begin{figure}
\hspace{4cm}\includegraphics[width=0.4\textwidth]{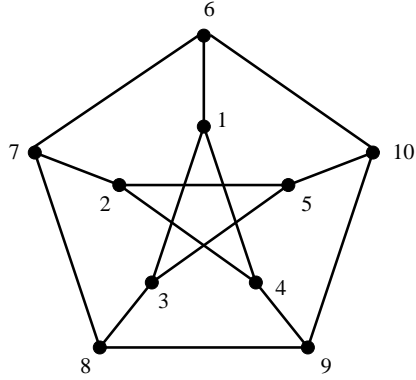}
\caption{\label{fig:PetersenGraph} The Petersen Graph. The
depicted labeled graph is not LC-equivalent to the graph which is
obtained from it by exchanging the labels at each end of the five
"spokes", i.e., the graph isomorphism which permutes the vertices
$1,2,3,4$ and $5$ with $6,7,8,9$ and $10$, respectively. However,
the lists of Schmidt ranks (or, equivalently, the connectivity
functions) of these graphs coincide.}
\end{figure}

We note that the Schmidt rank list does \emph{not} provide a {\em
complete set of invariants}\index{local  invariants!for graph
states} that would characterize all equivalence classes under
LC-operations. For the {\em Petersen graph}\footnote{This
counter-example has first been discovered in ref.~\cite{Flaass96}.}\index{Petersen graph (state)} shown in
fig.~\ref{fig:PetersenGraph} and the isomorphic graph, which is
obtained from it by exchanging the labels at each end of the five
"spokes", no local complementations exists that transforms one
graph into the other, although the Schmidt rank lists for both
graphs coincide.

In \cite{CliffordInvariants} a complete set of polynomial LU
invariants for graph states  is obtained (in terms of binary
trees). As stated in eq.~(\ref{SR2}) the set of invariants
$\{\text{SR}_A(G)\}_{A\subseteq V}$, corresponding to polynomial
invariants of degree $k=2$, can for stabilizer states be
formulated in terms of the dimension of a subspace of the
stabilizer $\mathcal{S}$. It was shown in
\cite{CliffordInvariants} that, similarly, the invariants of
degree $k\geq 3$ correspond to dimensions of  certain subspaces of
the $k-1$-fold direct product $\mathcal{S}^{\times (k-1)}$.
ref.~\cite{CliffordInvariants} also provides a finite complete set
of polynomial invariants for the smaller group $G=\mathcal{C}_1^N$
of local Clifford unitaries (see sec.~\ref{Def_LC}). For graph
states these LC-invariants of degree $k$ are again given by the
dimension of a subspaces in $\mathcal{S}^{\times (k-1)}$. Note
that the algebra of polynomial LC-invariants is in general larger
than the algebra of polynomial LU-invariants. In
\cite{CliffordInvariants} it is shown that the set of polynomial
LU-invariants for the degree $k=2$ and $k=3$ are equivalent to the
corresponding sets of LC-invariants. These results support the
conjecture (see sec.~\ref{Def_LC} and sec.~\ref{Local_Equivalence})
that for graph states the notion of LU-equivalence and
LC-equivalence coincide. \index{local invariants!polynomial|)}

\index{equivalence classes|(} In the remainder of this section we
will now discuss  LU-equivalence for graphs with $N\leq 7$
vertices \footnote{The classification of non-equivalent graph
states naturally generalizes to the case of stabilizer states or
codes, for which the graph represents a particular standard form
(see sec.~\ref{Def_Stab_States} and sec.~\ref{Application_QEC}).
For the quest of good error correcting (self-dual) codes such a
classification has recently attracted some attention. A similar
classification can also be found in ref.~\cite{Glynn02,Hoehn03} and has
been extended to graphs with up to $N=12$ vertices in
refs.~\cite{database,Glynn04}.}.
We have examined the graph states of
all non-isomorphic, connected graphs with up to seven vertices.
More precisely, from the set of all possible graphs with $7$
vertices ($2^{\tbinom{7}{2}} \approx 2\times 10^{6}$
possibilities), we  consider the subset of $1252$ graphs on up to
$7$ vertices that are non-isomorphic with respect to graph
isomorphisms. Note that a graph isomorphism physically comes up to
an exchange of particles. We furthermore restrict to those $996$
states that correspond to connected graphs\index{connected graph
(state)}\index{disconnected graph (state)}. This is because a
state $|G\rangle$ corresponding to a disconnected graph $G$ is
simply the tensor product $|G\rangle=|G_1\rangle^{A_1}\otimes
\ldots \otimes |G_M\rangle^{A_M}$ of the graph states
$|G_i\rangle$ corresponding to the connected components $G_i$ of
$G$, where $(A_1,\ldots,A_M)$ $(M\leq N)$ is some partitioning of
the vertex set $V$. Thus all entanglement properties of the
composite state $|G\rangle$ are essentially determined by the
entanglement properties of its components $|G_i\rangle$. In
particular, such a $(A_1,\ldots,A_M)$-product state is LU- or
LC-equivalent to some other graph state $|G'\rangle$ iff
$|G'\rangle$ allows for a decomposition
$|G'\rangle=|G'_1\rangle^{A_1}\otimes \ldots \otimes
|G'_M\rangle^{A_M}$ with respect to the same partitioning
$(A_1,\ldots,A_M)$ with components  $|G'_i\rangle$ that are LU- or
LC-equivalent to the respective states  $|G_i\rangle$ for
$|G\rangle$. Of the $996$ isomorphism-classes of corresponding
graph states, $46$ classes have turned out to be not invariant
under local unitary operations.

\begin{table}
\begin{center}
\begin{tabular}{||c||c|c||c|c||}
\hline 
    &  & non-isomorphic &  & non-isomorphic \\
 N & non-isomorphic & and non-LC-equivalent & non-isomorphic & and non-LC-equivalent \\
  & graphs  & graphs  & connected graphs  &  connected graphs\\
  \hline \hline
1      &    1     &     1     &       1    &    1    \\
2      &    2     &     2     &       1    &    1    \\
3      &    4     &     3     &       2    &    1    \\
4      &    11     &     6     &       6    &    2    \\
5      &    34     &     11     &       21    &    4    \\
6      &    156     &     26     &       112    &    11    \\
7      &    1,044     &     59     &      853   &    26    \\
8      &    12,346    &     182     &        11,117    &    101    \\
9      &    274,668     &     675     &       261,080    &    440    \\
10      &  12,005,168     &     3,990    &       11,716,571    &    3,132    \\
11      &    1,018,997,864     &     45,144     &       1,006,700,565    &    40,457    \\
12      &    165,091,172,592     &     1,323,363     &       164,059,830,476    &    1,274,068    \\
\hline 
\end{tabular}
\end{center}
\vspace{-0.5cm} \caption{\label{numbers_of_GS} In the first column
the number of graph states with $N=1,\ldots,12$  vertices is
listed that are non-isomorphic under graph isomorphisms. The third
column instead contains the corresponding number of non-isomorphic
graph states that correspond to connected graphs. The values for
both columns are taken from \cite{online ency}, where they can be
found under the sequence number {\em A000088} and {\em A001349}.
Similarly column No.~2 and No.~4 contain the corresponding numbers
of graph states that are not equivalent under graph isomorphism
and LC-operations. The values in the second column were computed
in ref.~\cite{database} together with a database of representatives for
each equivalence class. By checking the list of Schmidt ranks we
have shown that the values in the second and fourth column for
$N\leq 7$ vertices coincide with the numbers of non-isomorphic
graph states when considering the larger group of LU- or
SLOCC-operations. The values for $N=8,9,10,11,12$ in both columns
were again taken from ref.~\cite{database}, where also
a database of
representatives for each equivalence class can be found.}
\end{table}
Within each of these classes all graph states are equivalent
modulo local unitaries {\em and} additional graph isomorphisms.
Thus, if we exclude the graph isomorphisms, as e.g. in quantum
communication scenarios, the number of inequivalent classes of
graph states is even larger (see Tab.~\ref{numbers_of_GS}).

In List {\bf A} and {\bf B} of Table~\ref{TablePage} we give a
list of simple representatives of each equivalence class together
with a table summarizing some interesting properties of these
states. For this we have generated the of $996$ non-isomorphic,
connected graphs with the MATHEMATICA package  described in
sec.~\ref{Introduction} and tested for local equivalence
considering only LC-unitaries (see sec.~\ref{BinaryRepr}). By
considering the Schmidt rank with respect to all possible
bi-partitions, the corresponding lists of Schmidt ranks for each
representative turned out to be different even if we allow
arbitrary permutations of the vertices. This shows that the found
sets of locally invariant graph states are minimal even with
respect to the larger group of all LU. \vspace{0.5cm}

\begin{table}
\begin{tabular}{cc}
\hspace{-1cm}\includegraphics[width=0.45\linewidth]{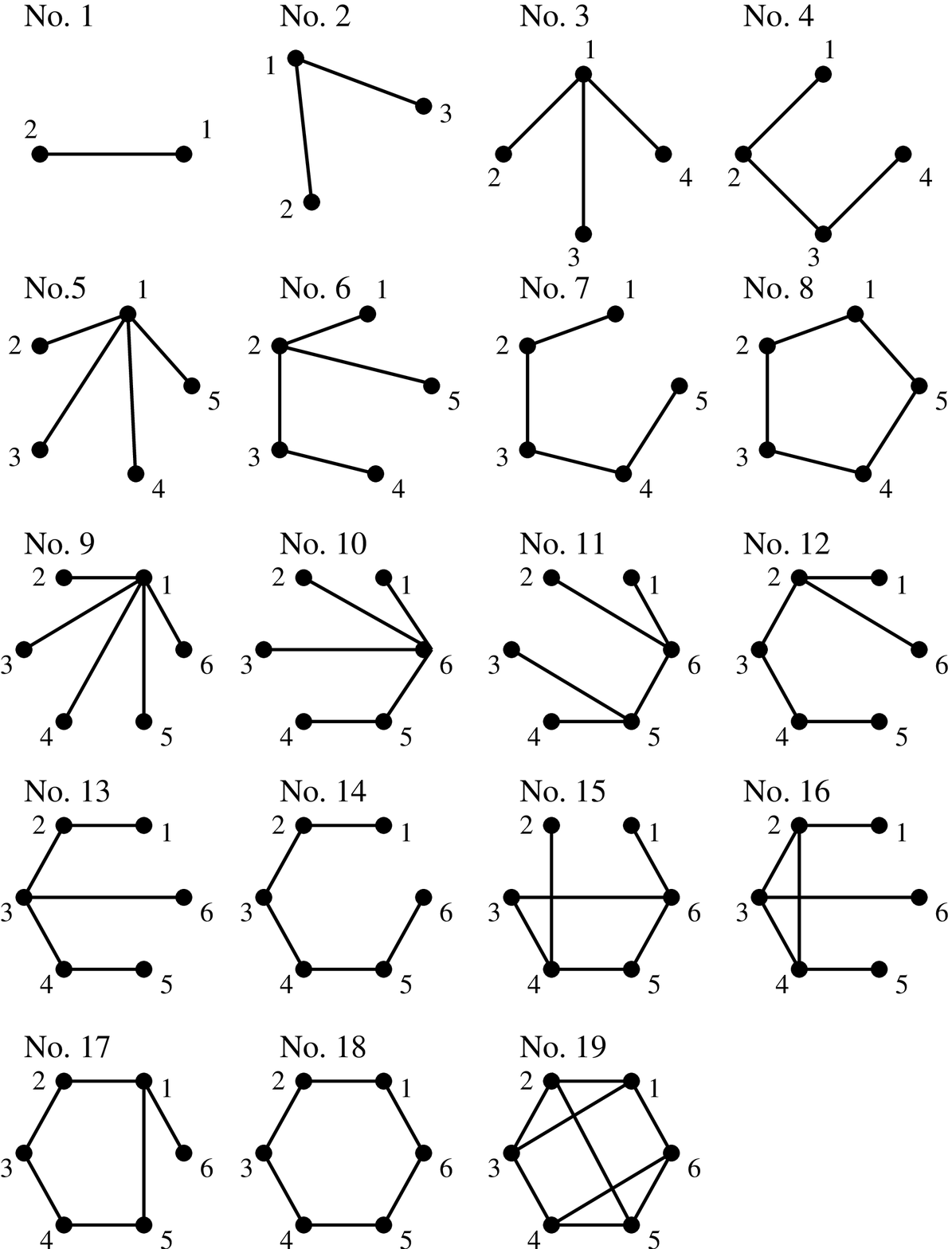} &
\raisebox{4cm}{\footnotesize
\begin{minipage}{0.75\linewidth}
\begin{tabular}{|ccccccccc|}
\hline
No.\ & $|\text{LUclass}|$ & $|V|$ & $|E|$ & $ \text{SR}_\text{max}$ & $\text{PP}$ &  $RI_3$ & $RI_2$ & $2-col$ \\
\hline
1   & 1        & 2     &  1    &   1 &  1 &   &   & yes \\
2   & 2        & 3     &  2    &   1 &  1 &   &   & yes \\
3   & 2        & 4     &  3    &   1 &  1 & & (0,3)  & yes \\
4   & 4        & 4     &  3    &   2 &  2 & & (2,1)  & yes \\
5   & 2        & 5     &  4    &   1 &  1 & & (0,10)  & yes \\
6   & 6        & 5     &  4    &   2 &  2 & & (6,4)  & yes \\
7   & 10       & 5     &  4    &   2 &  2 & & (8,2)  & yes \\
8   & 3        & 5     &  5    & 2 & 3 & & (10,0)  & no \\
9   & 2        & 6     &  5    &   1 &  1 & (0,0,10)& (0,15)  & yes \\
10  & 6        & 6     &  5    &   2 &  2 & (0,6,4)& (8,7)  & yes \\
11  & 4        & 6     &  5    &    2 & 2 & (0,9,1)& (8,7)  & yes \\
12  & 16       & 6     &  5    &   2 &  2 & (0,9,1)& (11,4)  & yes \\
13  & 10       & 6     &  5    &   3 &  3 & (4,4,2)& (12,3)  & yes \\
14  & 25       & 6     &  5    &   3 &  3 & (4,5,1)& (13,2)   & yes \\
15  & 5        & 6     &  6    &   2 &  2 & (0,10,0)& (12,3)   & yes \\
16  & 5        & 6     &  6    &    3 & 3 & (4,6,0)& (12,3)   & yes \\
17  & 21       & 6     &  6    &    3 & 3 & (4,6,0)& (14,1)   & yes \\
18  & 16       & 6     &  6    &    3 & 3 & (6,4,0)& (15,0)   & yes \\
19  & 2        & 6     &  9    &   3& 4 & (10,0,0)& (15,0)   & no \\
\hline
\end{tabular}
\end{minipage}
}
\\
\hspace{-1cm}\includegraphics[width=0.45\linewidth]{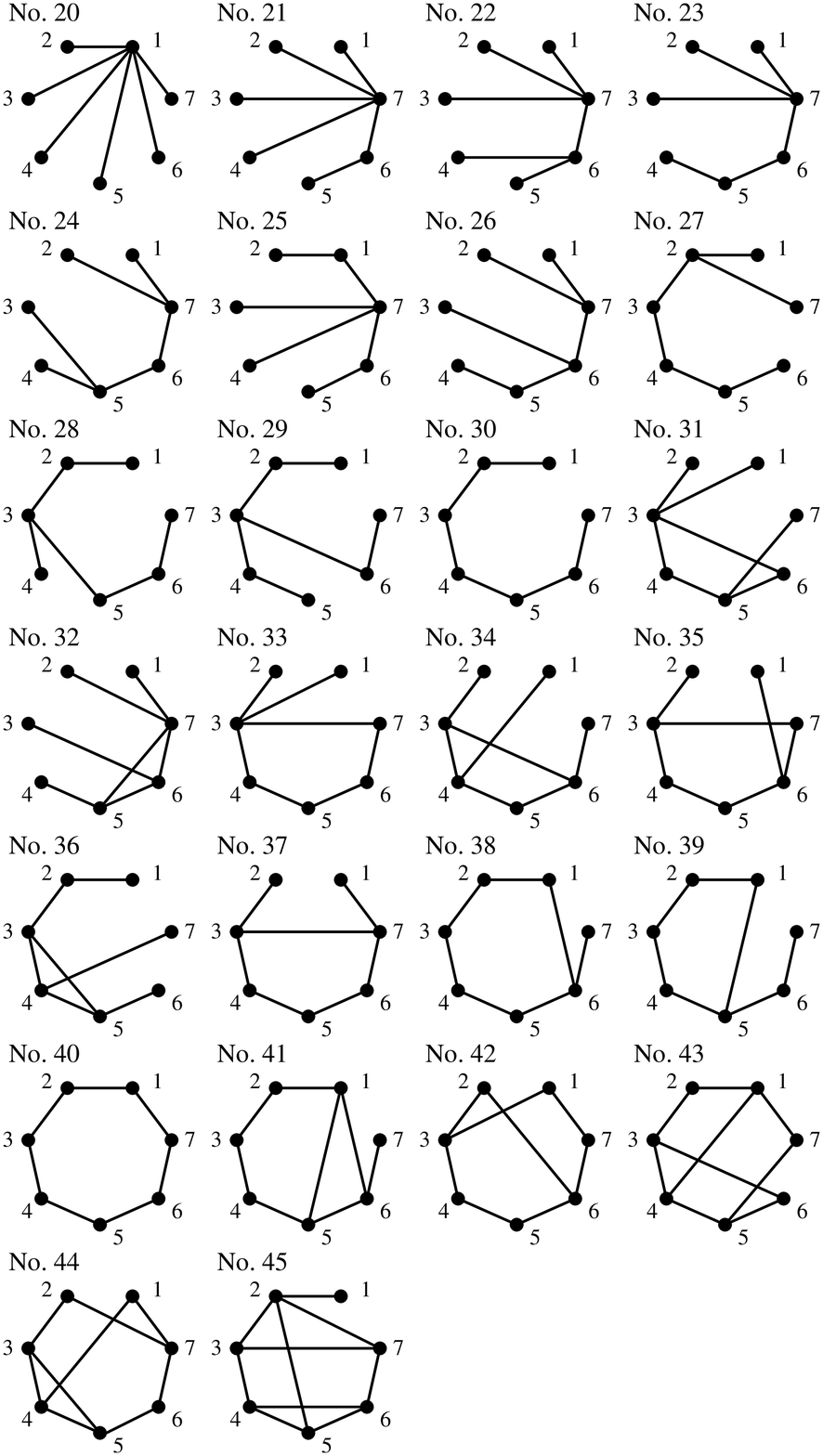} &
\raisebox{5.3cm}{\footnotesize
\begin{minipage}{0.77\linewidth}
\begin{tabular}{|ccccccccc|}
\hline
No.\ & $|\text{LUclass}|$ & $|V|$ & $|E|$ & $ \text{SR}_\text{max}$ & $\text{PP}$ &  $RI_3$ & $RI_2$ & $2-col$ \\
\hline
20  & 2        & 7     &  6    &     1 &   1 & (0,0,35)& (0,21)   & yes \\
21  & 6        & 7     &  6    &     2 &   2 & (0,20,15)& (10,11)   & yes \\
22  & 6        & 7     &  6    &     2 &   2 & (0,30,5)& (12,9)   & yes \\
23  & 16       & 7     &  6    &     2 &   2 & (0,30,5)& (14,7)   & yes \\
24  & 10       & 7     &  6    &     2 &   2 & (0,33,2)& (15,6)   & yes \\
25  & 10       & 7     &  6    &     3 &   3 & (12,16,7)& (16,5)   & yes \\
26  & 16       & 7     &  6    &     3 &   3 & (12,20,3)& (16,5)   & yes \\
27  & 44       & 7     &  6    &     3 &   3 & (12,21,2)& (17,4)   & yes \\
28  & 44       & 7     &  6    &     3 &   3 & (16,16,3)& (18,3)   & yes \\
29  & 14       & 7     &  6    &     3 &   3 & (20,12,3)& (18,3)   & yes \\
30  & 66       & 7     &  6    &     3 &   3 & (20,13,2)& (19,2)  & yes \\
31  & 10       & 7     &  7    &     2 &   2 & (0,34,1)& (16,5)   & yes \\
32  & 10       & 7     &  7    &     3 &   3 & (12,22,1)& (16,5)   & no \\
33  & 21       & 7     &  7    &     3 &   3 & (12,22,1)& (18,3)   & no \\
34  & 26       & 7     &  7    &     3 &   3 & (16,18,1)& (18,3)   & yes \\
35  & 36       & 7     &  7    &     3 &   3  & (16,19,0)& (19,2)   & no \\
36  & 28       & 7     &  7    &     3 &   3 & (20,14,1)& (18,3)   & no \\
37  & 72       & 7     &  7    &     3 &   3 & (20,15,0)& (19,2)   & no \\
38  & 114      & 7     &  7    &     3 &   3   & (22,13,0)& (20,1)   & yes \\
39  & 56       & 7     &  7    &     3 &   4 & (24,10,1)& (20,1)   & no \\
40  & 92       & 7     &  7    &     3 &   4& (28,7,0)& (21,0)   & no \\
41  & 57       & 7     &  8    &     3 &   4& (26,9,0)& (20,1)   & no \\
42  & 33       & 7     &  8    &     3 &   4& (28,7,0)& (21,0)   & no \\
43  & 9        & 7     &  9    &     3 &   3 & (28,7,0)& (21,0)   & yes \\
44  & 46       & 7     &  9    &     3 &   4& (32,3,0)& (21,0)   & no \\
45  & 9        & 7     &  10   &     3 &   4& (30,5,0)& (20,1)   & no \\
\hline
\end{tabular}
\end{minipage}
}
\end{tabular}
\begin{minipage}{1.15\linewidth}
\hspace{-1cm} \caption{ \label{TablePage} List {\bf A}: List of
connected graphs $N=2,3,4,5,6$ vertices that are not equivalent
under LU transformations and graph isomorphisms. List {\bf B}:
List of connected graphs with seven vertices that are not
equivalent under LU transformations and graph isomorphisms. The
corresponding tables list for each equivalence class the number of
vertices $|V|$ and edges $|E|$, the maximal Schmidt rank
$\text{SR}_\text{max}$, the Pauli persistency $\text{PP}$  (see
sec.~\ref{EntMeas_GS}), the rank index $RI_3$ and $RI_2$ (for
splits with 2 or 3 vertices in the smaller partition), the number
of non-isomorphic but LU equivalent graphs $|\text{LUclass}|$ and
the two-colorable property $2-col$.}
\end{minipage}
\end{table}

We have also listed the sizes of the corresponding equivalence
classes under LU and graph isomorphisms, as well as whether
two-colorable representatives exist. By the rank index given in
List {\bf A} and {\bf B} of Table~\ref{TablePage}, we simply
compressed the information contained in the Schmidt rank list with
respect to all bi-partite splittings, counting how many times a
certain rank occurs in splittings with either two or three
vertices in the smaller partition. For example, the rank index
$RI_3=(20,12,3)$ of graph number $29$ means that the rank $3$
occurs 20 times in all possible $3$-$4$-splits, the rank $2$
twelve times and the rank $1$ only three times. Similarly, because
of $RI_2=(18,3)$ the rank $2$ ($1$) occurs $18$ ($3$) times in
all $2$-$5$-splits of the graph number $29$. As it can be seen
from Tab.~\ref{SR_List_Comparison}, although the classes of graph
states in fig.~\ref{fig:LUruleExample1} and
fig.~\ref{fig:LUclassExample2}  have different Schmidt rank lists
and thus are non-LU-equivalent, both classes have the same rank
index $RI_2=(2,1)$, since the rank indices are invariant under
arbitrary permutations of the vertices. Thus no graph in
fig.~\ref{fig:LUruleExample1} is locally equivalent to any graph
in the equivalence class represented in
fig.~\ref{fig:LUclassExample2}. But both belong to the same
equivalence class represented by graph No.~4 in List~{\bf A} when
considering both local unitary transformations {\em and} graph
isomorphisms. In fact a permutation of vertices $2$ and $3$ maps
graph No.~4 in fig.~\ref{fig:LUclassExample2} onto graph No.~1 in
fig.~\ref{fig:LUruleExample1}.

In Tab.~\ref{numbers_of_GS} we have summarized the number of
equivalence classes for connected and all (i.e., possibly
disconnected) graphs and compared them with the corresponding
values when disregarding LC-equivalence. From a quantum
information point of view, which only considers states up to
LU-equivalence, the class of non-equivalent graph states provides
a significant reduction of the set of all graph states. The table
nevertheless shows that the obtained set of non-LU-equivalent
graph states is still sufficiently rich to form an interesting
subclass of states to serve as a starting point for the study of
multi-party entanglement. Besides the question of local
equivalence we will see in the remainder of this thesis that many
other interesting entanglement properties of graph states have a
concise and efficient translation in terms of the underlying
graph. This allows for an exemplary study of multi-party
entanglement in the regime of many parties.

\index{equivalence classes|)} \index{equivalence under!local
Clifford unitaries (LC)|)} \index{equivalence under!local
unitaries (LU)|)} \index{equivalence under!stochastic local
operations and classical communication (SLOCC)|)}


\section{Entanglement in graph states}\label{EntanglementGS}

As discussed in the previous sections, graph states provide an interesting class of multi-partite states that are relatively easy to survey even in the regime of many parties. Since the graph essentially encodes a preparation procedure of the state, we will now examine the question how the entanglement in a graph state is related to the topology of its underlying graph. More precisely we address the issue of quantifying and characterizing the entanglement of graph states.

We start in sec.~\ref{Bell&Witness} with a review of results from \cite{Gue04,To04} about the `non-classicality' of graph states and how the entanglement present in these states can be experimentally verified considering Bell inequalities or entanglement witnesses.  Then, classical correlations and entanglement between pairs of particles are discussed in sec.~\ref{Corr}. The main part of this section is finally devoted to the quantification of entanglement in graph states in terms of the Schmidt measure, which will be introduced in sec.~\ref{EntMeas_GS}. We present bounds and rules that render the evaluation of this measure feasible for some interesting classes of graph states and discuss some examples of practical interest.

\subsection{Bell inequalities and entanglement witnesses}\label{Bell&Witness}

The notion of entanglement in quantum mechanics as it was posed in ref.~\cite{EPR35,schroedinger35} by Einstein, Podolsky, Rosen and Schr\"odinger in the year 1935 has -- since Bell's reformulation \cite{Bell64} in 1964 -- frequently been used synonymous with `{\em non-classical correlations}'\index{non-classical correlations}, although today it is well-known \cite{We89,Po95} that one has to consider a finer distinction. Up to some so-called `detection loophole', first experiments \cite{As81} were able to verify that some quantum states can indeed reveal correlations which cannot be predicted by some {\em local hidden variable}\index{local (realistic) hidden variable (LHV) model} (LHV) models. In these models any observable has a predetermined value ({\em realism}), regardless of whether it is measured or not. Moreover, the choice of which observable is measured does not `affect the other parties' ({\em locality}). The two constraints can be phrased in terms of {\em Bell inequalities}, which bound the possible correlations that can be explained within these LHV models. For a precise formulation of the concepts of LHV description and the derivation of the corresponding Bell inequalities in the multi-party setting, we refer the reader to refs.~\cite{Me90,We01a,Zu01} and for a brief review to ref.~\cite{Per99,We01b}. Different Bell inequalities can also be regarded as {\em entanglement witnesses} \cite{Ter00,Lew00} for different types of entanglement in a multi-party entangled state. These witnesses can be quite useful to detect entanglement in the vicinity of graph states. In the following we will shortly review the results of \cite{Gue04,To04} on Bell inequalities and entanglement witnesses for graph states.

But let us first consider an extension of the GHZ-argument \cite{GHZ89} that rules out a LHV description of the spin statistics for GHZ-states, to the case of general graph states \cite{Sc04, Gue04}: The non-trivial graph state with two qubits is LC-equivalent to the singlet state and thus violates the original Bell inequality proposed by Bell in
ref.~\cite{Bell64}. For any connected graph state on more than three vertices any connected subgraph on three vertices $a,b,c$ gives rise to a contradiction within any possible explanation of the observed correlations between spin measurements at different particles. Consider, for example, the case where all three vertices $a,b,c$ are pairwise adjacent, i.e., $\{a,b\}, \{b,c\}, \{a,c\} \in E$. Due to the {\em non-commutativity} of the spin observable algebra one easily computes that the product of the corresponding correlation operators $K_a=\sigma_x^a\sigma_z^{N_a}$, $K_a=\sigma_x^b\sigma_z^{N_b}$ and $K_a=\sigma_x^c\sigma_z^{N_c}$ yields
\be K_a K_b K_c \,=\, - \sigma_x^a \sigma_x^b \sigma_x^c \sigma_z^{N_a+N_b+N_c} \; .\ee

\begin{wrapfigure}[9]{r}{0.3\textwidth}
\vspace{-1.2cm}
\includegraphics[width=0.3\textwidth]{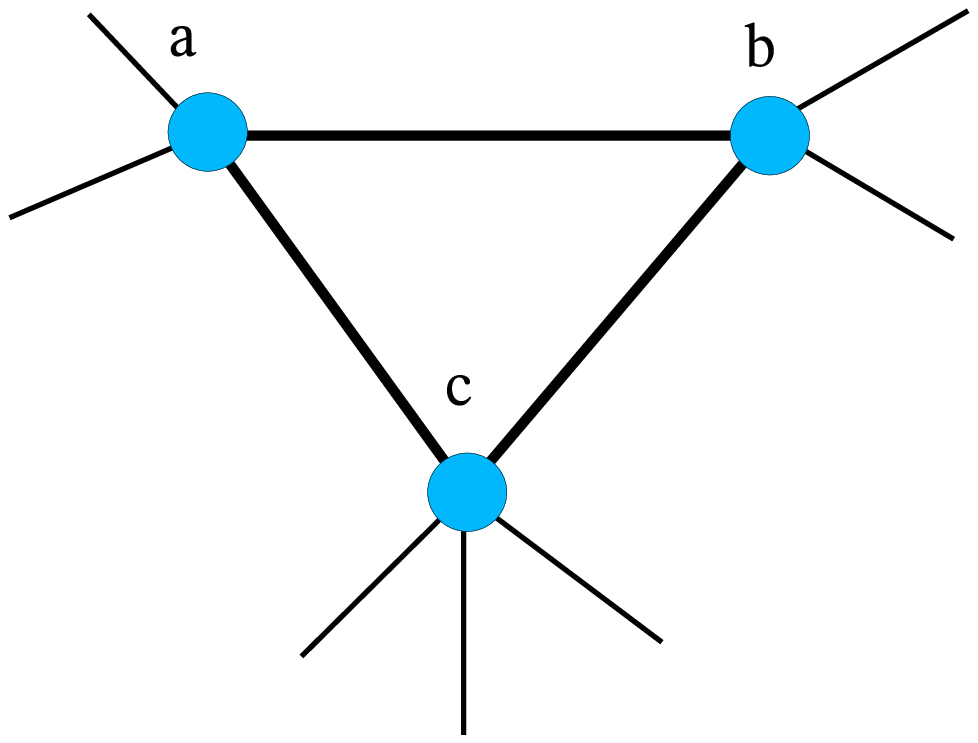}
\caption{\label{NonLocFig} Any connected subgraph on three vertices gives rise to a violation of local realism.}
\end{wrapfigure}
As discussed in sec.~\ref{DefOfGS_Stab} these stabilizer elements provide constraints to the four different measurement settings
\bea
(I)\hspace{0.2cm}  m_x^a m_z^{N_a} = 1 & & (III)\hspace{0.2cm}  m_x^c m_z^{N_c} = 1 \nonumber \\
(II)\hspace{0.2cm}  m_x^b m_z^{N_b} = 1 & & (IV) \hspace{0.2cm} - m_x^a m_x^b m_x^c m_z^{N_a+N_b+N_c} = 1\; ,\nonumber
\eea
where $m_x^a=\pm 1$ and $m_z^a=\pm 1$ denote the measurement outcomes if the qubit at vertex $a$ is measured in spin-$x$- or $z$-direction. More precisely any LHV model, which assigns predetermined\footnote{That means independently of the chosen measurement setting in (I)-(IV).} values $m_x$, $m_z$ to measurement outcomes to $x$-and $z$-measurements at the different vertices in $N_a\cup N_b\cup N_c$ with some probability, must be such that these assignments obey all the four equations (I)-(IV). Due to the {\em commutativity} for the multiplication of the measurement results the product of eq.~(I), (II) and (III) gives
\be m_x^a m_x^b m_x^c m_z^{N_a+N_b+N_c} = 1\ee
and thus a contradiction with eq.~(IV). A similar argument holds for the case where only two pairs of the three vertices $a,b,c$ are adjacent. Thus we have obtained \cite{Sc04, Gue04}:

{\proposition[\bf Non-classicality of graph states]\index{GHZ argument} Any graph state corresponding to a connected graph violates local realism. More precisely, for a connected graph state with more than two vertices any connected subgraph on three vertices $a,b,c$ yields a contradiction when trying to explain the correlations between the different Pauli-spin-observables present in the reduced state $\rho_G^{A}$ (on the subset $A=N_a\cup N_b\cup N_c$) by means of some LHV model.
}

More quantitatively one can compare (convex) functions of the correlations in a state obtained for different measurement settings with the results which can be found for these settings when considering all LHV models. Upper bounds to these functions give rise to so-called Bell inequalities\index{Bell inequality|(}, which the correlations arising from different measurement settings have to obey, if they can be explained by a LHV model. In fact for $N$ parties and two dichotomic\footnote{I.e., the respective observables have two measurement outcomes $\pm 1$.} observables at each party the set of states that allow for a LHV description, can be characterized by only one inequality \cite{We01a,Zu01}, which is a generalization of the CHSH inequality \cite{CHSH69} for the two-party setting. For the multi-party setting the different Bell inequalities can in general capture different types of entanglement according to different partitionings. For any non-trivial graph state Bell inequalities with three dichotomic measurements per site have been derived in ref.~\cite{Gue04} that are maximally violated by this graph state:
{\proposition[\bf Bell inequalities for graph states] A (connected) graph state $|G\rangle$ with stabilizer $\mathcal{S}$ maximally violates\footnote{I.e., $\left|\sum_{\sigma \in\mathcal{S}} \langle G | \sigma | G\rangle  \right| = 2^N$. } the Bell inequality \be\label{BellIneq} \left|\sum_{\sigma \in\mathcal{S}} \langle \sigma\rangle  \right| \leq \mathcal{C} \; ,\ee where $\mathcal{C}$ is the maximum of the absolute value of the mean value $\langle \sum_{\sigma \in\mathcal{S}} \sigma \rangle$ taken over all deterministic\footnote{Due to convexity it suffices to consider deterministic LHV models, which assign definite values $\pm 1$ to all observables (with unit probability).} LHV models. In the case that $\mathcal{C}\geq 2^{N-1}$ the Bell inequality detects only states that are NPT with respect to every partitioning. \index{partial transposition $\rho^{T_A}$}\index{NPT (negative partial transpose)}}

ref.~\cite{Gue04} also provides general rules for the computation of the value $\mathcal{C}$, which a priori demands to check an exponentially (with N) increasing number of LHV models. Numerical results for graphs with up to $10$ vertices show that Bell inequalities for rings and chains give rise to a large relative violation $\frac{2^N}{\mathcal{C}}$ for these states while the  Bell inequalities for the corresponding GHZ states yield a small relative violation.\index{Bell inequality|)}\index{entanglement criterion!Bell inequality}

This Bell inequality can be expressed in terms of an entanglement witness\footnote{An entanglement witness $\mathcal{W}$ is an observable with a positive or zero expectation value $\text{tr}(\mathcal{W}\rho)\geq 0$ for {\em all} separable states and a negative expectation value $\text{tr}(\mathcal{W}\rho)< 0$ for {\em some} entangled states. In a measurement of this witness the entanglement present in the latter states can thus be detected (`witnessed').} of the type \be\label{Witness0} \mathcal{W}\,:=\, \frac{\mathcal{C}}{2^N}\mathbf{1}_V-|G\rangle\langle G|\; ,\ee i.e., a state $\rho$ violates the eq.~(\ref{BellIneq}) iff  \be \langle \mathcal{W} \rangle = \text{tr}(\mathcal{W}\rho) < 0\; .\ee
To detect entanglement itself and not `non-classicality' with respect to LHV descriptions, the constant $\mathcal{C}$ can be chosen according to a maximization over the smaller set of separable states only\footnote{Note that there exists non-separable states that still allow for a LHV description \cite{We89}.}.
A concrete procedure to measure this witness in an experiment can be found by decomposing $\mathcal{W}$ into a sum of locally measurable operators. But, for an arbitrary graph state, the witness in eq.~(\ref{Witness0}) seems to be decomposable only into an exponentially (with $N$) increasing number of local measurement settings \cite{To04}.
T\'oth and G\"uhne therefore proposed more practical entanglement witnesses and Bell inequalities for graph states that in many cases, such as CSS states, can be evaluated in only two measurement settings \cite{To04,To05,To05b}\index{two-colorable graph (state)}\index{CSS state}\index{bi-partite graph (state)}:

{\proposition[\bf Entanglement witnesses for graph states]\index{entanglement witness}\index{entanglement criterion!witness}   Let $|G\rangle$ be a graph state corresponding to a connected graph. Then
\be\label{Witness1} \mathcal{W}_1^{ab}\, :=\,\mathbf{1}_V- K_a - K_b \ee
is an entanglement witness for the $|G\rangle$ that detects entanglement in the reduced state $\rho_G^{A}$ ($A=N_a\cup N_b \cup\{a,b\}$) with only two measurement settings and thus can rule out {\em full separability} of the total graph state. The entanglement witness \be\label{Witness2} \mathcal{W}_2 \,:= \,(N-1) \mathbf{1}_V- \sum_{a\in V} K_a  \ee detects {\em genuine multi-party entanglement}. If $G$ is $M$-colorable, then the evaluation of the witness eq.~(\ref{Witness2}) requires at most $M$ local measurement settings.\index{coloring}
}

For further entanglement witnesses that are particularly robust against global white noise and that can be derived for some special cases such as GHZ states and linear cluster states, we refer the reader again to ref.~\cite{To04,To05}.
In practice it is quite useful that the witness eq.~(\ref{Witness2}) for genuine entanglement gives also lower bounds to the fidelity $\langle G | \rho | G \rangle $ with the ideal graph state, when $\rho$ denotes the outcome of a actual preparation procedure for the graph state. This provides an efficient method to verify that, in a given experiment, the entanglement is really present in a form that is sufficiently close to a desired graph state. An alternative approach to detect multi-partite entanglement, which is particularly suited for implementations in optical lattices and magnetic micro-traps, can be found in ref.~\cite{MA04}.

\subsection{Two-particle correlations and localizable entanglement}\label{Corr}

In this section we consider the entanglement properties of the reduced state $\rho_G^{\{a,b\}}$ of two qubits $a$ and $b$ that is obtained after tracing out or disregarding the information about the remaining particles $c\in V\setminus\{a,b\}$ in a graph state $|G\rangle$. As discussed in sec.~\ref{Reduced_GS} the reduced state
\be \rho_{\{a,b\}}=\sum_{\sigma\in\mathcal{S}_{\{a,b\}}} \sigma \ee
is essentially given by those stabilizer elements $\sigma\in\mathcal{S}_{\{a,b\}}$ that act non-trivially only on the qubits $a$ and $b$. Let us first examine the {\em classical correlations}\index{correlation function $\text{Q}^{ab}_{ij}$} between two non-isolated vertices $a$ and $b$  in a graph state $|G\rangle$
\be\label{CorrelationFunc} Q^{ab}_{ij} := \langle G|\sigma_i^a\otimes \sigma_j^b|G \rangle  \, - \, \langle G|\sigma_i^a |G \rangle  \, \langle G| \sigma_j^b|G \rangle   \hspace{0.7cm} i,j=1,2,3\; .\ee
Note that these {\em correlation functions} $\text{Q}^{ab}_{ij}$ only depend on the reduced state $\rho_G^{\{a,b\}}$ of the particles $a$ and $b$ and are given by
\be\label{CorrelationFunc_1}
 \text{Q}^{ab}_{ij}=\text{tr}\left(\rho_G^{\{a,b\}} \sigma_i^a \sigma_j^b\right)
  = \left\{ \begin{array}{ccc} 1 & \text{if} & \sigma_i^a \sigma_j^b  \in \mathcal{S} \\  0 & \text{if} & \sigma_i^a \sigma_j^b  \notin \mathcal{S} \end{array} \right.\; ,
\ee
since the expectation values e.g. $\langle G|\sigma_i^a |G \rangle = \text{tr}\left(\rho_G^{a} \sigma_i^a \right) = \text{tr}\left(\frac{1}{2}\mathbf{1}_a \sigma_i^a \right)$ vanish for both (non-isolated !) vertices $a$ and $b$.
For the {\em maximal classical correlation}\index{maximal classical correlation $\text{Q}_\text{max}^{ab}$} between two vertices $a$ and $b$ in a graph state $|G\rangle$
\be\label{MaxCorrelation} \text{Q}_\text{max}^{ab} = \max_{i,j = 1,2,3} \, |\text{Q}_{ij}^{ab}| \ee
we find that it vanishes whenever the neighborhoods $N_a\setminus b$ and $N_b\setminus a$ of the two vertices with respect to the remaining graph are non-empty and distinct.
{\proposition[{\bf Two-party classical correlation}]\index{two-particle correlations}\label{2partyclassCorr}
For two non-isolated\footnote{If one vertex is isolated all correlation functions vanish, since the corresponding state is a product state e.g. $|G\rangle=|+\rangle^a|G\setminus a\rangle^{V \setminus a}$.} vertices $a,b \in V$ in some graph $G=(V,E)$ we have
\be
 \text{Q}_\text{max}^{ab}\, =\, \left\{ \begin{array}{ccl} 0 & \text{if} &  (N_a\setminus b),( N_b \setminus a) \neq \emptyset \; \text{and} \; (N_a\setminus b)\neq( N_b \setminus a)\\  1 & & \text{otherwise} \end{array} \right. \; .
\ee
}

{\em Proof:} According to eq.~(\ref{CorrelationFunc_1}) we have to show that all stabilizer elements $\sigma \in \mathcal{S}$ have support $\text{supp}(\sigma)$ on more than the two vertices $a$ and $b$ iff $(N_a\setminus b)\neq( N_b \setminus a)$  and $(N_a\setminus b),( N_b \setminus a) \neq \emptyset$. Since all generator elements are generated by different combinations of the correlation operators $\prod_{c\in C} K_c$ ($C\subseteq V$), the sufficiency can be derived as follows. In order to generate a stabilizer element $\sigma=\prod_{c\in C} K_c$ with $\text{supp}(\sigma)\subseteq \{a,b\}$ at most the two correlation operators on vertices $a$ and $b$ can be considered, i.e., $C \subseteq \{a,b\}$, because any other correlation operator $K_c$ for $c\in V\setminus \{a,b\}$ leads to a non-vanishing Pauli operator $\sigma_x^c$ or $\sigma_y^c$ on the central vertex $c$ outside of $\{a,b\}$. Moreover any support $\text{supp}(K_a)\setminus \{a,b\} \neq \emptyset$ or $\text{supp}(K_b)\setminus \{a,b\} \neq \emptyset$ of the correlation operators $K_a$ and $K_b$ outside of $\{a,b\}$ can only be compensated if both vertices have the same neighbors outside of $\{a,b\}$, i.e., $ (N_a\setminus b)=( N_b \setminus a)$. In this case we are left with the two possibilities
\be K_a K_b = \left\{ \begin{array}{ccc} \sigma_y^a \sigma_y^b & \text{if} & \{a,b\}\in E \\ \sigma_x^a \sigma_x^b & \text{if} & \{a,b\}\notin E  \end{array} \right.\;  .
\ee
\proofend

Although a graph state might contain some non-vanishing {\em classical} two-party correlations it does generally not include any entanglement between any two qubits, unless -as discussed below- the remaining parties are allowed to assist the revealing of such entanglement.
{\proposition[{\bf Two-party `quantum correlation'}]\index{non-classical correlations}
For any vertices $a,b \in V$ the reduced state $\rho_G^{\{a,b\}}$ of some graph state $|G\rangle$ is separable unless the graph contains the isolated edge\footnote{If $G$ contains an isolated edge $\{a,b\}$ the state $|G\rangle$ decomposes into a pure Bell state on the vertices $\{a,b\}$ and some other graph state on the remaining vertices.} $\{a,b\}$.
}

{\em Proof:}
If $G$ does not contain the edge $\{a,b\}$  as an isolated edge then according to Proposition~\ref{reduced_GS} the reduced state $\rho_G^{\{a,b\}}$ is either the rank-$2$-projector $\rho_G^{\{a,b\}}=\frac{1}{2}\left(P_{12} +\sigma_z^{C} P_{12} \sigma_z^{C}\right)$ ($P_{12}:= |G[\{1,2\}]\rangle\langle G[\{1,2\}]|$) for some set $C= \{1\},\{2\},\{1,2\}$ or it is a rank-$4$-projector and thus the maximally mixed state $\rho_G^{\{a,b\}}=\frac{1}{4}\mathbf{1}_{ab}$. In the latter case the above statement is trivial. That also the rank-$2$-projectors correspond to separable states can be derived from the fact that they are PPT (which means that their partial transpose is positive) according to eq.~(\ref{PTofa}), which is a sufficient condition for separability in $2\times 2$-systems.
\proofend

Although there is {\em per se} no entanglement between arbitrary two particles in a connected graph state, such entanglement can be revealed between any two parties $a$ and $b$, if the remaining parties are allowed to perform local measurements \cite{Briegel01}. More generally, the notion of {\em localizable entanglement}\index{localizable entanglement $\text{LE}^{ab}$} $\text{LE}^{ab}(\rho)$ was introduced in refs.~\cite{Frank,Verstraete04b,Popp04} for multi-spin states $\rho$, defined as the maximal amount\footnote{
E.g. in terms of its concurrence.} of entanglement that can be created (or {\em localized}), on average, between two spins at position $a$ and $b$ by performing local measurements on the other spins. For a general state $\rho$  it has been shown in ref.~\cite{Frank} that the localizable entanglement $\text{LE}^{ab}$ is related to the maximal classical correlation $Q_\text{max}$ and the {\em entanglement of assistance}\footnote{The entanglement of assistance extends the concept of localizable entanglement in that it allows also joint measurements to be performed on the other spins.} $\text{AE}^{ab}$\index{entanglement of assistance $\text{AE}^{ab}$} as measured by the concurrence \cite{DiVincenzo98,Laustsen03} :
\be \text{Q}_\text{max}^{ab}(\rho)\;\leq\; \text{LE}^{ab}(\rho)\;\leq \;\text{AE}^{ab}(\rho) \hspace{1cm}  \; .\ee
Despite of the separability of the reduced states, measurements on the remaining particles can nevertheless create maximal entanglement between any two vertices in a `connected' graph state, which corresponds to maximal localizable entanglement in this case.
{\proposition[{\bf Localizable entanglement}]\label{locEntGS}
Consider any two vertices $a,b \in V$ in a graph state corresponding to a {\em connected} graph $G=(V,E)$\index{connected graph (state)}.
In all measurement branches of the following protocol \cite{Briegel01} a maximal entangled state is created between the vertices $a$ and $b$:
\begin{enumerate}
\item Choose any path $(a_0=a,a_1,\ldots,a_{n-1},a_n=b)$ connecting the vertices $a$ and $b$.
\item Measure the spin of all vertices except $a_i$ in $z$-direction.
\item Measure the spin of all vertices $a_i$ for $i=1,\ldots, n-1$ in $x$-direction.
\end{enumerate}
Thus the {\em localizable entanglement $\text{LE}^{ab}$} of a `connected' graph state is maximal.
}

{\em Proof:}
If the graph is connected there exists a path connecting any two vertices $a,b \in V$.
According to Proposition~\ref{Pauli_Measurement} the $\sigma_z$-measurements in the second step simply remove all vertices but those on the path $(a_0=a,a_1,\ldots,a_{n-1},a_n=b)$. Note that the different local unitaries according to the different $\sigma_z$-measurement outcomes consists only of $\sigma_z$-operators and thus do not alter the measurement direction of the subsequent $\sigma_x$-measurements in the 3.~step. The same proposition also implies that a sequence of $\sigma_x$ measurements on the inner vertices $a_i$ ($i=1,\ldots, n-1$) of this path removes these inner vertices but keeps the connectivity between the two neighboring particles $a_{i-1}$ and $a_{i+1}$ in this chain. E.g. an $\sigma_x$-measurement at vertex $a_1$ of the initial chain $G_0$ yields the shorter chain $G_1=(G_0\setminus a_1) \cup \{a_0,a_2\}$ on the remaining vertices $(a_0=a,a_2,\ldots,a_{n-1},a_n=b)$. The LC-unitaries corresponding to the different measurement outcomes again do not rotate the subsequent measurement directions. Thus we `inductively' arrive at the final graph $G_{n-1}$ corresponding to the maximally entangled state $|G_{n-1}\rangle $ of eq.~(\ref{Graph_Bell_State}).

\proofend

Concluding, the results of this section indicate that the entanglement present in graph states is not based on {\em bi-partite} `quantum correlation' but the entanglement is rather {\em delocalized} among all particles.


%
\subsection{Quantifying entanglement}\label{EntMeas_GS}

We have seen that graph states are entangled quantum states that exhibit complex structures of genuine
multi-particle entanglement. The main aim of this section is to apply the quantitative theory of multi-particle entanglement to the study of correlations in graph states.
Needless to say, despite considerable research effort, there is no known computable entanglement measure that grasps all aspects of multi-particle entanglement in an appropriate manner, if there is any way to fill such a phrase with meaning. Several entanglement measures for multi-particle systems have yet been suggested and their properties studied \cite{Schmidt,Wei03,Tangle,Plenio,Meyer,Barnum,Fat04}.

In this section the underlying measure of entanglement is taken to be the Schmidt measure \cite{Schmidt}, which is a proper multi-particle entanglement monotone that is tailored to the characterization of such states. As holds true for any known measure of multi-particle entanglement, its computation is exceedingly difficult for general  states, yet for graph states this task becomes feasible to a very high extent. We present various upper and lower bounds for the Schmidt measure in graph theoretical terms, which largely draw from stabilizer theory. These bounds allow for an evaluation of the Schmidt measure for a large number of graphs of practical importance.

\index{Schmidt measure $\text{E}_S$|(}\index{entanglement measure}

The Schmidt measure has been employed to quantify the degree of entanglement, as a generalization of the Schmidt rank in the bi-partite setting \cite{Schmidt}. This measure is sufficiently coarse to be accessible for systems consisting of many constituents and to allow for an appropriate discussion of multi-particle entanglement in graph states.

Any state vector $|\psi\rangle\in {\bf H}_{1} \otimes ...\otimes {\bf H}_{N}$
of a composite quantum system with $N$ components
can be represented as
\begin{equation}\label{SchmidtM}
    |\psi \rangle = \sum_{i=1}^R \xi_i |\psi_i^{1}\rangle \otimes\ldots\otimes |\psi_i^{N}\rangle,
\end{equation}
where $\xi_i\in{\mathbbm{C}}$ for $i=1,...,R$, and $|\psi_i^{n}\rangle \in {\bf H}_{n}$ for $n=1,...,N$.
The {\em Schmidt measure}\index{Schmidt measure $\text{E}_S$} associated with a state vector $|\psi\rangle$ is then defined as
\begin{equation}
 \text{E}_S(|\psi\rangle ) = \log_2 (R_\text{min}),
\end{equation}
where $R_\text{min}$ is the minimal number $R$ of terms in the sum of eq.~(\ref{SchmidtM}) over all linear decompositions into product states. It can be extended to the entire state space (and not only the extreme points)
via a convex roof extension\footnote{Note that every positive function $E_\text{pure}$ defined on the set of pure states that vanishes exactly on product states (see property (i)) and is non-increasing under SLOCC (see property (ii)), can be extended to the entire state space by \[E(\rho):=\text{inf} \,\{\,\sum_i \lambda_i E_\text{pure}(|\psi_i\rangle) \; |\; \rho = \sum_i \lambda_i |\psi_i\rangle \langle\psi_i| \,\text{with} \, \lambda_i\geq 0 \,\sum_i \lambda_i=1\}\;. \] Note that infimum of the average entanglement can be regarded as taken with respect to all preparation procedures of the mixed state. The entanglement measure $E$ can be proven to (a) vanish exactly on all separable states and is (b) convex as well as (c)  non-increasing under SLOCC operations.}. It should be noted that the Schmidt measure $ \text{E}_S$ is {\em discrete}, e.g. in the case of two-level systems ${\bf H}_{n}=\mathbb{C}^2$ it can only take the values \be  \text{E}_S(|\psi\rangle ) \in \{\text{log}_2(m) \,|\, m=1,\ldots,2^N\}\; ,\ee
and thus fails to be {\em continuous}, which requires some care when extending the measure to general mixed states via the convex roof construction. However, since the set of graph states is discrete itself, the case does not occur that two graph states become arbitrarily close in terms of some distance measure but not with respect to their Schmidt measure. In any case the Schmidt measure is a general entanglement monotone with respect to general local operations and classical communication (LOCC), which  typically leave the set of graph states.
If the Schmidt measure of a state vector $|\psi\rangle$ is evaluated with respect to a partitioning $(A_1,...,A_M) $, it will be appended,
\begin{equation}
 \text{E}_S^{(A_1,...,A_M)  }(|\psi\rangle),
\end{equation}
in order to avoid confusion. For a graph $G=(V,E)$, the partitioning with $M=N$ and $A_n=\{n\}$ will be referred to as {\em finest partitioning}. If no upper index is appended to the Schmidt measure, the finest partitioning will be implicitly assumed.

Among the properties that are important for the rest of this section are the following \cite{Schmidt,MarcPhD}:
{\proposition[Schmidt measure]\index{entanglement measure} $ $
{\em
\begin{itemize}
\item [{\bf (i)}] $ \text{E}_S$ {\em vanishes on product states} (only), i.e.
\be\label{E of product states}
 \text{E}_S(| \psi\rangle)= 0 \hspace{1cm}\Longleftrightarrow \hspace{1cm}|\psi \rangle =|\psi^{1}\rangle \otimes \ldots \otimes |\psi^{N}\rangle \; .
\ee
\item [{\bf (ii)}]\index{entanglement measure!monotonicity} $ \text{E}_S$ is {\em non-increasing under SLOCC} \cite{Schmidt}, i.e.
\be\label{E under SLOCC}
 |\psi\rangle \longrightarrow_{\text {SLOCC}} |\psi'\rangle \hspace{1cm}\Longrightarrow \hspace{1cm}  \text{E}_S(|\psi'\rangle ) \leq  \text{E}_S(|\psi\rangle ) \; .
\ee
Similarly for the LU-equivalence we find
\begin{equation}\label{E under LU}
  |\psi\rangle \longleftrightarrow_{\text {LU}} |\psi'\rangle \hspace{1cm}\Longrightarrow \hspace{1cm}  \text{E}_S(|\psi'\rangle ) =  \text{E}_S(|\psi\rangle ) \; .
\end{equation}
\item [{\bf (iii)}] $ \text{E}_S$ is {\em non-increasing under a coarse graining of the partitioning}, i.e.
\begin{eqnarray} \label{E under coarse graining}
(A_1,...,A_M)\leq (B_1,...,B_{M'})  \;\Longrightarrow \;   \text{E}_S^{(A_1,...,A_M)}(|\psi\rangle) \geq  \text{E}_S^{(B_1,...,B_{M'})}(|\psi\rangle) \; .
\end{eqnarray}
Thus if two components are merged in order to form a new component, then the Schmidt measure can only decrease.
\item[{\bf (iv)}]\index{entanglement measure!subadditivity} $ \text{E}_S$ is {\em sub-additive}, i.e., 
\be\label{E sub-additivity}
  \text{E}_S^{(A_1,...,A_M,B_1,...,B_{M'})} \left( | \psi_1 \rangle \otimes | \psi_2 \rangle \right)  \leq   \text{E}_S^{(A_1,...,A_M)} \left( | \psi_1 \rangle \right) +  \text{E}_S^{(B_1,...,B_{M'})} \left( | \psi_2 \rangle \right)\; .
\ee
\item[{\bf (v)}] {\em For any bi-partition $(A,B)$} $ \text{E}_S$ coincides with the {\em Schmidt rank}\index{Schmidt rank $\text{SR}_A$} $ \text{E}_S(|\psi\rangle) = \text{SR}_{A} (|\psi\rangle)= \log_2({\text{rank}}  (\text{tr}_A[|\psi\rangle\langle\psi|])) $. In particular, $ \text{E}_S$ is additive within a given bi-partitioning, i.e., if $A=A_1\cup A_2$  and $B=B_1 \cup B_2$,
then
\be\label{E additivity for bi-partitions}
 \text{E}_S^{(A,B)}(|\psi_1\rangle \otimes |\psi_2\rangle ) = \text{E}_S^{(A_1,B_1)}(|\psi_1\rangle) +  \text{E}_S^{(A_2,B_2)}(|\psi_2\rangle) \;.
\ee
\end{itemize}
}}

It should be noted that for general pure states of multi-partite quantum systems the Schmidt
measure is -- as any other measure of multi-partite entanglement -- exceedingly difficult to compute \cite{He04}. In the following we will provide lower and upper bounds for the Schmidt measure of graph states in graph theoretic terms, which will coincide in many cases, and will then apply these rules, and calculate the Schmidt measure for some of graphs and graph classes that are of interest for applications..
\index{Schmidt measure $\text{E}_S$|)}

\index{Schmidt measure $\text{E}_S$!for graph states|(}

\index{Schmidt measure $\text{E}_S$!lower bound}\index{maximal Schmidt rank $\text{SR}_\text{max}$|(}\index{Schmidt rank $\text{SR}_A$}
Let us first derive a {\em lower bound} to the Schmidt measure, namely the {\em maximal Schmidt rank} $\text{SR}_\text{max}$
\be \text{SR}_\text{max}(G)\, :=\, \max_{A\subseteq V}\,\text{SR}_A (G)  \; .\ee
If one maximizes over all bi-partitionings $(A,B)$ of a graph $G=(V,E)$, then according to eq.~(\ref{E under coarse graining}) one obtains a lower bound for the Schmidt measure with respect to the finest partitioning.

Since the  Schmidt ranks  $\text{SR}_A (\psi)$ for the different bi-partitions are already entanglement monotones with respect to $(A,B)$-local SLOCC-operations, it is straightforward to see that $\text{SR}_\text{max}$ is a proper though also discrete entanglement measure\footnote{$\text{SR}_\text{max}$ has properties {\bf (i)} -- {\bf (iv).}} that captures the notion of maximal bi-partite entanglement contained in a multi-party entangled state. In fact this measure was also considered in ref.~\cite{Vidal03,Vidal04} in the context of an efficient simulation of a quantum algorithm on a classical computer. There it was shown that, if through a pure-state quantum computation all underlying pure states of the $N$-qubit quantum register have polynomial bounded `Schmidt rank', i.e.,  $2^{\text{SR}_\text{max}}\leq \text{poly}(N)$, then this quantum computation can be {\em efficiently simulated by a classical algorithm}\index{quantum computation} requiring only polynomial increasing memory space and computational time. The classical algorithm allowing for a simulation of quantum computations with only {\em slightly entangled} quantum register uses an efficient decomposition of the pure state as in eq.~(\ref{SchmidtM}) that allows to directly read off and manipulate the Schmidt coefficients for different bi-partitionings through the quantum computation. This algorithm proved to be very useful also for an efficient simulation of the dynamics in one-dimensional quantum many-body systems due to its tight connection to existing methods such as {\em density matrix renormalization group (DMRG)}\index{density matrix renormalization group (DMRG)}\cite{Sc05,Daley04} and opens a way to even simulate systems in higher dimensions, where these methods so far have not seemed to be very suitable \cite{AdvancedDMRG,Ve042d}.

For graph states the maximal Schmidt rank $\text{SR}_\text{max}$ actually coincides with continuous entanglement measures such as the {\em entropy of entanglement}\index{entropy of entanglement ${S}_{A}$}\index{entanglement measure!entropy ${S}_{A}$} or the {\em purity of the reduced density matrices}\index{purity}\index{entanglement measure!purity}. From eq.~(\ref{reduced_GS_2}) one can compute that the entropy or the purity of the reduced density matrices for $|G\rangle$ according to a bi-partition $(A,B)$ gives
\be \text{SR}_A(G) \,=\,- \text{tr} [\rho^A_G\log_2(\rho^A_G)] \,=\, \log_2(\text{tr} [(\rho^A_G)^2])  \;. \ee  This again expresses the fact that, for a non-empty graph, $|G\rangle$ is a `maximally' $(A,B)$-entangled
state vector with $2^{ \text{E}_S^{(A,B)}}$ Schmidt coefficients.

Finally, the Schmidt rank of a graph state is {\em closely related to error correcting properties of a corresponding graph code}\index{stabilizer code}\index{quantum error correcting code (QEC)} (see sec.~\ref{Application_QEC}). Let $A$ be a partition, according to which $|G \rangle $ has maximal
Schmidt rank. Then, according to ref.~\cite{Schlinge02a}, choosing a subset $X\subseteq A$, the graph code, which encodes an input on vertices $X$ in output on vertices $Y=V\setminus X$ according to $G$, detects the error configuration $E=A\setminus X$, i.e., any errors occurring on only one half of the vertex set $E$ can be corrected.
In particular, all {\em strongly error correcting graph codes} in ref.~\cite{Schlinge02a} must have Schmidt measure $N/2$. The following proposition gives at least a sufficient condition \cite{He04,MarcPhD} when a partition has a maximal Schmidt rank with respect to the corresponding bi-partite split.

{\proposition[\bf Maximal Schmidt rank]\label{sufficient crit for max rank}
A {\em sufficient criterion} for a bi-partite split $(A,B)$ to have  maximal Schmidt rank is that the graph $G_{AB}$ corresponding to the edges between the partitions contains no cycles, and that the smaller partition contains at most one leaf with respect to the subgraph $G_{AB}$. If $G_{AB}$ is not connected, then it is sufficient that the above criterion holds for every connected component of $G_{AB}$.
}

Note that a {\em leaf}\index{leaf in a graph} is a vertex of degree 1, i.e., a vertex to which exactly one edge is incident \cite{Graph}.

\index{maximal Schmidt rank $\text{SR}_\text{max}$|)}

\index{persistency|(}
\index{Pauli persistency $\text{PP}$|(}
\index{Schmidt measure $\text{E}_S$!upper bound}

For the upper bound we consider a sequence of local projective measurements that finally completely disentangles the state vector $| \psi \rangle$ in each of the measurement results. Let $m$ denotes the number of measurement results with non-zero probability. Clearly, any of the states resulting from the different measurement outcomes is a product state and thus the whole measurement procedure gives rise to a decomposition of the initial state $| \psi \rangle$ as in eq.~(\ref{SchmidtM}). Thus we obtain the upper bound
\begin{equation}\label{Persistency}
     \text{E}_S(|\psi \rangle) \leq \log_2 (m)\; .
\end{equation}
In particular, for any sequence of measurements in the Pauli basis $\sigma_x$, $\sigma_y$ or $\sigma_z$ that yields an
empty graph, the number of local measurements in this sequence gives an upper bound on the Schmidt measure of the corresponding graph state.  This is because --apart from the trivial case of a $\sigma_x$-measurement at an isolated vertex--, both measurement results $\pm1$ of a local Pauli measurement are attained with probability $1/2$ and yield locally equivalent graph state vectors  $|G'\rangle$ and $|G''\rangle$. More generally we find
\begin{equation}\label{E under projective measurement}
     \text{E}_S(|G'\rangle ) \leq  \text{E}_S(|G\rangle ) \leq  \text{E}_S(|G'\rangle ) \,+\, 1\; .
\end{equation}

In the following we will call the minimal number of local Pauli measurements to disentangle a graph state its {\em Pauli persistency}  $\text{PP}(G)$. The notion of {\em persistency} was introduce in ref.~\cite{Briegel01} in the context of general projective measurements in order to study the stability of entanglement in cluster states with respect to local measurements. Since each $\sigma_z$ measurement simply deletes all edges incident to a vertex, any subset $V'\subseteq V$ of vertices in a graph $G$, to which any edge of $G$ is incident, allows for an disentangling sequence of local measurements. In graph theory those vertex subsets are called {\em vertex covers}\index{vertex cover}\index{minimal vertex cover $\text{VC}$} $\text{VC}(G)$.
Thus we have found the upper bounds
\be  \mathbf{ \text{E}_S}(|G\rangle)\,\leq\, \text{\bf PP}(G) \,\leq\, \text{\bf VC}(G) \; .\ee

\begin{wrapfigure}[16]{r}{0.5\textwidth}
\vspace{-0.8cm}\includegraphics[width=0.5\textwidth]{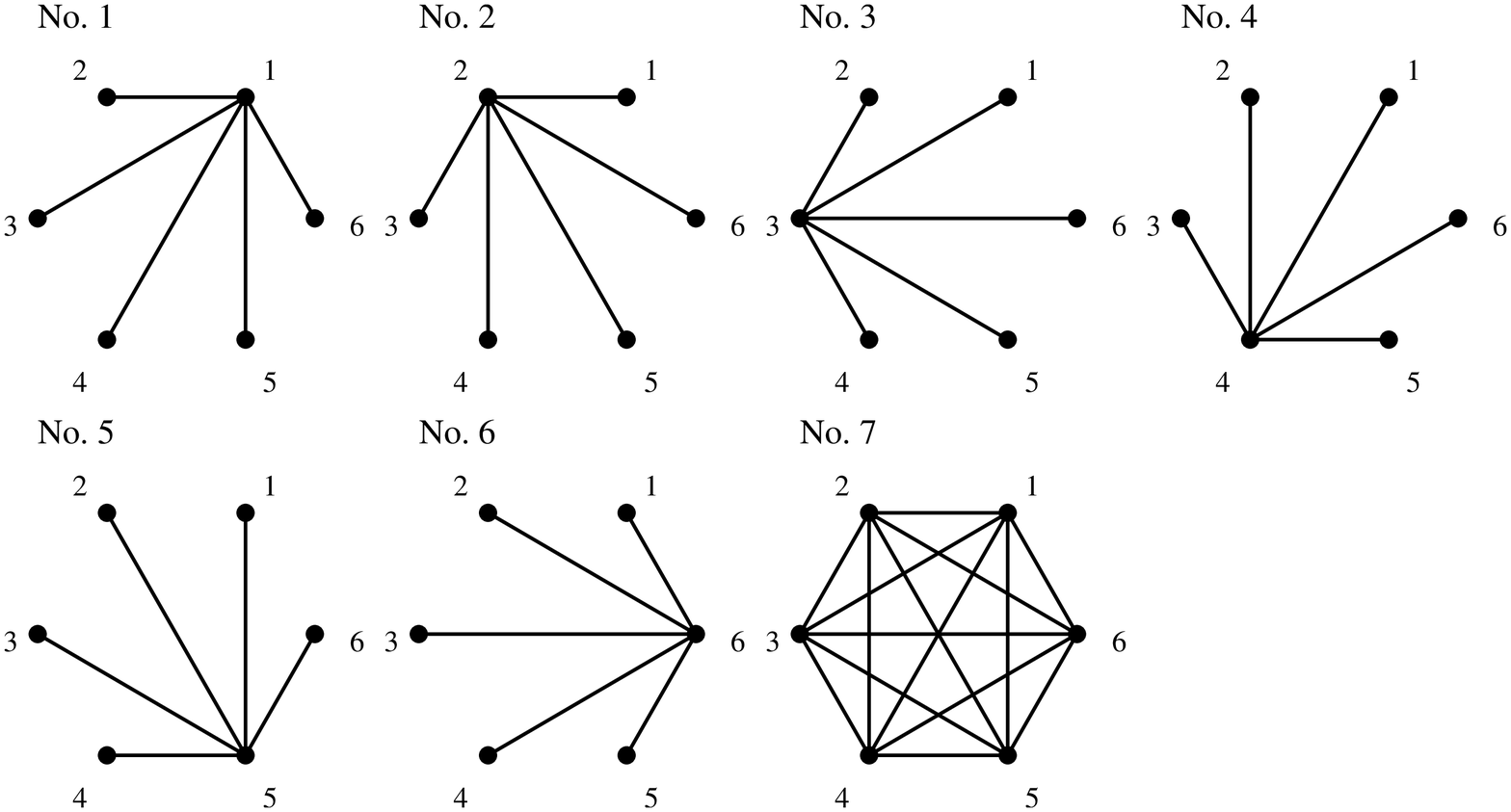}
\caption{\label{fig:LUclassExample3} A single $\sigma_y$-measurement at an arbitrary vertex in the complete graph No.
7 suffices to disentangle the corresponding state. Similarly, a single $\sigma_z$-measurement at the central vertex in the graphs No.\ 1--6 or a single $\sigma_x$-measurement at the non-central vertices is a disentangling measurement. This is due to the fact  that all graphs (No.\ 1--7) are locally equivalent by LC-unitaries, which transform the measurement basis correspondingly.}
\end{wrapfigure}
For graphs with many edges a combination of $\sigma_z$ and $\sigma_y$ will give better bounds than restricting to $\sigma_{z}$ measurements only. For example, according to the measurement rules in Proposition~\ref{Pauli_Measurement}, any complete graph (in which all vertices are adjacent) can be disentangled by just one $\sigma_y$-measurement at any vertex (see fig.~\ref{fig:LUclassExample3}). As we have seen, this corresponds to the fact that these graph states are LC-equivalent to the GHZ-type graph states (see also sec.~\ref{GHZ_GS}), in which every vertex is adjacent to the same central vertex.

\index{persistency|)}
\index{Pauli persistency $\text{PP}$|)}

Let us briefly summarize the relevant bounds for our further considerations in a proposition.

{\proposition[\bf Bounds to the Schmidt measure]
For any graph state $|G\rangle$ the Schmidt measure $\text{E}_A$ is bounded from below by the maximal Schmidt rank  $\text{SR}_\text{max}$ and from above by the Pauli persistency $\text{PP}$ or the minimal vertex cover $\text{VC}$, i.e.
\be\label{Bounds_SM} \text{\bf SR}_\text{\bf max}(G)\,\leq\, \text{\bf E}_S(|G\rangle)\,\leq\, \text{\bf PP}(G) \,\leq\, \text{\bf VC}(G) \; .\ee
}
An application of the LC-rule (see Proposition~\ref{loc}), of course, does not change the Schmidt measure. But also other local changes to the graph, such as the deletion of edges or vertices, have only a bounded effect on the Schmidt measure \cite{He04}:

{\proposition[\bf Edge-/Vertex rule]\label{vertexr}\index{Schmidt measure $\text{E}_S$!vertex rule}
\index{Schmidt measure $\text{E}_S$!edge rule}
\hspace{5cm}
\begin{itemize}
      \item By {\em deleting or adding edges} between two vertices of a graph $G$ the Schmidt measure of the resulting graph $G'$ can at most decrease or increase by $1$;
      \item If a {\em vertex is deleted}, the Schmidt measure of the resulting graph $G'$ decreases, but at most by $1$.
\end{itemize}
}

\begin{figure}[th]
\begin{center}
\includegraphics[width=0.44\textwidth]{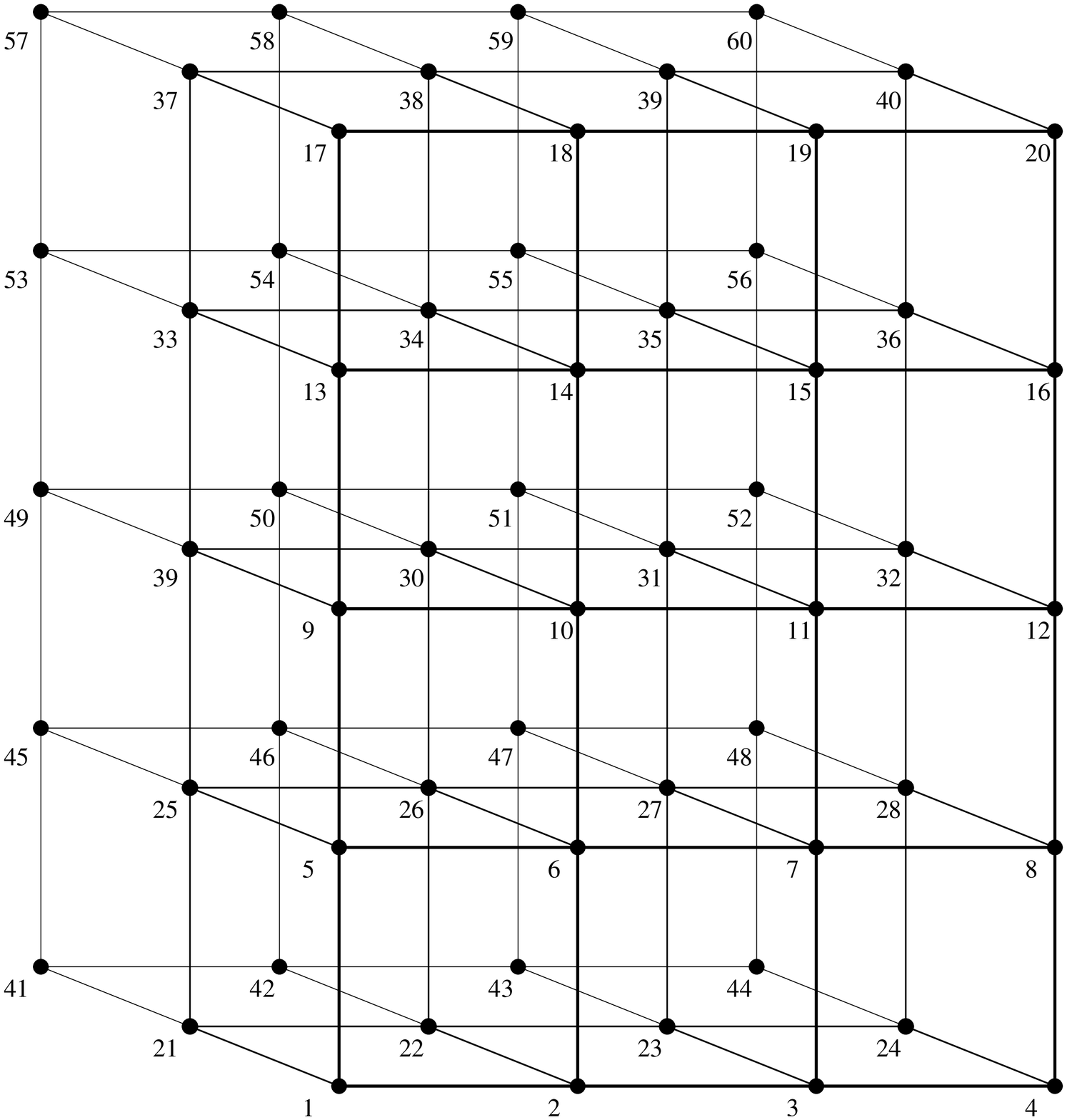}\hspace*{0.03\textwidth}\includegraphics[width=0.44\textwidth]{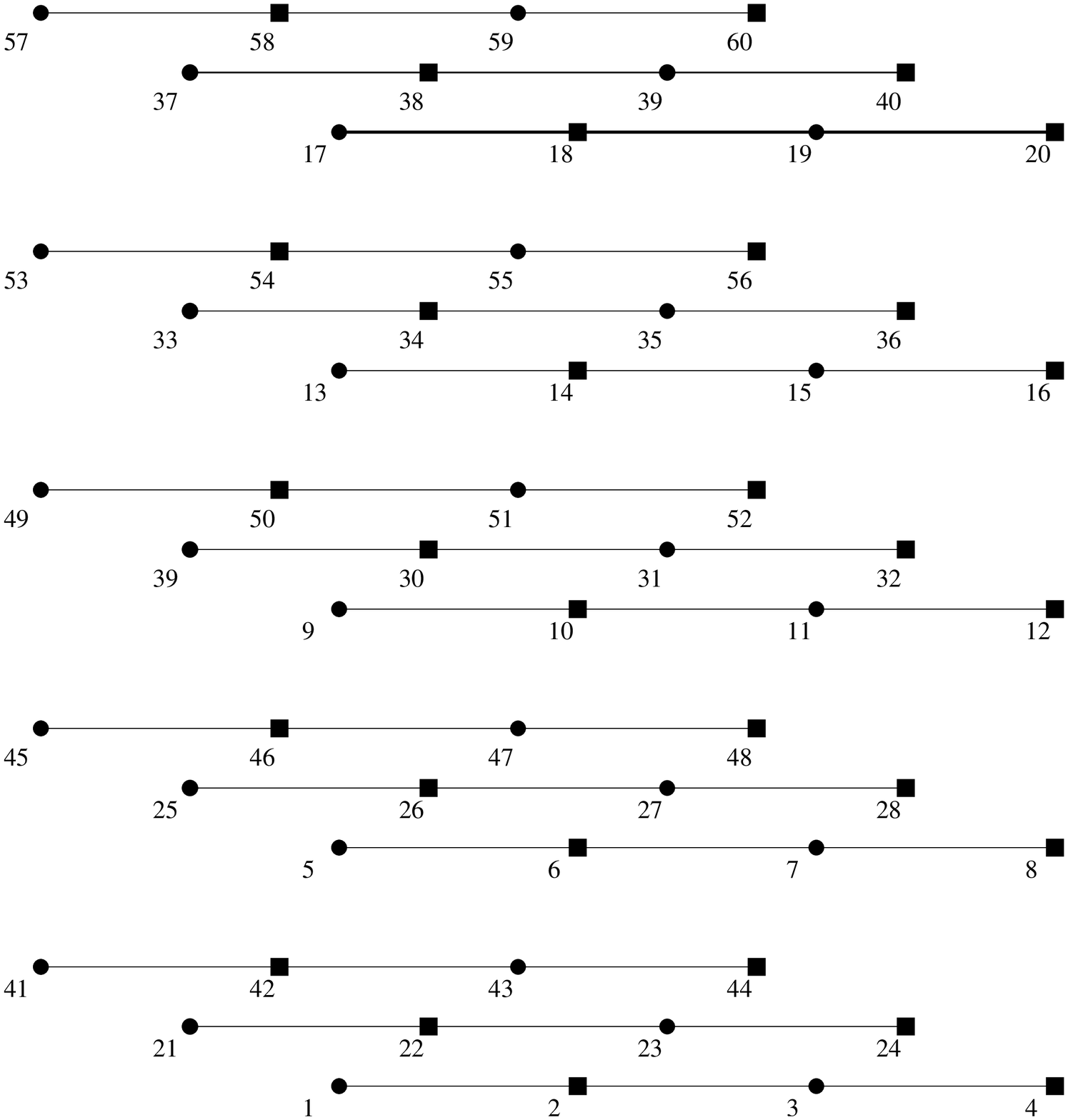}
\end{center}
\vspace{-0.5cm}
\caption{\label{fig:3DExampleEven}  An example for the $(4,5,3)$-cluster state and the graph corresponding to the adjacency matrix $\mathbf{\Gamma}^{AB}$ (see eq.~(\ref{Gamma for bi-partition})) for a bipartitioning $(A,B)$ with maximal Schmidt rank $\text{SR}_A  = \text{SR}_\text{max}$.  Here the vertices in $A$ are depicted by small boxes $\vrule height4pt width3pt depth0pt$ . }
\end{figure}

Let us now apply these findings to evaluate the Schmidt measure for some important classes of graph states:

{\proposition[\bf Examples]\label{E_SExamples}
\hspace{5cm}
\begin{itemize}
     \item The Schmidt measure for any multi--partite {\bf GHZ states} is $1$.
     \item The Schmidt measure of a {\bf 1D-, 2D-, 3D-cluster} state is $\lfloor \frac{N}{2} \rfloor$.
     \item The Schmidt measure of an entangled {\bf ring} with an even number $N$ of vertices is given by
      $N/2$.
     \item The Schmidt measure of a {\bf tree}  is the size of its minimal vertex cover $\text{VC}$.
\end{itemize}
}

\begin{figure}[th]
\begin{center}
\includegraphics[width=0.85\textwidth]{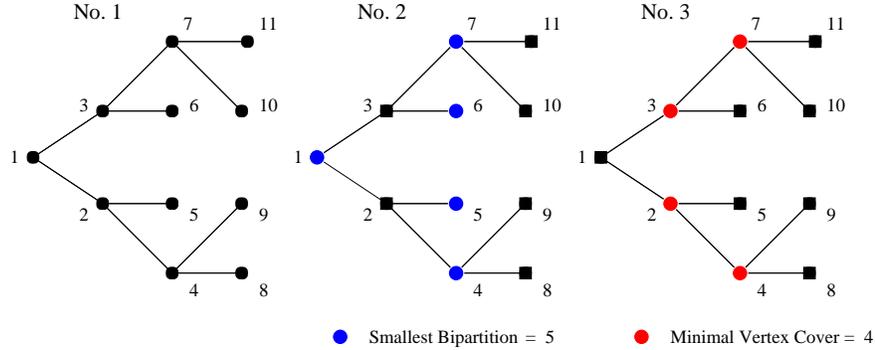}
\end{center}
\vspace{-0.5cm}
\caption{\label{fig:TreeExample1} The graph No.\ 1 represents a tree. Its bi-partitioning $(A,B)$, for which  in graph
No.\ 2 the vertices in $A$ are depicted by disks, neither is a minimal vertex cover nor yields maximal partial rank. Instead the set of vertices $A$, represented by large disc's in graph No.\ 3, is a minimal vertex cover with maximal partial rank.}
\end{figure}
\begin{wrapfigure}[12]{r}{0.5\textwidth}
\vspace{-0.7cm}
\includegraphics[width=0.45\textwidth]{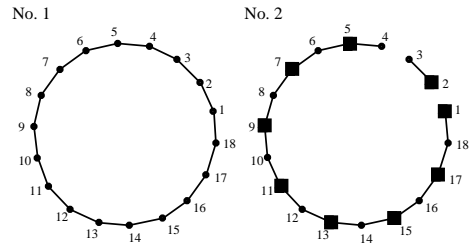}
\caption{\label{fig:EvenRingExample} Graph No.~1 is an entangled ring on $18$ vertices.
Graph No.~2 represents the graph corresponding to the adjacency matrix $\mathbf{\Gamma}^{AB}$ (see eq.~(\ref{Gamma for bi-partition})) for a bipartitioning $(A,B)$ with maximal Schmidt rank $\text{SR}_A  = \text{SR}_\text{max}$. The vertices of partition $A$ are depicted by boxes.}
\end{wrapfigure}

In all these cases, the upper and lower bounds, i.e., the maximal Schmidt rank and the Pauli persistency, coincide \footnote{Since the maximal Schmidt rank for any state can be at most $\lfloor  \frac{N}{2} \rfloor$, the Schmidt measure of those cases, where upper and lower bound coincide, is also bounded by this number. Thus all graph states which allow for a determination of the Schmidt measure along these lines have Schmidt measure of at most $\lfloor  \frac{N}{2} \rfloor$.} Although we refer to ref.~\cite{He04} for a detailed proof, for a even ring or a cluster state with at least one side of `even length', we have indicated the partition $A$ with maximal Schmidt rank $\text{SR}_A=\text{SR}_\text{max}$ (see Proposition~\ref{sufficient crit for max rank}) in fig.~\ref{fig:3DExampleEven} or fig.~\ref{fig:EvenRingExample} respectively.

Fig.~\ref{fig:TreeExample1} gives an example for a tree for which the Schmidt measure does not
coincide with the size of the smaller bi-partition, the upper bound according to
Proposition \ref{E_SExamples}.

\index{equivalence classes}
We have also computed the lower and upper bounds to the Schmidt measure, the Pauli persistency and the maximal partial rank, for the non equivalent graphs in List {\bf A} and {\bf B} in Table~\ref{TablePage}. They are listed in the corresponding tables in Sec~\ref{Local_Equivalence}.

For connected graphs the Schmidt rank $0$ cannot occur for any bi-partite splitting $(A,B)$, since this would correspond to an empty graph $G_{AB}$ between the partitions. Because the rank index is invariant under permutations of the partitions according to graph isomorphisms it provides information about whether two graph states are equivalent under local unitaries {\em plus} graph isomorphisms as treated in sec.~\ref{Local_Equivalence}. But note
that graph number $40$, $42$ and $44$ are examples for non-equivalent graphs with the same rank index.
Nevertheless, comparing the list of Schmidt ranks with respect to all bi-partitions in detail shows that no
permutation of the vertex set exists (especially none which is induced by a proper graph isomorphism on both sides), which would cause a permutation of the corresponding rank list, such that two of the graphs could be locally equivalent.

For $295$ of $995$ non-isomorphic graphs the lower and upper bound differs and that in these cases the Schmidt measure also  non-integer values in $\text{log}_2\{1,...,2^{N} \}$ are possible. Moreover note that only graph number $8$ and $19$ have maximal Schmidt rank with respect to all bi-partite splits. Entanglement here is distributed symmetrically between all parties, which makes it "difficult" to disentangle the state by few measurements. From this one can understand why the gap between the lower and upper bound occurs in such cases. As discussed above, from all graph codes with less than $7$ vertices only these two are candidates for strongly error detecting graph codes introduced in ref.~\cite{Schlinge02a}.

\begin{wrapfigure}[13]{r}{0.4\textwidth}
\vspace{-0.5cm}
\includegraphics[width=0.4\textwidth]{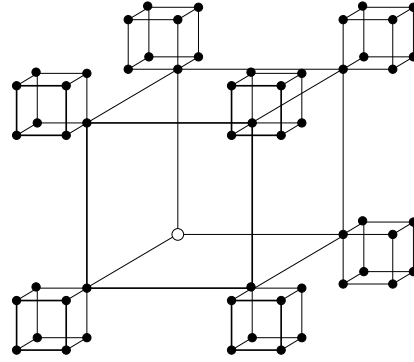}
\caption{\label{fig:bee} \footnotesize Resource graph state for the concatenated $[7,1,3]$-CSS code.}
\end{wrapfigure}
In the remainder  we will briefly discuss two examples which were already introduced in sec.~\ref{GS_Examples} in the context of quantum error correction and the one-way model for quantum computation, namely the graph that is used to realize the QFT on three qubits in the one-way quantum computer. The vision behind this is to flesh out the notion of entanglement as an algorithmic resource, as it has been put forward in ref.~\cite{OneWay1,OneWay2,OneWay3,OneWay5}.
\vspace{0.1cm}

{\em Example 1: Concatenated $[7,1,3]$-CSS-code.}\index{CSS code}\index{quantum error correcting code (QEC)}\index{stabilizer code}

As discussed in sec.~\ref{Application_QEC} the graph $G$ depicted in Fig.~\ref{fig:bee} represents an
encoding procedure for the concatenated $[7,1,3]$-CSS-code. The corresponding graph state has Schmidt measure $28$. For encoding, seven $\sigma_x$-measurements at all vertices of the inner square except $\circ$ have to be performed. In this resulting graph obtained without measuring the vertex $\circ$ represents the resource for the
alternative  encoding procedure. It has maximal Schmidt measure $25$, whereas the corresponding $0$ and $1$ code words have Schmidt measure $24$. They can be obtained with probability $1/2$ from $|G'\rangle$ by a $\sigma_z$-measurement at the vertex $\circ$.

{\em Example 2: Quantum Fourier Transform (QFT) on $3$ qubits.}\index{quantum Fourier transformation (QFT)}

\begin{wrapfigure}[21]{r}{0.4\textwidth}
\vspace{-0.8cm}
\begin{center}
\includegraphics[width=0.4\textwidth]{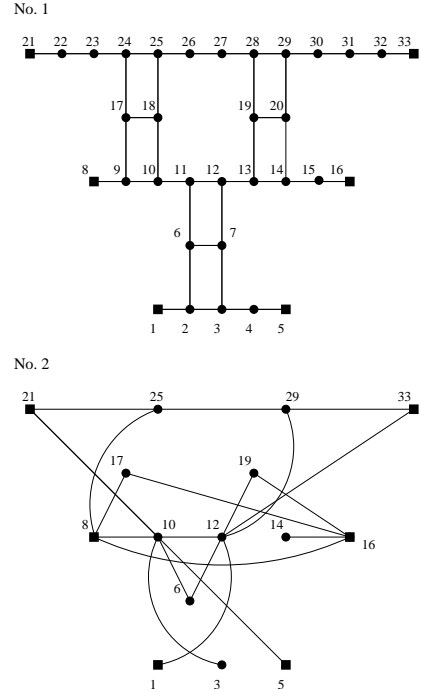}
\end{center}
\caption{\label{fig:3QFT}\footnotesize Graphs associated with the QFT on 3 qubits in the one-way quantum computer. The boxes denote the input (left) and output (right) vertices.
}
\end{wrapfigure}
The graph No.\ 1 in fig.~\ref{fig:3QFT} is a
simple example of an entangled graph state as it occurs in the
one-way computer of refs.~\cite{OneWay1,OneWay2,OneWay3,OneWay5}. This specific
example represents the initial resource (part of a cluster)
required for the quantum Fourier transform QFT on 3 qubits
\cite{OneWay1,OneWay2,OneWay3,OneWay5}. It has Schmidt measure $15$, where the partition
\begin{equation}
        A=\{2, 4, 7, 9, 11, 13, 15, 18, 20, 22, 24, 26, 28, 30, 32\}
        \nonumber
\end{equation}
is a minimal vertex cover with maximal Schmidt rank. In the process
of performing the QFT, all vertices except the output vertices
$5,16,33$ are measured locally. During this process, the
entanglement of the resource state (with respect to every
partitioning) can only decrease.  Similar as with the graph state vector
$|G'\rangle$ obtained from fig.~\ref{fig:bee}, graph No.\ 2
represents the input-independent resource needed for the essential
(non-Clifford) part of the QFT protocol \cite{OneWay1,OneWay2,OneWay3,OneWay5}. It has
Schmidt measure $5$, where the partition $A=\{2, 9, 10, 11, 15\}$
now provides a minimal vertex cover with maximal Schmidt rank.
\vspace{-0.0cm}
\index{Schmidt measure $\text{E}_S$!for graph states|)}


\section{Weighted graph states}\label{WeightedGS}\index{weighted graph state \texttt{"|}$G\rangle$|(}\index{weighted graph}

In this section we extend the concept of a graph describing some state to the class of weighted graph states \cite{He04,Du05a}. We return to the definition of graph states in terms of the underlying interaction pattern. In sec.~\ref{DefOfGS_Int} it was shown that any interaction pattern, for which the temporal order of the interactions between the particles is irrelevant and thus can be described by a graph, can only contain interactions $H_{ab}$ that are up to local $z$-rotations Ising-type interactions or likewise are given by the phase gate Hamiltonian
$H_{ab} = |1\rangle^a\langle1|\otimes |1\rangle^b\langle1|$. In order to describe the state by a simple graph we have fixed the interaction time for the particles to a phase $\pi$ in the previous sections.
We will now allow the particles to interact according to the same Hamiltonian $H_{ab}$ but for different interaction times $\varphi_{ab}$. This corresponds to the situation of a disordered system as it occurs e.g. in a spin glass or semi--quantal Boltzmann gas as described below. The interaction pattern is now summarized by a weighted graph, in which every edge is specified by a phase $\varphi_{ab}$ corresponding to the time the particles $a$ and $b$ have interacted.
 The {\em weighted graph state} $|G\rangle$ is thus given by
\be
|G\rangle = \prod_{\{a,b\} \in E} U_{ab} |+\rangle^{V} \label{wgraph}
\ee
where the operations $U_{ab}$ depend on the interaction phases $\varphi_{ab}$:
\be
U_{ab}:= e^{-i \varphi_{ab} H_{ab}} = e^{-i \frac{\varphi_{ab}}{4} \left(\mathbf{1}^{k}-\sigma_z^{k}\right)\otimes\left(\mathbf{1}^{l}-\sigma_z^{l}\right)}.
\ee\index{Ising interaction $H^I_{ab}$, $U^I_{ab}$}
The corresponding adjacency matrix $\mathbf{\Gamma}$\index{adjacency matrix $\mathbf{\Gamma}$} for this weighted graph collects the weights $\Gamma_{ab}=\varphi_{ab}=\varphi_{ba}$ and gives rise to a concise decomposition of the weighted graph states with respect to the standard basis $|W\rangle_z=|W_1\rangle^1\cdots|W_N\rangle^N$ ($W=(W_1,\ldots,W_N)\in \mathbb{F}_2^N$) \cite{CDHB05}:
\be\label{WGSdecomp} |G\rangle \,= \, 2^{-\frac{N}{2}}\, \sum_{W\subseteq V} \prod_{\{a,b\} \in E} U_{ab} |W\rangle_z \, = \, 2^{-\frac{N}{2}}\, \sum_{W\subseteq V} e^{i \frac{1}{2} W \cdot \mathbf{\Gamma} \cdot W} |W\rangle_z\; .\ee
This easily follows for each basis state $|W\rangle_z$ by induction over the involved interaction unitaries $U_{ab}$ using the fact that \be  U_{ab}|W_a\rangle^a|W_b\rangle^b=e^{i\varphi_{ab}W_aW_b}|W_a\rangle^a|W_b\rangle^b=e^{i\frac{1}{2}\left(W_a\varphi_{ab}W_b + W_b\varphi_{ba}W_a\right)}|W_a\rangle^a|W_b\rangle^b\; . \ee
We remark that for any weighted graph states $|G\rangle$, the set of states \be|W\rangle = \sigma_z^W |G\rangle
\ee form a basis for $({\mathbb{C}}^2)^{V}$. This can most easily be seen by realizing that $\sigma_z^W$ commutes with $U_{ab}$, and $\sigma_z^W|+\rangle^{V}$ gives rise to a orthonormal basis consisting of pure product states. The application of $\prod_{\{a,b\} \in E} U_{ab}$ transforms this orthonormal basis into a basis of weighted graph states, where each of the basis states has equivalent entanglement properties and the states are connected by local unitary operations.

In contrast to a straightforward extension of the interaction picture to the case of weighted graph states, no such generalization of the stabilizer formalism in terms of generators within the Pauli group is possible. This implies that many results obtained in the previous sections do no longer hold for weighted graph states. Nevertheless one is still able to show \cite{CDHB05} that entanglement in a weighted graph state is closely related to the connectivity properties of the underlying graph. More precisely, the weighted graph state is entangled (non-separable) with respect to the bi-partition $(A,B)$ iff there are non-vanishing interaction phases $\varphi_{ab}>0$ for some $a\in A$ and $b\in B$ between these partitions\index{connected graph (state)}. This is because the rank of the reduced density matrix as computed in Proposition~\ref{ReducedRhoWGS} in sec.~\ref{WeightedGS} is larger than one\footnote{More precisely, for any $\varphi_{ab}\neq 0$ the rows according to index $\emptyset$ and $a\equiv\{a\}$ differ: We find for the matrix elements $\frac{C_{\emptyset \emptyset}}{C_{a \emptyset}} \neq \frac{C_{\emptyset a}}{C_{a a}}$, since $C_{\emptyset \emptyset}=C_{aa}=1$ and $C_{\emptyset a}= C_{a \emptyset} = e^{i\sum_{b\in N_a}\varphi_{ab}/2}\prod_{b\in N_a}\cos[\frac{\varphi_{ab}}{2}] $.} in this case. Moreover, between arbitrary two parties $A$ and $B$ entanglement can be created (localized) by performing local operations on the remaining vertices (i.e., in  $V\setminus (A\cup B)$) iff the parties are connected by a path with non-vanishing interaction phases\index{localizable entanglement $\text{LE}^{ab}$}. The localizable entanglement is not zero in these cases, since the protocol in Proposition~\ref{locEntGS} of sec.~\ref{Corr} can also be applied to weighted graph states in order to reveal entanglement between some vertices $a\in A$ and $b\in B$ \cite{CDHB05}.

Moreover for the following analysis of entanglement present in weighted graph states it is crucial that the reduced density matrices of these states can still be determined efficiently:

{\proposition[{\bf Reduced state for weighted graph states}]\label{ReducedRhoWGS}\index{reduced state $\rho_G^A$}
Let $A\subseteq V$ be a subset of vertices for a weighted graph $G$ and $B=V\setminus A$ the corresponding complement in V. Then the reduced state $\rho_G^A =\text{tr}_B(|G\rangle\langle G|)$ is given by \be\label{rhoversustilderho} \rho_G^A = \prod_{\{a,b\} \in E_A} U_{ab} \tilde \rho_G^A U_{ab}^\dagger\; ,\ee  where $E_A=E\cap (A\times A)$ denotes the set of edges within $A$ and
\be\label{ReducedWGS} \tilde\rho_G^A = \sum_{A_1,A_2\subseteq A} C_{A_1A_2} |A_1\rangle_z^A\langle A_2| \;.\ee
The matrix elements are
\bea\label{CorrReducedWGS} C_{A_1A_2} &=&   \frac{1}{2^{N}}\sum_{B'\subseteq B} e^{i (A_1-A_2) \cdot \mathbf{\Gamma}'\cdot B'}  \nonumber \\
&=&  e^{i \frac{1}{2} \sum_{b \in B} (A_1-A_2)\cdot \Gamma_b'}\prod_{b \in B} \cos\left[\frac{1}{2}(A_1-A_2)\cdot \Gamma_b'\right] \; .\eea
Here $\Gamma_b'$ denotes the $b$-th column of the matrix $\mathbf{\Gamma}'=\mathbf{\Gamma}_{AB}$ representing the edges $\{a,b\}$ with weights $\varphi_{ab}$ between the partitions $A$ and $B$ (see eq.~(\ref{Gamma for bi-partition})).
}

{\em Proof:}
From elementary facts\footnote{I.e., $\text{tr}_B [C^A\otimes U^B \rho D^A\otimes (U^B)^\dagger ] = C \,\text{tr}_B(\rho)\, D$ for arbitrary unitaries $U$ and matrices $C$, $D$.} about the partial trace $\rho_G^A =\text{tr}_B(|G\rangle\langle G|)$  it follows that, for the computation, we might as well apply the interaction unitaries $U_{aa'}$ in eq.~(\ref{wgraph}) acting solely on vertices in $A$ after the partial trace and neglect all unitaries acting only on $B$. Thus $\tilde\rho_G^A $ in eq.~(\ref{rhoversustilderho}) is the reduced density matrix corresponding to the graph $G_{AB}$ with adjacency matrix $\mathbf{\Gamma}'=\mathbf{\Gamma}_{AB}$. In order to determine $\tilde\rho_G^A $ we split up the basis vectors $|W\rangle$ in  eq.~(\ref{WGSdecomp}), e.g. $|W_1\rangle^V=|A_1\rangle^A|B_1\rangle^B$, according to the partitioning $(A,B)$ and compute
\bea \tilde\rho_G^A & = & \frac{1}{2^N} \, \text{tr}_B\,\left[ \sum_{W_1,W_2 \subseteq A} e^{i \frac{1}{2}\left( W_1 \cdot \mathbf{\Gamma}' \cdot W_1 - W_2 \cdot \mathbf{\Gamma}' \cdot W_2\right)}   |W_1\rangle_z\langle W_2|\right]\nonumber \\ &=&  \sum_{A_1,A_2\subseteq A} \left(\frac{1}{2^{N}}\sum_{B'\subseteq B} e^{i (A_1-A_2) \cdot \mathbf{\Gamma}'\cdot B'}\right) |A_1\rangle_z^A\langle A_2| \; .\eea
The matrix element $C_{A_1A_2}$ can be further simplified as stated in eq.~(\ref{CorrReducedWGS}):
\bea C_{A_1A_2} & = &\frac{1}{2^{N}}\sum_{B'\subseteq B} \prod_{b \in B'} e^{i (A_1-A_2) \cdot \mathbf{\Gamma}'_b} = \frac{1}{2^{N}}\prod_{b \in B} \left( 1 + e^{i (A_1-A_2) \cdot \mathbf{\Gamma}'_b}\right) \\
& = & e^{i \frac{1}{2} \sum_{b \in B} (A_1-A_2)\cdot \Gamma_b'}\prod_{b \in B} \cos\left[\frac{1}{2}(A_1-A_2)\cdot \Gamma_b'\right]
\eea
using $\frac{1}{2}(1+e^{i\phi})=e^{i\frac{\phi}{2}}\cos(\frac{\phi}{2})$.
\proofend

Since the entanglement properties between the partitions $(A,B)$ are invariant under LOCC, we can disregard the unitaries remaining on $A$ and thus directly examine $\tilde\rho_G^A$.  From eq.~(\ref{CorrReducedWGS}) one sees that the total effect of the interactions with particles in $B$ on the matrix elements (coherences) $C_{A_1A_2}$ consists in multiplying the effects for each individual particle $b\in B$. Thus $\tilde\rho_G^A$ can alternatively be obtained as the Hadamard product\footnote{I.e., componentwise matrix multiplication of matrices written in the standard basis.}\index{Hadamard product} of the reduced states $\tilde\rho_{G_b}^A$ due to sole effect of particle $b\in B$, i.e., according to the induced graph $G_b:=G[A\cup b]$ on the vertex $b$ and all vertices within $A$.
Note that the computational effort to calculate the reduced density matrix scales exponentially in $|A|$ but only linearly in the number $|B|$ of particles in the remaining system, while for general pure states an exponential increase (with $N$) in computation time and memory cost is required for both parties. Hence all quantities that depend on the reduced density operator can be determined efficiently. For instance, from $\rho_G^A$ of one and two qubits, we can calculate all two-point (and also higher order) correlation functions $\text{Q}_{ij}^{ab}$,
\be
\text{Q}_{ij}^{ab}=\langle\sigma_i^{a}\sigma_j^{b}\rangle- \langle\sigma_i^{a}\rangle\langle\sigma_j^{b}\rangle \hspace{1cm} i,j=1,2,3,
\ee
the entanglement of formation\footnote{For a two-qubit mixed state $\rho^{ab}$ the entanglement of formation $\text{E}_F(\rho)=f(C(\rho))$ is related to the concurrence $C(\rho):=\max\{0,\lambda_1-\lambda_2-\lambda_3-\lambda_4\}$, where $\lambda_i$ are the eigenvalues of the Hermitian matrix $(\sqrt{\rho}(\sigma_y^a\sigma_y^b\rho^*\sigma_y^a\sigma_y^b)\sqrt{\rho})^{1/2}$
in decreasing order, by some monotonically increasing function $f$ \cite{Wooters98}.} between pairs of particles, as well as lower and upper bounds on the localizable entanglement $\text{LE}^{ab}$ \cite{Frank} that was already discussed in sec.~\ref{Corr} for simple graph states. Note that the maximal classical correlation $\text{Q}_{\max}^{ab}$\index{maximal classical correlation $\text{Q}_\text{max}^{ab}$}\index{correlation function $\text{Q}^{ab}_{ij}$} between two particles is given by the largest singular value of the matrix $\text{Q}_{ij}^{ab}$ \cite{Frank}. Recall that the localizable entanglement $\text{LE}^{ab}$ is the maximum amount of entanglement that can be established between a pair of particles $a,b$, on average, by performing local measurements on all other particles. Moreover the relation $\text{Q}_\text{max}^{ab}\leq \text{LE}^{ab}\leq \text{AE}^{ab}$ holds \cite{Frank}, where $\text{AE}^{ab}$ is the concurrence of assistance\index{entanglement of assistance $\text{AE}^{ab}$} \cite{Laustsen03}.

With an efficient calculation of the reduced state\index{reduced state $\rho_G^A$} we can also determine the {\em entropy}\footnote{Note that for pure bi-partite states the entanglement of formation and distillation both are given by the entropy of entanglement $S_A$. } {\em of bi-partite entanglement}\index{entropy of entanglement ${S}_{A}$}\index{entanglement measure!entropy ${S}_{A}$}  \be {S}_{A}(|G\rangle) := S(\rho^A_G)\equiv - \text{tr} [\rho^A_G\log_2(\rho^A_G)] \ee between a small number of vertices $A$ and the rest, as well as the multi-partite entanglement measure $E_{\rm MW}$\index{entanglement measure!Meyer Wallach $\text{E}_{\rm MW}$} proposed in ref.~\cite{Meyer,Br03}. This measure $\text{E}_{\rm MW}$ is given by $\text{E}_{\rm MW} = 2[1-\frac{1}{N}\sum_{a\in V} \text{tr}({\rho^a_G}^2)]$ \cite{Meyer,Br03}.

\index{Valence Bond Solids (VBS)|)}

This method for calculating reduced density matrices can be readily extended to the case, in which the particles are initially prepared in an arbitrary product state $\bigotimes_{a\in V} \left(\alpha_a|0\rangle + \beta_a|1\rangle\right)$ instead of $|+\rangle^V$ or even to slightly entangled\footnote{I.e., pure states with small Schmidt rank with respect to the respective bi-partition.} initial states. Similarly the VBS--picture can be generalized to all states produced by the interaction Hamiltonian $H_{ab}$ acting on arbitrary product input states $|\phi_1\rangle^1\cdots |\phi_N\rangle^N$. However, further modifications of the involved states are required. The (unnormalized) VBS--like state is of the form $|\tilde \Psi\rangle=\bigotimes_{a,b}|\chi_{a^ib^j}\rangle$ with $|\chi_{a^ib^j}\rangle = U_{a^ib^j}|\sqrt[d_a]{\phi_a}\rangle^{a^i}|\sqrt[d_b]{\phi_b}\rangle^{b^j}$, where $|\phi\rangle =\alpha|0\rangle+\beta|1\rangle$ and$|\sqrt[n]{\phi}\rangle:=\sqrt[n]{\alpha}|0\rangle + \sqrt[n]\beta|1\rangle$. But, for the sake of simplicity, in the remainder of this article we restrict to weighted graph states, i.e., states arising from input states $|+\rangle^{V}$.

Since the class of weighted states comprises particles interacting for different interaction times, it provides an interesting model for the study of the entanglement dynamics in many--particle systems. In the remainder we will briefly review a few results about the static and dynamic entanglement properties of spin lattices \footnote{The investigation of entanglement properties of strongly interacting many body systems has proven to be a fruitful approach.
Clearly, the ground
states of interacting many-body systems at zero temperature are
correlated. These correlations are not only reflected by
scaling laws for two point correlation functions:
In fact, it turns out that characteristic scaling
laws concerning ground state entanglement can be established
reminding of the behavior of such
two point correlation functions
\cite{Nielsen,Osterloh,HC,Latorre,Frank,Vi03,Pl04}.
This observation refers on the one hand to
entanglement properties
of two distinguished constituents of a many-body system.
On the other hand, it holds for
{\it block entanglement} of a number $L$ of
consecutive constituents and the rest of an
infinite chain \cite{Latorre,HC}. Notably, the specifics of
the scaling of entanglement were shown to indicate
quantum phase transitions
\cite{Nielsen,Osterloh,Latorre,Frank,Vi03}.
Long--range correlations can even be found in systems with gapped Hamiltonians in the sense of a divergent characteristic length
of the {\it localizable entanglement}
\cite{Verstraete04b,Ve03b}.}. For studies on spin
gases, see refs.~\cite{Du05a,CDHB05}; these results are based on the efficient method to calculate the reduced density operators of a small number
$L\leq 10$ of arbitrary spins as described above.

In harmonic systems \cite{HC,Pl04}, it may be remarked,
also higher-dimensional
systems and systems of non-integer dimension
can be studied, leading in particular to
`area-theorems', i.e., statements on the relationship between
the degree of entanglement of a distinguished region of
the full lattice system and its boundary area. These statements that gapped harmonic
systems indeed imply the validity of an area theorem hold
true on harmonic systems even
defined on general graphs \cite{Gap}.

Let us start our considerations with the case where the graph has some lattice structure\index{lattice graph (state)}. Thus we consider $N$ spin 1/2 systems (qubits) with pairwise interactions, described by an Ising--type Hamiltonian (see eq.~(\ref{PhaseGate}))
\be\label{InteractionHamiltonian}
H=\sum_{a < b} f(a,b) \frac{1}{4}(\mathbf{1}-\sigma_z^a)\otimes(\mathbf{1}-\sigma_z^b).
\ee
We assume that the spins are arranged on a $d$--dimensional lattice with fixed geometry and are initially completely polarized in $x$--direction, i.e., $|\Psi_0\rangle = |+\rangle^{V}$. As indicated in sec.~\ref{WeightedGS} the methods we develop can also describe disordered systems with random coefficients $f(a,b)$ and can take arbitrary (product) input states into account. We are interested in (entanglement) properties of the state
\be\label{UnitaryEvolution}
|\Psi_t\rangle := e^{-itH}|\Psi_0\rangle.
\ee
We consider the situation where the coupling between spins obeys a certain distance law, in the sense that the coefficients $f(a,b)$, describing the strength of the coupling, only depend on the distance $r_{ab}:= \|a-b\|$ between particles $a$ and $b$, i.e., $f(a,b)=f(r_{ab})$. In the example of ions stored in microtraps \cite{Ci00b,Ja02} one finds for instance $f(r_{ab})=r_{ab}^{-3}$ \cite{Ja02}.

In the following the (bi-partite) entanglement between blocks of a small number $L\leq 10$ of neighboring spins as measured by the entropy of entanglement will be denoted by $S_L=S(\rho_L)$. Clearly, $0 \leq S_L \leq L$, where $S_L=L$ indicates maximal entanglement between the blocks\index{entropy of entanglement ${S}_{A}$}.
For different distance laws, we have investigated the scaling of block-wise entanglement $S_L$ in ref.~\cite{Du05a} and observed the following:
\begin{itemize}
\item The maximal two point correlations in a spin chain decay slower than exponential for all power laws $f(r_{ab})\propto r_{ab}^{-\alpha}$. Therefore, the {\em correlation length}\index{correlation length} $\xi$ and also the {\em entanglement length}\index{entanglement length} $\xi_E$ {\em diverge} \cite{Frank}. This indicates long--range quantum correlations for all power laws, as we find that only exponential fall--off functions $f(a,b)=e^{-\kappa r_{ab}}$ lead to a finite correlation length.
\item  In the limit $N\to \infty$ and $L \to \infty$, entropy $S_L$ saturates as a function of $L$ for power laws $f(r_{ab}) = r_{ab}^{-\alpha}$ with $\alpha>1$. This result generalizes to $d$--dimensional lattices. When considering blocks of $L$ particles contained in a $d$--dimensional ball, $S_L$ can at most grow like the volume of that ball, whereas we find that for $\alpha > d$ the upper bound on $S_L$ grows at most like the surface of the ball.
\item With the results of the previous section about the Schmidt rank one can also examine the special case of simple graph states analytically. Here, the interaction Hamiltonian has a fixed interaction length $\lambda$ and constant interaction strength $f(r_{ab}) = 1$ if $r_{ab} \leq \lambda$ and zero otherwise. One finds for the resulting states $|\Psi_{\pi}\rangle$, i.e., for $t=\pi$, that the entropy $S_L$ equals $L$ if the radius of the hypersphere is smaller than $\lambda$. Otherwise, $S_L$ scales essentially like the volume of a surface shell with thickness $\lambda$  that is $S_L \propto \lambda L^{(d-1)/d}$.
\item One can also consider the {\em dynamics} of entanglement that means the change of entanglement and correlations of the state $|\Psi_t\rangle$ with time. The scaling of the entropy with the block size $L$ is essentially still governed by the specific form of the distance dependence for any finite $t$, because infinitely remote regions still influence a subsystem in a similar way as discussed before. For large times $t$, more and more of the interaction phases $\varphi_{ab}=f(a,b)t$ start to oscillate (as they are effectively taken modulo $\pi$) and approach in the limit of large $t$ a (quasi)--random distribution. In the limit of an infinite chain and $t \to \infty$, the entropy of the reduced density operator of any finite group $A$ of particles is maximal, $S(\rho_A)=|A|$. This can be seen by considering the off diagonal elements of reduced density operators, which all contain infinite products of cosines of (sums of) {\em random} angles. All these products tend to zero for $N\to \infty$, leading to a maximally mixed state.

\end{itemize}

\begin{wrapfigure}[9]{r}{0.45\textwidth}
\vspace{-0.3cm}
\includegraphics[width=0.45\textwidth]{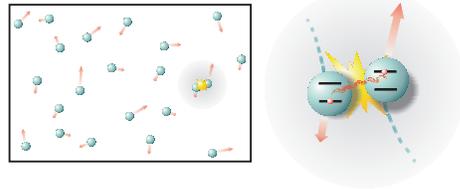}
\caption{\label{Fig:BoltzmannGas} Schematical drawing of a semi-quantal Boltzmann gas \cite{CDHB05}. }
\end{wrapfigure}
Finally we will slightly change the physical setup and consider a {\em spin gas} that is a system of interacting spins with coupling strengths that are now {\em stochastic} functions in time \cite{CDHB05}. One example is given by a {\em semi-quantal Boltzmann gas} of particles where each particle follows a classical trajectory but carries a quantum degree of freedom that is affected whenever two particles collide. During a collision of two particles, the internal degrees of freedom interact and can become entangled. This model was introduced in ref.~\cite{He04} and closely analyzed in ref.~\cite{CDHB05}.  There an ideal gas was considered in thermal equilibrium with elastic collisions, whose mean free path is comparable to the size of the enclosing volume. For a fixed interaction pattern of the particles the time dependence of the weighted graph state $|\Psi_t\rangle$, representing the internal degrees of freedom, is given by the adjacency matrix $\mathbf{\Gamma}(t)$ as a function of time. From this the corresponding statistical state can be obtained through an averaging process by assigning a probability to every collision history and assuming that the collision particles acquire an interaction phase $\varphi_{ab}$ inversely proportional to their relative velocities. Thus the weighted graph state, corresponding to the resulting statistical szenario, is specified by the density, the volume and the temperature of the underlying Boltzmann gas as well as by the initial spatial distribution of the particles, which was chosen to be homogeneous. Whereas for low temperatures $T$ the initial entanglement generation rate is proportional to $\sqrt{T}$, in the case of high temperatures the entanglement generation is governed by the slow collision events and proportional to $\sqrt{T^{-1}}$ \cite{CDHB05}. Moreover in the long time limit and for sufficiently many particles the state $|\Psi_\infty \rangle$ (equilibrium state) is maximally entangled with respect to all bi-partitions $(A,B)$, i.e., $S_A\simeq N_A$.

In \cite{CDHB05} another model is also discussed, where the gas particles hop randomly between different sites of a lattice. This {\em semi-quantal lattice gas}\index{lattice gas} allows the (numerical) study of highly correlated classical trajectories. This framework provides microscopic models to compare Markovian and non-Markovian as well as correlated and uncorrelated decoherence\index{decoherence} processes. Non-Markovian decoherence\footnote{Here the decoherence of the state for a set of particles is due to collisions with the remaining particles, after tracing out the environmental degrees of freedom.} of a set of particles is dominated by (repeated) collisions with the same set of environmental particles (high density), whereas Markovian noise is mainly driven by independent collisions with different particles (dilute gas). Similarly, correlated decoherence processes can be modeled by collisions of some particles with the same particle from the environment.

We also remark that the fact that for weighted graph states reduced density operators of (few) particles can be calculated efficiently can be used to develop a novel method for ground state approximation of strongly correlated quantum systems \cite{APDB05}. In particular, (superpositions) of locally transformed weighted graph states can be used to approximate the ground state of a strongly correlated system in any spatial geometry or dimension. As expectation values of local observables, including energy, can in this case be efficiently calculated, a (numerical) variation over weighted graph states can be performed and good approximation to the ground state may be obtained. The fact that weighted graph states exhibit rich entanglement features, including infinite correlation and entanglement length, maximal localizable entanglement and maximal block--wise entanglement, may even allow for a successful treatment of critical systems or systems in dimension $d\geq 2$.


\section{Graph states in the presence of decoherence}\label{GS_decoherence}
\index{graph diagonal state|(}

Let us now return to the case of simple graphs. Real implementations for graph states are subjected to decoherence\index{decoherence}\index{noisy graph states}. Thus the corresponding real state $\rho$ is in general some mixed state that is --depending on the quality of the preparation procedure-- more or less close\footnote{E.g. in terms of the fidelity $\langle G|\rho|G\rangle$.} to the ideal pure state $|G\rangle$.
First, we study the stability of entanglement in graph states when they are exposed to noise and discuss the lifetime of entanglement with respect to the number of particles in the system.
Finally, a method to overcome the effect of decoherence is introduced, namely {\it entanglement purification}, which was already mentioned in sec.~\ref{EPP_Blind}, where we also indicate possible applications to fault tolerant quantum computation (see also sec.~\ref{one-way-QC}) and to secret sharing. In this section we will focus on one of these applications and show how the purification protocols can be modified such that the distribution of the (two-colorable) graph states can remain unknown to the different parties ({\it blind purification}).

\index{graph state preparation}
Let us start with a few remarks underlying noise models that we consider in the following.
An arbitrary noise process acting on an $N$-qubit system that ideally is prepared in the pure graph state $\rho_G=|G\rangle\langle G|$, is frequently represented as a completely positive map $\mathcal{D}$ (CPM) \index{completely positive maps (CPM)} that can be decomposed in terms of Pauli matrices acting from left and right:
\be\label{ArbitrayDecohNqubits} \rho\; =\;\mathcal{D}(\rho_G) \;:=\; \sum_{\genfrac{}{}{0pt}{}{k_i,l_i =0}{(i=1,\ldots,N)}}^3\,\mathbf{E}_{\genfrac{}{}{0pt}{}{k_1,\ldots, k_N }{l_1,\ldots,l_N}}\, \sigma^1_{k_1} \cdots\sigma_{k_N}^N  \,\rho_G\, \sigma^1_{l_1} \cdots\sigma_{l_N}^N \;.\ee
Under certain conditions on the noise process \cite{GKS}, the completely positive maps $\mathcal{D}=\mathcal{D}_t$ are the solutions to a master equation describing the dynamics of the noise process in time. But, in the following, we keep our considerations on the level of CPMs, whose Kraus coefficient matrix may be time dependent, i.e.,  $\mathbf{E}_{\genfrac{}{}{0pt}{}{k_1,\ldots, k_N }{l_1,\ldots,l_N}}=\mathbf{E}_{\genfrac{}{}{0pt}{}{k_1,\ldots, k_N }{l_1,\ldots,l_N}}(t)$, if some dynamical description\footnote{For details we refer e.g. to ref.~\cite{Du05b}.} is imposed.

In \cite{Du05b} it is shown that by randomly choosing some Pauli matrix for each individual particle and applying them {\em before} and {\em after} the actual decoherence process occurs, an arbitrary noise process can be `depolarized' such that the overall noise process is described by a tensor $\mathbf{E}_{\genfrac{}{}{0pt}{}{k_1,\ldots, k_N}{l_1,\ldots, l_N}}$ that is diagonal, i.e.
\be\label{PauliDecohNqubits} \rho'\; =\;\mathcal{D}'(\rho_G) \;=\; \sum_{\genfrac{}{}{0pt}{}{k_i=0 }{(i=1,\ldots,N)}}^3\,\mathbf{E}_{k_1,\ldots, k_N }\, \sigma^1_{k_1} \cdots\sigma_{k_N}^N  \,\rho_G\, \sigma^1_{k_1} \cdots\sigma_{k_N}^N \;. \ee
We refer to such channels $\mathcal{D}'$ that are diagonal when decomposed with respect to the Pauli matrices, as ($N$-party) {\em Pauli channels}\index{Pauli channel}. For graph states we can make use of the following relations\footnote{These relation directly follow from the stabilizing properties of the correlation operators $K_a$ corresponding to the graph state $|G\rangle$.}
\be \sigma_x^a |G\rangle \;=\; \sigma_z^{N_a} |G\rangle \hspace{2cm} \sigma_y^a |G\rangle \;=\; \sigma_z^{N_a+a} |G\rangle\; \ee  in order to rewrite all $\sigma_x$- and  $\sigma_y$-operators in the decomposition eq.~(\ref{PauliDecohNqubits}) in terms of $\sigma_z$-operators. In this way one verifies that the resulting state $\rho'$ in eq.~(\ref{PauliDecohNqubits}) is diagonal in the graph state basis $|U\rangle_G =\sigma_z^U |G\rangle$\index{graph state basis \texttt{"|}$U\rangle_G$} (see Proposition~\ref{Graph state basis}). In the following we call such states {\em graph diagonal states}. In order to achieve the simplified standard form for the map describing the decoherence process, i.e., for arbitrary input states, one has to perform some operation {\em before} the actual noise process has affected the state. Considering noisy preparation procedures for graph states, a `before' does not make much sense and thus the resulting state $\mathcal{D}(\rho_G)$ can hardly be regarded as some standard form for an imperfectly prepared graph state $\rho$. But it was shown in ref.~\cite{Du03a} that the resulting standard form can nevertheless be obtained by performing another twirling operation {\em after} the (noisy) preparation procedure that gives rise to the same depolarization for the corresponding mixed state $\rho$ \cite{Du03a,MarcPhD}:
{\proposition[{\bf Graph-diagonal states as standard forms for mixed states}]\label{GraphTwirling}\index{twirling!in graph state basis} Any mixed state $\rho$ of $N$ qubits can be depolarized into some {\em graph diagonal state} \be \label{RhoG}
\rho' \;=  \; \sum_{U\subseteq V} \, \lambda_{U}\,  |U\rangle_G \langle U  | \ee for some graph $G$ with $N$ vertices by uniformly choosing the $2^N$ stabilizer elements $\sigma\in \mathcal{S}$ and applying them to the state, i.e., the corresponding twirling protocol is
\be \rho' \;=\; \frac{1}{2^N}\, \sum_{\sigma \in \mathcal{S}}\, \sigma \rho \, \sigma \; .\ee
}

For the rest of this section, we restrict to decoherence that arises if all qubits {\em individually} (or independently) are affected by noise described by the same Pauli channel
\begin{equation}\label{Dec_Pauli}
 {\cal D} (\rho) =  \sum_{i=0}^3 p_i\sigma_i \rho \sigma_i \hspace{0.5cm} \text{with} \hspace{0.5cm} (\sum_{i=0}^3 p_i =1)\, .
\end{equation}
These decoherence models are of particular interest in quantum information theory, especially in the study of fault-tolerant quantum computation, and contain for example:
\begin{itemize}
\item[1.] for \; $p_0=\frac{1+3p}{4}$\; and\; $p_1=p_2=p_3=\frac{1-p}{4}$\; the {\em depolarizing channel}\index{depolarizing channel}
\be \mathcal{D}(\rho) \;=\; p \, \rho + (1-p)\,\frac{1}{2}\mathbf{1}  \; ;\ee
\item[2.] for \; $p_0=\frac{1+p}{2}$,\, $p_1=p_2=0$\; and \;$p_3= \frac{1-p}{2}$\;the {\em dephasing channel}\index{dephasing channel};
\be \mathcal{D}(\rho) \;=\; p \, \rho + \frac{1-p}{2}\,(\rho + \sigma_z \rho \sigma_z)  \; ;\ee
\item[3.] for \;$p_0=\frac{1+p}{2}$, $p_2=p_3=0$\; and \;$p_1=\frac{1-p}{2} $\; the {\em bitflip channel}\index{bitflip channel}
\be \mathcal{D}(\rho) \;=\; p \, \rho + \frac{1-p}{2}\,(\rho + \sigma_x \rho \sigma_x)  \; .\ee
\end{itemize}

The corresponding coefficients $\lambda_U$ are given by the following proposition \cite{Du04b}.
{\proposition[{\bf Effect of individual Pauli channels on a graph state}]\label{IndivPauliChGS}
Under decoherence described by the same individual Pauli channel $\mathcal{D}$, the graph state $|G\rangle$ transforms into a mixed state $\rho = \prod_{a \in V} {\cal D}^{a}(\rho_G)\; $ that is diagonal in the graph state basis $|U\rangle_G $ .
The diagonal elements $\lambda_U$ in eq.~(\ref{RhoG}) can be computed to be of the form\footnote{In both expressions we have made use of the notational simplifications described in sec.~\ref{GS_Notations}. For example $\mathbf{\Gamma} U'$ denotes both the set and the binary vector that is obtained by the multiplication (modulo $2$) of the adjacency matrix $\mathbf{\Gamma}$ with the binary vector corresponding to the set $U'$.
}
\be
\label{PauliLambda}
\lambda_U \; = \; p_0^{|V|} \,  \sum_{U'\subseteq V} \, q_1^{|U'\setminus (\mathbf{\Gamma} U' + U)|}\, q_2^{|U'\cap (\mathbf{\Gamma} U' + U)|} \,q_3^{| (\mathbf{\Gamma} U' +U)\setminus U'|}\; ,
\ee
where $q_i:=\frac{p_i}{p_0}$ for $i=1,2,3$.
In the case of the depolarizing channel $(q:=q_1=q_2=q_3=\frac{1-p}{3p+1})$ this simplifies to
\begin{equation}
\label{DepolLambda}
\lambda_U = p_0^{|V|} \,  \sum_{U'\subseteq V} \, q^{|U'\,\cup\, (\mathbf{\Gamma} U' +U)|}\; .
\end{equation}
}

In the next subsection, we will analyze the time dependence of the entanglement properties of the decohered state $\rho(t)$ for different initial graph states $\rho(0)=|G\rangle\langle G |$. Although the results of the following subsections can be generalized to more general decoherence models, we will restrict our review to the special case of noise that effects each particle in the graph state individually. In other words the evolution at each qubit is described by the map ${\cal D}^a$ given by eq.~(\ref{Dec_Pauli}) with Pauli operators $\sigma_j$ acting on qubit $a$. We will be interested in the evolution of a given pure state $|\Psi\rangle$ of $N$ qubits under this decoherence model. That is, the initial state $|\Psi\rangle$ suffers from decoherence and evolves in time to a mixed state $\rho(t)$ given by
\be\label{decoh}
\rho(t) = {\cal D}^1 {\cal D}^2 \ldots {\cal D}^N |G\rangle\langle G| \; .
\ee
The depolarizing channel with $p(t)=e^{-\kappa t}$ is of particular interest, since the decohered state due to an arbitrary noise channel can be further depolarized to some state, which might
also be obtained directly by some depolarizing channel. Moreover, among the stated noise models the depolarizing channel is the only channel that is basis independent, i.e., invariant under unitary transformations.
We will frequently use the Pauli channel and will describe the entanglement properties of $\rho(t)$ in terms of the parameters $p_i$. Nevertheless one has to keep in mind that the time dependence itself is already included in the parameters $p_i=p_i(t)$.

\subsection{Stability of entanglement}\label{StabilityOfGS}

Let us now examine the stability of entanglement in graph states under the influence of decoherence. As we have seen, multi-particle entanglement is a central property for many practical applications in quantum information. For all these applications it is therefore of great interest, to determine the lifetime of entanglement when it is exposed to noise. A second motivation aims at a more fundamental direction. With nowadays available technologies it is possible to prepare and observe entanglement on microscopic scales; but it is also often argued that this task might become exceedingly difficult when considering a macroscopic number of particles. So this subsection addresses the question, whether multi-particle entanglement can be stable on a macroscopic level.
For the lifetime of entanglement it is not only necessary to specify the underlying decoherence model, but also the very notion of multi particle entanglement\footnote{The lifetime discussed in this chapter differs conceptually from the often used $T_1$- or $T_2$-decoherence (or -relaxation) rates in that the latter are rather related to the stability of quantum coherences and classical correlations rather than to the stability of entanglement. For some some interrelations between these notions see e.g. ref.~\cite{Tolkunov04}.} itself. This is mainly due to the fact that multi party entanglement is a subtle issue in quantum information theory (see e.g. \cite{Du99,multi-party}). Apart from some special cases, the existence of an entanglement measure that is satisfying for information theoretic purposes as well as applicable and calculable for mixed states, is still an open problem\footnote{We note that for systems with only a few number of qubits ($N\leq7$), quite recently \cite{Carvalho04} the effect of decoherence on GHZ- and W-states was studied in terms of an entanglement measure which is a generalization of the concurrence.}.
In the following we will therefore concentrate on the discussion of two qualitative entanglement criteria. Throughout the chapter we will consider $N$ two--level systems (qubits) with corresponding Hilbert space ${\cal H} =(\mathbb{C}^2)^{\otimes N}$. The $N$ particles are distributed among $N$ parties $1,\ldots ,N$. Starting with a pure GHZ or graph state we will consider the $N$-party separability and distillability properties of the decohered state $\rho(t)$ (see eq.~(\ref{decoh}))\index{lifetime of entanglement}:

On the one end of the scale the state $\rho(t)$ can still be
 {\em $N$-party distillable entangled}\index{distillable!N-party entangled}, as it is the case for the corresponding pure states in question. Hereby we call $\rho(t)$  $N$-party distillable, if any
 other true $N$-party entangled state $|\Phi\rangle$ can be obtained (distilled) asymptotically from multiple copies of $\rho$ under local operations and classical communication (LOCC) \cite{Du99,Th02}:
\be\label{Ndistillable}
\rho^{\otimes k} \;\longrightarrow_{\text{LOCC}} \; |\Phi\rangle\langle \Phi|\; .
\ee
We remark that in the multi--copy case all true $N$--party entangled states are equivalent since they can be transformed into each other by LOCC. That is, the condition that any true $N$ party entangled state can be created can be replaced by the condition that some $N$--party entangled state, e.g. the initial pure state, can be created. Disregarding the practicability of the underlying distillation protocol, the state $\rho(t)$ is then as useful as any other entangled state and therefore can in principle be regarded as a universal resource for quantum information processing such as quantum communication.

On the other end of the scale, $\rho(t)$ might have also become completely separable or classical in the sense that it can be described by a classical mixture of product states, i.e., $\rho$ is  {\em $N$-party separable}\index{separable states!N-party (completely)}, if
\be\label{Nseparable}
 \rho (t) = \sum_k\, p_k\, \rho_k^{1}\otimes \rho_k^{2}\otimes \ldots \otimes \rho_k^{N} \; .
\ee
If a state is completely separable, it is no longer entangled with respect to any partitioning. In between these two extremal cases, $\rho(t)$ can contain different types of {\em blockwise entanglement}\index{blockwise entanglement}.
We can consider different partitionings of particles into $M$ groups ($M\leq N$), where each group forms a subsystem with a higher dimensional state space and consists of several particles. {\em $M$-party distillability [separability]}\index{separable states!M-party} can then be defined {\em with respect to a given partitioning} in a similar way, where the notion of {\em local} operation has to be adapted accordingly. We will call $\rho(t)$ {\em $M$-party distillable}\index{distillable!M-party entangled}, if there exists at least one partitioning, with respect to which $\rho(t)$ is $M$-party distillable.

Based on the notion of $M$--party separability and distillability, one can define the lifetime of entanglement. A $N$--party state $|\Psi\rangle\langle\Psi|$ which is subjected to decoherence for time $t$ evolves into a mixed state $\rho(t)$. The lifetime of $N$--party distillable entanglement is given by the time after which the state $\rho(t)$ looses the property of $N$--party distillability. This implies that lower bounds on the lifetime of distillable entanglement can be obtained by showing that the state $\rho(t)$ is distillable, while an upper bound can be obtained by proving non--distillability of $\rho(t)$. When considering partitions of the system into $M$ groups, the lifetime of $M$--party entanglement with respect to a given partition is defined accordingly. We refer to the lifetime of $M$--party entanglement as the time after which $\rho(t)$ is non--distillable with respect to {\em all} $M$--party partitions. In a similar way, one can define a lifetime with respect to the separability properties of $\rho(t)$.

In order to determine entanglement properties of the mixed states in question, we will continuously make use of the partial transposition criterion\index{partial transposition $\rho^{T_A}$|textbf} \cite{Peres96,Ho97}, an entanglement criterion which provides necessary conditions for distillability and separability. The partial transposition is defined for bi-partite systems only, while a system can in general consist of several parties. Making use of the concept of partitionings of the system, in particular considering all bi-partitionings, one can use the partial transposition criteria also for multi-partite states. Let $A$ denote a subset of $m$ parties $a_1, \ldots ,a_m$. In general, given an operator $X$ acting on $\mathbb{C}^{d_A}\otimes\mathbb{C}^{d_B}$, we define the partial transpose of $X$ with
respect to the first subsystem in the basis
$\{|1\rangle,|2\rangle,\ldots,|d_A\rangle\}$, $X^{T_A}$, as follows:
\be
X^{T_A} := \sum_{i,j=1}^{d_A}\sum_{k,l=1}^{d_B}
\langle i,k|X|j,l\rangle \; |j,k\rangle\langle i,l|.
\ee
A Hermitian operator $X$ has a non--positive [positive] partial transpose\index{PPT (non-negative partial transpose)}\index{NPT (negative partial transpose)}
(NPT) [(PPT)] if $X^{T_A}$ is not positive [positive] respectively. That is, $X^{T_A}$ is NPT if there exist some
$|\Psi\rangle$ such that $\langle\Psi|X^{T_A}|\Psi\rangle <0$.

The positivity of the operator $\rho^{T_A}$ gives a necessary criterion for separability\index{entanglement criterion!partial transposition}, whereas the non-positivity of $\rho^{T_A}$ is necessary for the distillability of the density operator $\rho$. In particular, if a bi-partite density operator is PPT, then it is certainly not  distillable \cite{Ho97}. This implies \cite{Du99} that if a multi-particle density operator $\rho$ is PPT with respect to at least one bi-partite partition, then $\rho$ is certainly not $N$--party distillable. On the other hand, positivity of all bi-partite partitions is a necessary condition for $N$--party separability.  In the case of two dimensional systems $\mathbb{C}^2\otimes\mathbb{C}^2$ the PPT [NPT] criterion is necessary {{\it and} sufficient for separability [distillability] \cite{Peres96,Ho96}. A detailed discussion of the application of the partial transposition criteria to multi-partite systems can be found in ref.~\cite{Du99}.

Although the computation of the spectrum for the partial transposition of a general density matrix requires a numerically demanding diagonalization procedure\footnote{Note that the size of the spectrum grows exponentially with the number of particles.}, for graph diagonal states this spectrum can be directly determined from the spectrum $(\lambda_U)_{U\subseteq V}$ of the graph diagonal state itself \cite{Du04b}:
{\proposition[{\bf Partial transposition for graph diagonal states}]\index{partial transposition $\rho^{T_A}$}\label{PTofGSLemma}
 For any graph diagonal state $\rho$ (\ref{RhoG}) the {\em partial transposition} $\rho^{T_A}$ with respect to some partition $A$ is again diagonal in the (same) graph state basis $|U\rangle_G$.
In order to compute the corresponding eigenvalues, let $\mathbf{\Gamma}'=\mathbf{\Gamma}_{AA^c}$ denote the adjacency matrix of the graph between the partition $A$ and its complement $A^c$ (see eq.~(\ref{Gamma for bi-partition})).
Then:
\begin{eqnarray}
\label{PTofRhoG}
  \rho^{T_A}  & = &  \sum_{U \subseteq V} \, \lambda'_U \, |U \rangle_G\langle U|  \; \; \text{with}\\
\lambda'_U & = &  \frac{|\text{ker}\,\mathbf{\Gamma}'|}{2^{|A|}}\, \sum_{\genfrac{}{}{0pt}{}{(X,Y) \in }{ (\text{ker}\,\mathbf{\Gamma}')^{\bot} \times (\text{Im}\,\mathbf{\Gamma}')}} \,
(-1)^{\langle X , A_Y  \rangle} \, \lambda_{\left( U + X + Y\right)} \nonumber \; ,
\end{eqnarray}
where $A_Y \in A$ is any set with $\mathbf{\Gamma}' A_Y = Y$, and the kernel $\text{ker}$ or the orthocomplement $\bot$ are taken with respect to the subspace $\mathcal{P}(A)$ spanned by the sets in $A$.
}

Let us give two examples for formula (\ref{PTofRhoG}):
If $\mathbf{\Gamma}'$ is invertible, then $\text{ker}\, \mathbf{\Gamma}' =\{0\}$ and $(\text{ker}\, \mathbf{\Gamma}')^\bot =\mathcal{P}(A)$ holds.
 Moreover $(\ref{PTofRhoG})$ can be simplified by parameterizing $\text{Im}\,\mathbf{\Gamma}'$ with $Y=\mathbf{\Gamma}' A_2$, where $A_2 \subseteq A$:
\be
\lambda'_U  =   \frac{1}{2^{|A|}}\, \sum_{A_1,A_2 \subseteq A} \,
(-1)^{\langle A_1 , A_2  \rangle} \, \lambda_{\left( U + A_1 + \mathbf{\Gamma}' A_2\right)} \; .\ee
If $A=\{a\}$ for a non-isolated vertex $a \in V$ the eigenvalues of the partial transposition with respect to $A$ are
\begin{equation}\label{PTofa}
\lambda'_U  =   \frac{1}{2}\, \left( \lambda_U + \lambda_{U + N_a} + \lambda_{U + a} - \lambda_{U + N_a + a}\right) \; .
\end{equation}
Similarly for the partial transposition with respect to the split $A=\{a,b\}$ versus rest, where $a,b \in V$  are two non-adjacent vertices
 with linearly independent neighbor sets $N_a$ and $N_b$, one obtains:
\begin{equation}\label{PTofab}
\lambda'_U   =   \frac{1}{4}\, \left( \sum_{X \in \mathcal{M}_+} \, \lambda_{U+X} - \sum_{X \in \mathcal{M}_-} \, \lambda_{U+X} \right) \; ,
\end{equation}
where
\begin{eqnarray}
\mathcal{M}_+ & = & \{ \emptyset, a, b, a+b, N_a, N_b, N_a+N_b, a+N_b,  b+N_a, a+b+N_a+N_b\}\; \;\text{and} \cr
\mathcal{M}_- &= &\{ a+ N_a,b+ N_b, a+N_a+N_b, b+N_a+N_b,  a+b+N_a, a+b+N_b\} \; .\nonumber
\end{eqnarray}
If $a$ and $b$ are adjacent the same formula holds but with neighbor sets $N'_a=N_a\setminus b$ and $N'_b=N_b\setminus a$ restricted to $A^c$.

This procedure to compute the eigenvalues of the partial transposition described in (iii) does not require the diagonalization of a $2^N\times2^N$-matrix and therefore allows the evaluation of the PPT criteria with respect to different partitions, as long as the vector consisting of the initial eigenvalues $\lambda_U$ (which is already of length $2^N$) is small enough to be stored and -in the case that it occurs as a result of Pauli channel- as long as this vector can also be computed fast enough.

\begin{wrapfigure}[15]{r}{0.5\textwidth}
\vspace{-0.1cm}
\includegraphics[width=0.5\textwidth]{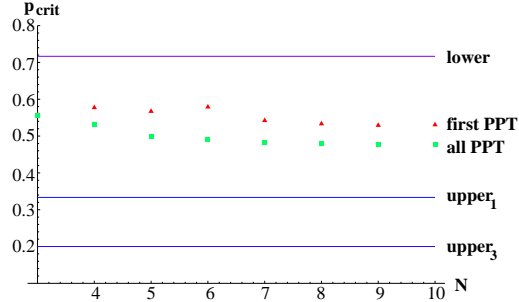}
\caption{\label{fig:DepolRing} \small For the case of particles in rings of size $N\leq 10$, which individually decohere according to the same depolarizing channel with parameter $p$:  the critical value $p_{\text{crit}}$, after which the first [last] partition becomes PPT $\triangle$ [$\Box$], the lower and upper bounds according to ref.~\cite{Du04b}.  }
\end{wrapfigure}
Consider, for illustration, rings up to size $N=10$ suffering from decoherence due to the depolarizing channel, which are examined with the help of the partial transpose with respect to all possible partitions. Fig.~\ref{fig:DepolRing} depicts the critical value for $p$, after which the state $\rho$ first becomes PPT with  respect to some partition, which implies that at this point the state $\rho$ is certainly no longer $N$-party distillable. For  fig.~\ref{fig:DepolRing} the critical value $p_{\text{crit}}$ has also been computed, after which the state $\rho$ has become PPT with respect to all partitions, i.e., after which $\rho$ contains at most
 bound entanglement with respect to any partition.  In contrast to the case of $N$-party GHZ states,
for which the one-versus-rest partition is the first to become PPT, the numerical results for small $N$ indicate that in rings this split seems to be most
stable against decoherence due to noise described by individual depolarizing channels, and that the smallest eigenvalue of the partial transposition with respect to these one-versus-rest splits $\{a\}$ is given by $\lambda_{N_a+a}$.

If the initial state is a GHZ state, e.g. in the star graph representation with center at qubit $1$, a direct decomposition with respect to the logical basis $(|0\rangle,|1\rangle)$ is sometimes more advantageous for the computation of the partial transposition and its eigenvalues.
In particular, a certain subclass of such GHZ diagonal states\index{GHZ diagonal state}\index{GHZ state}, namely the one where $\lambda_{U}=\lambda_{U+1}$ for all $U\subseteq V$ except $\emptyset$ and $\{1\}$, allows for a very precise determination\footnote{\label{GHZDfootnote}More precisely, starting with a pure star graph state $|G\rangle$ this class of states $\rho$ is given by \be \rho \;=\; \lambda_0 |G\rangle\langle G| + \lambda_{1} |1\rangle_G\langle 1| + \sum_{U\subseteq V\setminus 1} \lambda_U \left( |U\rangle_G\langle U| + |U+1\rangle_G\langle U+1| \right) \nonumber \ee with $\Delta=\lambda_0-\lambda_1\geq0$ and thus determined by $2^{N-1}$ parameters. In \cite{Du99} it is shown that (i) any mixed state $\rho'$ can be depolarized to this class of states $\rho$ by means of random local operations (leaving $\lambda_0\equiv \lambda'_\emptyset$, $\lambda_1\equiv \lambda'_{\{1\}}$ and all $\lambda_U=1/2 (\lambda_{U} +\lambda_{U+1})$ invariant), (ii) $\rho$ is {\em PPT with respect to the bi-partite split $(A,B)$}\index{PPT (non-negative partial transpose)} iff $\Delta\leq 2\lambda_{A\setminus 1}$, (iii)  $\rho$ is {\em separable with respect to the partition $(A_1,\ldots,A_M)$} iff all bi-partite splits $(A,B)\geq (A_1,\ldots,A_M)$ are PPT, (iv) a maximally entangled state is distillable between the pair $a$ and $b$ iff all bi-partite splits $(A,B)$ such that $a\in A$ and $b\in B$ are NPT\index{NPT (negative partial transpose)} and (v) a maximally entangled $M$-party GHZ state is distillable between the vertices $a_1,\ldots, a_M$ iff all bi-partite splits $(A,B)$, such that neither all $a_1,\ldots, a_M\in A$ nor all $a_1,\ldots, a_M\in B$, are NPT.} of its entanglement properties \cite{Du99}.

This class of GHZ-diagonal states is, for example, naturally obtained, if the qubits of the initial GHZ state are effected individually by white noise. In this case, most of the above entanglement properties of the decohered mixed state $\rho(t)$ can be determined analytically \cite{Si02,Du04b,Band04}:

{\proposition[\bf GHZ states in the presence of individual white noise]\hspace{4cm}

Consider a GHZ state that is exposed to decoherence described by individual depolarizing channels $\mathcal{D}_t$ with parameter $p(t)=e^{-\kappa t}$. Then the resulting mixed state $\rho(t)=\mathcal{D}_t^V\,|GHZ\rangle\langle GHZ|$ is $N$-party distillable [separable] iff the partial transpose $\left(\rho_G\right)^{T_A}$ w.r.t. all possible partitions $A$ are non-positive [positive]. More precisely,
\begin{itemize}
\item[1.] $\rho(t)$ remains {\em $N$-party distillable entangled} as long as the {\em most fragile} splits,\\ $1$-versus-{\small $(N-1)$} particles, remain NPT.
\item[2.] $\rho(t)$ becomes {\em $N$-party separable} as soon as the  {\em most stable} splits,\\ $\frac{\lfloor N \rfloor}{2}$-versus-$\frac{\lceil N \rceil}{2}$ particles, become PPT.
\item[3.] The {\em lifetime of $N$-party distillable entanglement} {\bf decreases} as the number of particles  $N \rightarrow \infty$.
\end{itemize}
The {\em $M$-party distillability [separability]} of the resulting mixed state $\rho(t)$ with respect to a fixed partitioning is determined by the subsystem that contains the smallest number of particles, since the corresponding partial transposition is the first one to become positive. In particular:
\begin{itemize}
\item[4.] The maximum number $M$ of subsystems that remain entangled, i.e., the effective size of {\em $M$-party distillable entanglement} {\bf decreases} in time.
\item[5.] As $N \rightarrow \infty$ any partitioning into groups of size $m$ leads to vanishing [finite]        {\em lifetime of the corresponding $M=\frac{N}{m}$-party entanglement}, if the size $m$ of each group is finite [tends to $\infty$].
\end{itemize}
{\em Encoding} the qubits of a GHZ state by some quantum error correcting code and performing error correction at the end of the noise process, the lifetime of [blockwise] entanglement between the {\em logical qubits} can be {\bf increased} effectively. Thus the lifetime $M$-party entanglement in encoded GHZ states can also remain finite on a macroscopic level as long as the level of encoding (e.g. concatenations) is sufficiently large.
}

Thus, we find that without error-correction the lifetime of $N$-party entanglement in GHZ-states vanishes on  a macroscopic level. This result can also be extended to more general decoherence models \cite{Du04b,Band04}. In the remainder of this subsection we now discuss the lifetime of $N$-party entanglement in graph states and show\footnote{An analysis of long-range entanglement in the thermal state of a three-dimensional cluster state can be found in ref.~\cite{Raussen04}.} that for a significant subclass, such as the cluster states, the lifetime of distillable entanglement is, in contrast to GHZ-states, essentially independent of $N$.

To this aim, we establish a lower bound on the lifetime of graph states by considering an explicit entanglement distillation protocol. The distillability of a $N$--party entangled state can be shown by providing a procedure that allows one to generate maximally entangled pairs shared between any pair of neighboring particles. This is sufficient, as any $N$--party entangled states can be then produced by local operations from these entangled pairs. The distillability of such pairs serves only as a tool to show $N$--party distillability, and no conclusions about the nature of entanglement contained in the state can be drawn. In particular, one should not conclude that entanglement contained in a cluster state were in some sense only bi-partite.

For decoherence of individual particles described by Pauli channels, one can make use of the following facts \cite{Du04b}: (i) measuring all but two particles $a,b$ in the eigenbasis of $\sigma_z$ results into the creation of another graph state with only a single edge $\{a,b\}$ (see Proposition~\ref{Pauli_Measurement} in sec.~\ref{Pauli measurements}); (ii) the action on a specific graph state of any operator $O$ which is a tensor product of Pauli operators can be equivalently described by an operator $O'$ consisting of only $\sigma_z$ operators acting on the same graph state. This implies that a Pauli--diagonal map ${\cal D}^k$ acting on qubit $k$ of a graph state $|G\rangle$ can be described by a map ${\cal M}$ whose Kraus operators contain only $\sigma_z$ operators acting on qubit $k$ and its neighbors. This follows from $\sigma_x^{a}|U\rangle_G = (-1)^{U_a}\sigma_x^{a} K_a |U\rangle_G$, where $S_x^{a} := \sigma_x^{a} K_a$ is an operator which contains only products of $\sigma_z$ operators at neighboring particles of particle $a$, and similarly for $\sigma_y$.

Hence measurements of $\sigma_z$ on all but particles $a,b$ commute with the action of maps describing the decoherence process. The resulting state $\rho_{ab}$ of particles $a,b$ is only influenced by noise acting on particles $a,b$ and their neighbors $N_a,N_b$. The measurements effectively decouple particles $a,b$ from the remaining particles. Distillability of $\rho_{ab}$ can be determined by employing the partial transposition criterion, where for a two qubit system negative partial transposition already ensures distillability. As $\rho_{ab}$ is only determined by noise operators acting on particles $a,b$ and their neighbors, this already implies that for all graph states with constant degree (e.g. cluster states), the distillability of $\rho_{ab}$ will {\em not} depend on the size of the system $N$. In fact, one finds a threshold value for parameters describing the decoherence process that only depends on the local degree of the graph. To be precise, the influence of independent neighbors and joint neighbors is slightly different. We remark that in certain cases, e.g. for GHZ states, the local degree itself may depend on $N$. If this is however not the case, the lower bound on the lifetime of distillable entanglement obtained in this way is constant and shows no scaling with the size of the system $N$.

\begin{wrapfigure}[16]{r}[0.1\textwidth]{0.5\textwidth}
\vspace{-0.4cm}
\includegraphics[width=0.5\textwidth]{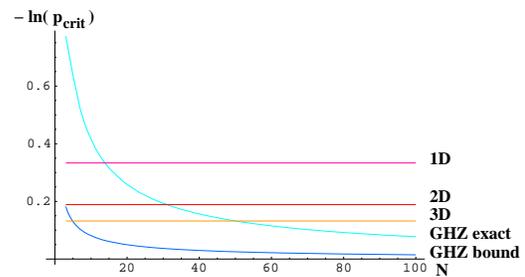}
\caption{\label{fig:Lower} \small Under individual coupling due to the same depolarizing channel the lower bounds on $\kappa t$ to the lifetime of distillable $N$-party entanglement for the 1D- , 2D-, 3D-cluster state remain constant for arbitrary system sizes $N$. For the $N$-party GHZ state the lower bound as well as the exact value for $\kappa t$, until which GHZ state remains distillable entangled, strictly decreases and goes to zero as $N\rightarrow \infty$. }
\end{wrapfigure}
For local depolarizing channel, one finds for instance that $\rho_{ab}$ is distillable entangled if $p^{|N_a|+1} +  p^{|N_a+N_b|} + p^{|N_b|+1} > 1$ \cite{Du04b}. Solving this polynomial inequality yields: for GHZ states:  $\kappa t > ln(2)/N$, in agreement with the analytic results for GHZ states; for 1D, 2D, 3D cluster states: $\kappa t > 0.3331, 0.1886, 0.1318$ respectively. The results are summarized in fig.~\ref{fig:Lower}.

The results can be extended to any noise model with some finite range, i.e., all Kraus operators act non--trivially only on a finite, localized number of qubits. Also in this case, for graph states corresponding to graphs with a constant degree there exists a lower bound on the lifetime of distillable entanglement that is finite and constant, independent of the total number of particles $N$.

Several {\em upper bounds} on the distillable entanglement in general graph states can also be provided \cite{Du04b}, which again do not depend on the size of the system as long as the maximal degree remains constant. These bounds are obtained by employing the partial transposition criterion, or by considering the noisy interactions resulting from interactions between adjacent particles in the graph and phase noise acting on them. In the second case, an upper bound on the lifetime is obtained by ensuring that the resulting CPMs are not capable to generate entanglement.  Considering blockwise entanglement a detailed analysis for general graph states has not been accomplished yet. Nevertheless, the scaling behavior of $M$-party entanglement  is restricted to a range between the upper and lower bounds, which in the case of cluster and similar graph states are independent of the number $N$ of particles. In this sense also the scaling behavior of blockwise entanglement in these states must be essentially independent of the size of the system.
Finally, most of the above results can also be extended to the class of weighted graph states (see sec.~\ref{WeightedGS}).

To summarize, we found that under quite general assumptions about the underlying noise model the lifetime of true  $N$-party-entanglement in GHZ-states decreases with the size of the system, whereas for cluster and similar graph states the lifetime of N-party entanglement is essentially  independent of the size of the system. These results suggest {\em a remarkable robustness of certain classes of macroscopic entangled states --namely all graph states with constant degree-- under various decoherence processes.}

\subsection{Entanglement purification}\label{EPP}\index{entanglement purification}

Although the results of the previous section suggest a robust scaling behavior of the entanglement in graph states with respect to system size, we nevertheless need methods to protect theses states from noise. Many applications make use of pure graph states, and hence a method to generate high--fidelity approximations to such pure graph states is of significant importance. The reasons that in a realistic situation one ends up with mixed states rather than pure states are manifold. For instance, the qubits constituting the graph state may interact with uncontrollable degrees of freedom of the environment, leading to decoherence; in distributed scenarios which have to be considered in the context of certain multi-party communication settings, the multi-particle entangled states are distributed through noisy quantum channels. In this section, we will discuss ways to maintain or recover the entanglement of such (noisy) graph states. In particular, we will describe {\em multi-party entanglement purification protocols} that allow one to create, from many identical copies of noisy entangled graph states, few copies with increased fidelity. The fidelity can, under the idealized assumption of perfect local control operations, be made arbitrarily close to one. Even in the presence of noisy local control operations, a significant enhancement of the fidelity is possible. The entanglement purification protocols show in fact a remarkable robustness against noise, where channel errors of the order of several tens of percent are correctible, and errors in local control operations of the order of percent are tolerable.

In ref.~\cite{Du03a} multi-particle entanglement purification
protocols for all two--colorable graph states have been developed.
These protocols are generalizations of a protocol for GHZ--state
purification introduced in ref.~\cite{Mu98} and further developed in
ref.~\cite{Ma01}. In the following we will briefly discuss the recurrence protocols
\index{recurrence protocol} for the purification of general
two--colorable graph states \index{two colorable graph states}. We remark that also hashing and breeding protocols are not discussed here,
which operate jointly on many copies and which allow for purification with
a finite yield were developed in refs.~\cite{Du03a, Lo04}. Given a noisy graph state $\rho'$
corresponding to a two colorable graph $G$, one can always transform
$\rho'$ to a standard form $\rho$ diagonal in the graph states basis
corresponding to $|G\rangle$ without changing the diagonal elements,
\begin{equation}
\rho= \sum_W \lambda_W |W\rangle\langle W|,
\end{equation}
where $|W\rangle =\sigma_z^{W}|G\rangle$ and $\lambda_W ={\rm tr} (|W\rangle\langle W| \rho')= {\rm tr} (|W\rangle\langle W| \rho)$ (recall that $\{|W\rangle\}$ form a basis). The transformation to this standard form is achieved by applying randomly the local operations corresponding to the correlation operators $K_a$ of $|G\rangle$ \cite{Du03a}. For notational convenience, we distinguish the two sets $A$ and $B$ corresponding to the two colors in the two--coloring of the graph, with indices $W_A$ and $W_B$, and identify $|W_A,W_B\rangle \equiv |W\rangle$.

We consider a recurrence purification protocol that consists of two sub--protocols, $P1$ and $P2$, each of which acts on two identical copies $\rho_1=\rho_2=\rho$, $\rho_{12}:=\rho_1\otimes \rho_2$. The basic idea consists in transferring (non--local) information about the first pair to the second, and reveal this information by measurements. In sub--protocol $P1$, all parties who belong to the set $A$ apply local CNOT operations  to their particles, with the particle belonging to $\rho_2$ as source, and $\rho_1$ as target. Similarly, all parties belonging to set $B$ apply local CNOT operations to their particles, but with the particle belonging to $\rho_1$ as source, and $\rho_2$ as target. The action of such a multilateral CNOT operation on two graph states can be determined within the stabilizer formalism and is given by \cite{Du03a}
\begin{eqnarray}
|W_A,W_B\rangle|V_A,V_B\rangle \longmapsto |W_A,W_B + V_B\rangle |V_A + W_A,V_B\rangle,
\end{eqnarray}
where $W_B + V_B$ again denotes bitwise addition modulo 2. A measurement of all particles of $\rho_2$ follows, where the particles belonging to set $A$ [$B$] are measured in the eigenbasis $\{|0\rangle_x,|1\rangle_x\}$ of $\sigma_x$ [$\{|0\rangle_z,|1\rangle_z\}$ of $\sigma_z$] respectively. The measurements in sets $A$ [$B$] yield results $(-1)^{\xi_a}$ [$(-1)^{\zeta_b}$], with $\xi_a,\zeta_b \in\{0,1\}$. Only if the measurement outcomes fulfill $(\xi_a+\sum_{b\in N_a}\zeta_b){\rm mod}2=0 ~\forall a$ \,--which implies $W_A + V_A={\bf 0}$--\, the first state is kept. In this case, one finds that the remaining state is again diagonal in the graph--state basis, with new coefficients
\begin{eqnarray}
\tilde\lambda_{U_A,U_B}=\sum_{\{(W_B,V_B)|W_B+ V_B = U_B\}}\frac{1}{2K} \lambda_{U_A,W_B}\lambda_{U_A,V_B},
\end{eqnarray}
Here $K$ is a normalization constant such that ${\rm tr}(\tilde \rho)=1$, indicating the probability of success of the protocol.
In sub-protocol $P2$ the role of sets $A$ and $B$ are exchanged, and a similar expression for the action of the protocol on initial graph diagonal states can be derived \cite{Du03a}.  The total purification protocol consists in a sequential application of sub--protocols $P1$ and $P2$. While sub--protocol $P1$ serves to gain information about $W_A$, sub--protocol $P2$ reveals information about $W_B$. Typically, sub--protocol $P1$ increase the weight of  all coefficients $\lambda_{{\bf 0},W_B}$, while $P2$ amplifies coefficients $\lambda_{W_A,{\bf 0}}$. In total, this leads to the desired amplification of $\lambda_{{\bf 0},{\bf 0}}$, where the fixed point of the protocol is $\lambda_{{\bf 0},{\bf 0}}=1$, i.e., iterative applications lead to the distillation of states with fidelity arbitrarily close to unity.

A numerical analysis of the purification regime of these recurrence protocols was performed in ref.~\cite{Du03a}. Results analogous to the robustness of entanglement under decoherence are found for the purification regime of the protocols. That is, for GHZ states the acceptable amount of channel noise (per particle), such that purification is still possible, decreases with increasing particle numbers. For cluster states, or more generally all two colorable graph states with constant degree, this threshold value turns out to be independent of the particle number $N$. Channel errors of the order of several tens of percent are correctable.
The applicability and performance of the protocol was also analyzed when taking noisy local control operations into account. In this case, no maximally entangled states can be produced, but still the fidelity of the states can be increased provided the local operations are not too noisy.  The reachable fidelity and fixed point is determined by the noise in local control operations. Again, for GHZ state the threshold value for local control operations such that purification is possible is more stringent for higher particle numbers, while errors of the order of several percent are tolerable in the case of cluster--state purification of arbitrary size \cite{Du03a}. This remarkable robustness of entanglement purification schemes opens the way for applications based on high fidelity graph states in real world scenarios.

\subsection{Multipartite secure state distribution}\index{secure state distribution}

Based on the multi-party entanglement purification protocol described in the previous section, a modification of the scheme was found that allows one to purify two--colorable graph states in such a way that all (but one) of the involved parties do not know ---and have no means to learn--- which state they are trying to purify. Such a process is the basic tool to achieve secure state distribution (see \cite{Du05c}) and constitutes a novel quantum primitive with possible applications in distributed secure applications. The goal is to generate a high fidelity approximation to a secretly chosen graph state among spatially separated agents. This is done by distributing multi-party entangled states from a central station via noisy quantum channels to local agents, and increasing the fidelity of the states via entanglement purification. It is important that at no stage of the protocol, the involved local agents should be able to learn about the identity of the state they are processing, and at the end they should be able to hand over a high fidelity approximation of this state to end--users (who have placed the order for this state). The secrecy in this context is of particular importance, as different graph states represent different resources for security (and other) applications, and the end--user may want to keep the ressources unknown to the other parties or to potential eavesdroppers.

The basic idea of the modification is to keep the identity of the state at any stage of the protocol secret, and generate effectively states that are --from the point of view of the local agents-- described by completely mixed states. To this aim, random local basis changes using Pauli operations are independently applied to the individual copies before distributing the states and hence before purification starts. This ensures not only that initial states are randomized (recall that already application of $\sigma_z$ operations at different locations produces all possible basis states), but also that measurement outcomes and measurement statistics are randomized. The purification protocol can be adopted to account for these additional basis changes. There are a number of technical details regarding the exact protocol to ensure unconditional security, and we refer the interested reader to ref.~\cite{Du05c} for details. We remark that two alternative ways to achieve the secure distribution of two--colorable graph states were also proposed in ref.~\cite{Du05c}, one based on the purification of Bell pairs with subsequent teleportation, the other based on the purification of enlarged graph states where randomization (and secrecy) is achieved by additional entanglement with some central station. The three different protocols are applicable in different regimes, where generally direct multi-particle entanglement purification turns out to allow for higher target fidelities than schemes based on bi-partite purification.

\section{Summary}

Graph states form a rich class of entangled states that exhibit
key aspects of multi-partite entanglement. At the same time, they
can be described by a number of parameters that grows only
moderately with the system size. They have a variety of
applications in quantum information theory, most prominently as
algorithmic resources in the context of the one-way quantum
computer, but also in other fields such as quantum error
correction and multi-partite quantum communication, as well as in
the study of foundational issues such as non-locality and
decoherence.

In this review, we have given a tutorial introduction into the
theory of graph states. We have introduced various equivalent ways
how to define graph states, and discussed the basic notions and
properties of these states. The focus of this review has been on
their entanglement properties. These include aspects of
non-locality, bi-partite and multi-partite entanglement and its
classification in terms of the Schmidt measure, the distillability
properties of mixed entangled states close to a pure graph state,
as well as the robustness of their entanglement under decoherence.
We have also reviewed some of the known applications of graph
states, as well as proposals for their experimental
implementation. Some of the latter material, specifically about
implementations, should thus be taken as preliminary and
reflecting only the current state of research.

\section*{Acknowledgements}

We would like to thank a number of colleagues for fruitful
discussions in the context of this work during the past
few years. The (most likely incomplete) list of people includes
S. Anders, H. Aschauer, S. Benjamin, J. Calsamiglia, D. Browne, J.~I. Cirac,
J. Dehaene, D. Gross, O. G\"uhne, E. Hostens, L. Hartmann,
A. Miyake, M.~B. Plenio, E. Rains, D. Schlingemann, S. Perdrix,
T. Stace, G. T\'oth, F. Verstraete, and P. Zoller,
as well as other people with whom we had the pleasure
to exchange our ideas on conferences or meetings.

This work was supported by the Austrian Science Foundation (FWF); by the
European Union through projects QUPRODIS, PROSECCO, OLAQUI, SCALA and QAP;
by the \"Osterreichische Akademie der Wissenschaften through project APART;
by the Deutsche Forschungsgemeinschaft (DFG); by EPSRC; by the Microsoft
Research Foundation;  by the European Research Councils (EURYI Scheme);  by
MURI under grant No.~DE-FG03-92-ER40701; by the NSF under contract number
PHY-0456720; and by the Flemisch fund for scientific research
(FWO) through projects G.0120.03 and G.0452.04.


\addcontentsline{toc}{section}{\protect\numberline{} References}

\end{document}